\documentclass[journal]{IEEEtran}

\usepackage{graphicx, subfig}
\usepackage{multirow}
\usepackage{hyperref}
\hypersetup{colorlinks,linkcolor=blue}
\usepackage{longtable}
\usepackage{color}
\usepackage{enumerate}
\usepackage{mdwlist}
\usepackage{times}
\usepackage{comment}
\usepackage{mathtools}
\usepackage{amsmath}
\usepackage{amssymb}
\usepackage{tabulary}
\usepackage{booktabs}
\usepackage{ulem}
\usepackage{algorithmic, algorithm}
\usepackage{float}
\usepackage{verbatim}
\usepackage[sort, numbers]{natbib}
\usepackage{ulem}
\usepackage{caption}

\newtheorem{lemma}{Lemma}
\newtheorem{corollary}{Corollary}

\newtheorem{theorem}{Theorem}

\newtheorem{remark}{Remark}
\newtheorem{assumption}{Assumption}

\newcommand*{\scalefactorone}{0.32}

\newcommand*{\scalefactortwo}{0.3}

\newcommand{\INDSTATE}[1][1]{\STATE\hspace{-#1\algorithmicindent}}

\DeclarePairedDelimiter{\ceil}{\lceil}{\rceil}

\makeatletter
\newcommand{\setword}[2]{%
  \phantomsection
  #1\def\@currentlabel{\unexpanded{#1}}\label{#2}%
}
\makeatother

\newcommand\oprocendsymbol{\hbox{$\square$}}
\newcommand\oprocend{\relax\ifmmode\else\unskip\hfill\fi\oprocendsymbol}

\begin{document}
\title{A Distributed Version of the Hungarian Method\\ for Multi-Robot Assignment}

\author{Smriti Chopra,
        Giuseppe Notarstefano,
        Matthew Rice,
        and Magnus Egerstedt%
\thanks{Smriti Chopra, Matthew Rice and Magnus Egerstedt are with the Department
of Electrical and Computer Engineering, Georgia Institute of Technology, Atlanta,
Georgia, USA; {\tt\small Emails: smriti.chopra@gatech.edu, mrice32@gatech.edu,
  magnus@ece.gatech.edu.} 
Giuseppe Notarstefano is with the Department of Engineering, Universit del
Salento, Lecce, Italy; {\tt\small Email: giuseppe.notarstefano@unisalento.it}

This result is part of a project that has received funding from the European
Research Council (ERC) under the European Union's Horizon 2020 research and
innovation programme (grant agreement No 638992 - OPT4SMART).
}
}%
\maketitle

\begin{abstract}
  In this paper, we propose a distributed version of the Hungarian Method to
  solve the well known assignment problem. In the context of multi-robot
  applications, all robots cooperatively compute a common assignment that
  optimizes a given global criterion (e.g. the total distance traveled) within a
  finite set of local computations and communications over a peer-to-peer
  network. As a motivating application, we consider a class of multi-robot
  routing problems with ``spatio-temporal'' constraints, i.e. spatial targets
  that require servicing at particular time instants. As a means of demonstrating the theory developed in this paper, the robots cooperatively find online, suboptimal routes by applying an iterative
  version of the proposed algorithm, in a distributed and dynamic setting. As a concrete
  experimental test-bed, we provide an interactive ``multi-robot orchestral''
  framework in which a team of robots cooperatively plays a piece of music 
  on a so-called orchestral floor.
\end{abstract}

\section*{Nomenclature}

In the following table, we present the nomenclature used in this paper.

\begin{table}[htbp]\caption*{Table of Notation}
\centering %
\begin{tabular}{r c p{6.5cm} }
\toprule
$V$ & $\triangleq$ & Vertex partitioning of two disjoint sets $R$ (robots) and $P$ (targets) respectively, denoted by $V = (R,P)$\\   
$E^i_{orig}$ & $\triangleq$ & Edge set containing the edges between a robot $i \in R$ and every target in $P$\\   
$w^i_{orig}$ & $\triangleq$ & Weight function corresponding to the edge set $E^i_{orig}$\\   
$G^i_{orig}$ & $\triangleq$ & Robot $i$'s original information, denoted by the bipartite weighted graph $G^i_{orig} = (V, E^i_{orig}, w^i_{orig})$\\   
$E^i_y$ & $\triangleq$ & Robot $i$'s equality subgraph edges\\   
$E^i_{cand}$ & $\triangleq$ & Robot $i$'s candidate edges\\   
$E^i_{lean}$ & $\triangleq$ & Edge partitioning of two disjoint sets $E^i_y$ and $E^i_{cand}$ respectively, denoted by $E^i_{lean} = (E^i_y, E^i_{cand})$\\   
$w^i_{lean}$ & $\triangleq$ & Weight function corresponding to the edge set $E^i_{lean}$\\   
$G^i_{lean}$ & $\triangleq$ & Robot $i$'s information, denoted by the bipartite weighted graph $G^i_{lean} = (V, E^i_{lean}, w^i_{lean})$\\
$y^i$ & $\triangleq$ & Robot $i$'s vertex labeling function\\
$\gamma^i$ & $\triangleq$ & Robot $i$'s counter value\\
$\mathbb{G}^i$ & $\triangleq$ & Robot $i$'s full state, given by $\mathbb{G}^i = (G^i_{lean}, y^i, \gamma^i)$\\   
$M^i$ & $\triangleq$ & Robot $i$'s Maximum Cardinality Matching \\   
$V^i_c$ & $\triangleq$ & Robot $i$'s MInimum Vertex Cover\\   
\bottomrule
\end{tabular}
\end{table}

\section{Introduction}
Assignment problems are an integral part of combinatorial optimization, with
wide applicability in theory as well as practice \cite{{koopmans:ejes1957a},
  {papadimitriou:1998a}, {pentico:ejor2007a}, {burkard:2009a}}. Various techniques have been proposed for solving such problems (see \cite{{bertsekas:mp1981a},{derigs:aor1985a}, {balinski:or1985a}} for early references).
For scenarios involving multiple mobile robots, assignment problems often
comprise of finding a one-to-one matching between robots and tasks, while
minimizing some assignment benefit. Moreover, a frequent requirement is the need for a distributed framework, 
since an infrastructure that supports a centralized
authority is often not a feasible option (prohibitively high cost for global computation
and information). It is preferable
that robots coordinate with one another to allocate and execute individual
tasks, through an efficient, distributed mechanism - a feat often challenging
due to the limited communication capabilities and global knowledge of each
robot. 

In this paper, we address such assignment problems with \textit{linear} objective functions, formally called Linear Sum Assignment Problems (LSAPs) \cite{burkard:1999a}, under a distributed setting in which robots communicate locally with ``adjacent
neighbors'' via a dynamic, directed information-exchange network. Among centralized algorithms, the Hungarian
Method \cite{kuhn:nrl1955a} was the first to compute an optimal solution to the LSAP in
finite time, and as such, forms the basis of our proposed distributed algorithm.

In cooperative robotics, assignment problems often form building blocks for more complex tasks, and have been widely investigated in the literature \cite{{mataric:ar2003a}, {gerkey:ijrr2004a}, {ji:2006a},{smith:2007a}}.
In particular, auction-based (market) algorithms are a very popular approach towards task assignments (see \cite{dias:pi2006a} for a survey). Such algorithms require robots to bid on tasks, rendering them
more or less attractive based on the corresponding prices computed
\cite{{bertsekas:i1990a}, {gerkey:rait2002a}}.
A generic framework and a variety of bidding rules for auction-based multi-robot
routing have been proposed in \cite{lagoudakis:2005a}.
An auction algorithm for dynamic allocation of tasks to robots in the presence
of uncertainties and malfunctions has been proposed and tested in
\cite{nanjanath:ras2010a}.
Auction algorithms, though computationally efficient, usually require a
coordinator or shared memory.
In \cite{zavlanos:2008a}, the authors develop an auction algorithm without such
constraints, and apply it towards multi-robot coordination in \cite{michael:2008a}. In particular, the agents obtain updated prices, required for accurate bidding, in a multi-hop fashion using only local information. The authors prove that the algorithm converges to an assignment that maximizes the total assignment benefit within a linear approximation of the optimal one, in $\mathcal{O} (\Delta n^3 \ceil*{{\frac{\max\{c_{i.j}\} - \min\{c_{i.j}\}}{\epsilon}}})$ iterations.
In \cite{choi:rit2009a}, a market-based decision strategy is proposed for decentralized task selection, and a consensus routine, based on
local communications, is used for conflict resolutions and agreement on the winning bid values.
In \cite{luo:2011a}, an auction algorithm is proposed to provide an almost optimal solution to the
assignment problem with set precedence constraints. The algorithm is first
presented in a shared-memory scenario, and later extended via a consensus
algorithm, to a distributed one.

Game-theoretic formulations for solving vehicle-target assignment problems are discussed \cite{{arslan:jdsmc2007a}, {marden:or2013a}}, where robots are viewed as self-interested decision makers, and the objective is to optimize a global utility function through robots that make individually rational decisions to optimize their own utility functions.
Among other decentralized techniques, coordination algorithms for task allocation that use only local
sensing and no direct communication between robots have been proposed in
\cite{lerman:ijrr2006a}. 
Additionally, consensus based approaches that typically
require the robots to converge on a consistent situational awareness before
performing the assignment have been explored in \cite{{alighanbari:2005a}, {tahbaz-salehi:2006a}, {dionne:2007a}}. Though such methods are robust, they are typically
slow to converge, and require the transmission of large amounts of data.
Distributed methods that solve linear programs, for instance, can also be
employed towards solving assignment problems \cite{{dutta:2008a}, {burger:a2012a}}, though they are computationally expensive, especially in
comparison to more streamlined algorithms, developed for the purpose of solving
assignment problems. 

In \cite{{giordani:2010a}, {giordani:cie2013a}}, the authors propose a distributed version of the Hungarian Method, similar to the contribution in this paper. They show that their algorithm converges in $\mathcal{O}(n^3)$ cumulative time, with $\mathcal{O}(n^3)$ number of messages exchanged among the robots, and no coordinator or shared memory. In particular, their algorithm involves root robots that (i) initiate message exchange with other robots in the network via a depth-first search (DFS), and (ii) synchronize the decision rounds (iterations, each containing multiple communication hops) across all robots.

The main distinctive feature of this paper is the redesign of the popular (centralized) Hungarian Method, under a distributed computation model, characteristic of traditional multi-robot applications - where every iteration of an algorithm ideally involves multiple anonymous agents performing two tasks: 1) exchanging information with their neighbors (via single-hop communication), and 2) executing identical computation routines. Our primary objective is not to improve convergence speeds, or information overheads of existing methods, but to remain comparable while providing a novel, distributed implementation of the centralized method. We prove that our algorithm converges in $\mathcal{O}(n^3)$ iterations, and show through simulation experiments with varying problem sizes, that the average convergence is much faster in practice, thus making our algorithm relevant to current literature.

The contribution of this paper is twofold: (i) As the main contribution, we
develop a distributed version of the Hungarian Method to enable
a team of robots to cooperatively compute optimal solutions to task
assignment problems (LSAPs), without any coordinator or shared memory. Specifically, each
robot runs a local routine to execute ad-hoc sub-steps of the centralized Hungarian
Method, and exchanges estimates of the solution with neighboring
robots. We show that in finite time (or in a finite number of communication rounds $\mathcal{O}(r^3)$ if executing synchronously, with $r$ being the total number of robots in the system), all robots
converge to a {\it common} optimal assignment (the LSAP can have multiple optimal solutions). Through simulation experiments over varying problem sizes, we characterize the average number of iterations required for convergence, as well as the computational load per robot. 
(ii) We demonstrate our proposed algorithm by extending it towards a class of
``spatio-temporal'' multi-robot routing problems previously introduced in
\cite{chopra:a2015a, chopra:pacc2014a}, now considered under a distributed and dynamic setting.
In essence, the robots find online, sub-optimal routes by solving a sequence of assignment problems iteratively, using the distributed algorithm for each instance. As a motivating application and concrete experimental test-bed, we develop the ``multi-robot
orchestral'' framework, where spatio-temporal routing is musically interpreted as ``playing a series of notes at particular time instants'' on a so-called orchestral floor (a music surface where planar positions correspond to distinct notes of different instruments). Moreover, we allow a user to act akin to a ``conductor'', modifying the music that the robots are playing in real time through a tablet interface. Under such a framework, we demonstrate the theory developed in this paper through simulations and hardware experiments.  

The remainder of this paper is organized as follows: In Section \ref{sec_centr_assign}, we briefly review the assignment problem, and the Hungarian Method used for solving it. In Section \ref{sec_prob_def}, we set-up the distributed version of the assignment problem central to this paper, while in Section \ref{sec_dist_algo}, we provide a description of our proposed algorithm. We discuss convergence and optimality in Section \ref{sec:convergence}, followed by the motivating application of spatio-temporal multi-robot routing in Section \ref{sec_motiv_eg}. 

\section{A review of the Linear Sum Assignment Problem and the Hungarian Method}
\label{sec_centr_assign}
In this section, we consider the Linear Sum Assignment Problem (LSAP) under a centralized setting \cite{burkard:1999a}, before we delve into its proposed distributed counterpart. We revisit some key definitions and theorems, used to express the general form of the {LSAP} in graph theoretic terms, and to understand the Hungarian Method employed for solving it.  

\begin{enumerate}[-]
\item{\textbf{Bipartite Graph:} A graph $G = (V,E)$, where the vertex set $V$ is decomposed into two disjoint sets of vertices $R$ and $P$ respectively, such that no two vertices in the same set are adjacent. In general, we say that the graph $G$ has bipartition $(R,P)$.}
\item{\textbf{Matching:} A set of edges without common vertices}.

\item{\textbf{Maximum Cardinality Matching:} A matching that contains the largest possible number of edges.}

\item{\textbf{Vertex Cover:} A set of vertices such that each edge is incident on at least one vertex of the set.}

\item{\textbf{Minimum Vertex Cover:} A vertex cover that contains the smallest possible number of vertices.}
\end{enumerate}

\begin{remark}
\label{note_1}
In a bipartite graph, the number of edges in a maximum cardinality matching
equals the number of vertices in a minimum vertex cover (by Konig's theorem
\cite{konig:a}). In fact, due to this inter-relation between a matching and a
vertex cover, algorithms used for finding a maximum cardinality matching ${M}$
(e.g. Hopcroft-Karp \cite{hopcroft:sjc1973a}), can be extended to finding a
corresponding minimum vertex cover $V_c\subset V$. \oprocend 
\end{remark}

\subsection{The Linear Sum Assignment Problem}
\label{LSAP}
Using the definitions presented above, we proceed to review the formal, graph theoretic interpretation of the {LSAP}.

\subsection*{\textbf{Minimum Weight Bipartite Matching Problem (P)}} 
\textit{``Given a graph $G = (V, E)$ with bipartition $(R, P)$ and weight function $w : E \rightarrow \mathbb{R}$, the objective is to find a maximum cardinality matching ${M}$ of minimum cost, where the cost of matching ${M}$ is given by $c({M}) = \sum_{e\in {M}} w(e)$''}. 

Without loss of generality, we can assume that $G$ is \textit{complete}\footnote{by adding edges with prohibitively large weights denoted by $\mathfrak{M}$.}, i.e. there exists an edge between every vertex $i\in R$, and every vertex $j\in P$, and \textit{balanced}\footnote{by adding dummy vertices and associated 0-weight edges.}, i.e. $|R| = |P| = |V|/2$. Hence, a maximum cardinality matching ${M}$ is always a perfect matching, i.e. $|{M}| = |V|/2$. Next, we review the \textbf{dual} of the above problem:

\subsection*{\textbf{ Dual of Minimum Weight Bipartite Matching Problem (D)}} 

\textit{``Given a graph $ G = (V, E)$ with bipartition $(R, P)$, a weight function $w : E \rightarrow \mathbb{R}$, and a vertex labeling function $y:V\rightarrow \mathbb{R}$, the objective is to find a \textit{feasible} labeling of maximum cost, where a feasible labeling is a choice of labels $y$, such that $w(i,j) \geq y(i) + y(j) \,\forall \,(i,j)\in E$, and the cost of the labeling is given by $c(y) = \sum_{i\in R} y(i) + \sum_{j\in P} y(j)$''}.

Moreover, given a feasible labeling $y$, an \textbf{equality subgraph} $G _{y}= (V,E_{y})$ is defined as a subgraph of $G$ where,
\begin{align}
\label{eqn_equality_graph}
E_{y} = \{(i,j)\,|\,y(i) + y(j) = w(i,j)\}
\end{align}
and the \textbf{slack} of an edge $(i,j)$ is defined as,
\begin{align}
\label{eqn_slack}
\mathrm{slack}(w,y,i,j) = w(i,j) - (y(i) + y(j))
\end{align}

\subsection{The Hungarian Method}
\label{subsec_hungmethsteps}

Now that we have discussed the {Minimum Weight Bipartite Matching Problem}, as well as its corresponding dual, we review a key theorem that provides the basis for the Hungarian Method \cite{kuhn:nrl1955a}, the first primal-dual algorithm developed for solving the {LSAP}. 

\begin{theorem}[Kuhn-Munkres]
\label{thm_hung}
Given a bipartite graph $G = (V, E)$ with bipartition $(R, P)$, a weight
function $w : E \rightarrow \mathbb{R}_{\geq 0}$, and a vertex labeling function
$y:V\rightarrow \mathbb{R}$, let $M$ and $y$ be feasible ($M$ is a perfect
matching and $y$ is a feasible labeling). Then $M$ and $y$ are optimal if and
only if $M\subseteq E_y$, i.e. each edge in $M$ is also in the set of equality
subgraph edges $E_{y}$, given by (\ref{eqn_equality_graph}).\oprocend
\end{theorem}

From this point onwards, for notational convenience, we will denote the
\textit{weighted, bipartite graph} by $G = (V, E, w)$, i.e. a tuple consisting of the vertex set
$V$, the edge set $E$ and the corresponding edge weight function $w$.

\begin{figure}
\subfloat[An example of a weighted, bipartite graph $G= (V,E,w)$.]{
\includegraphics[scale=\scalefactorone]{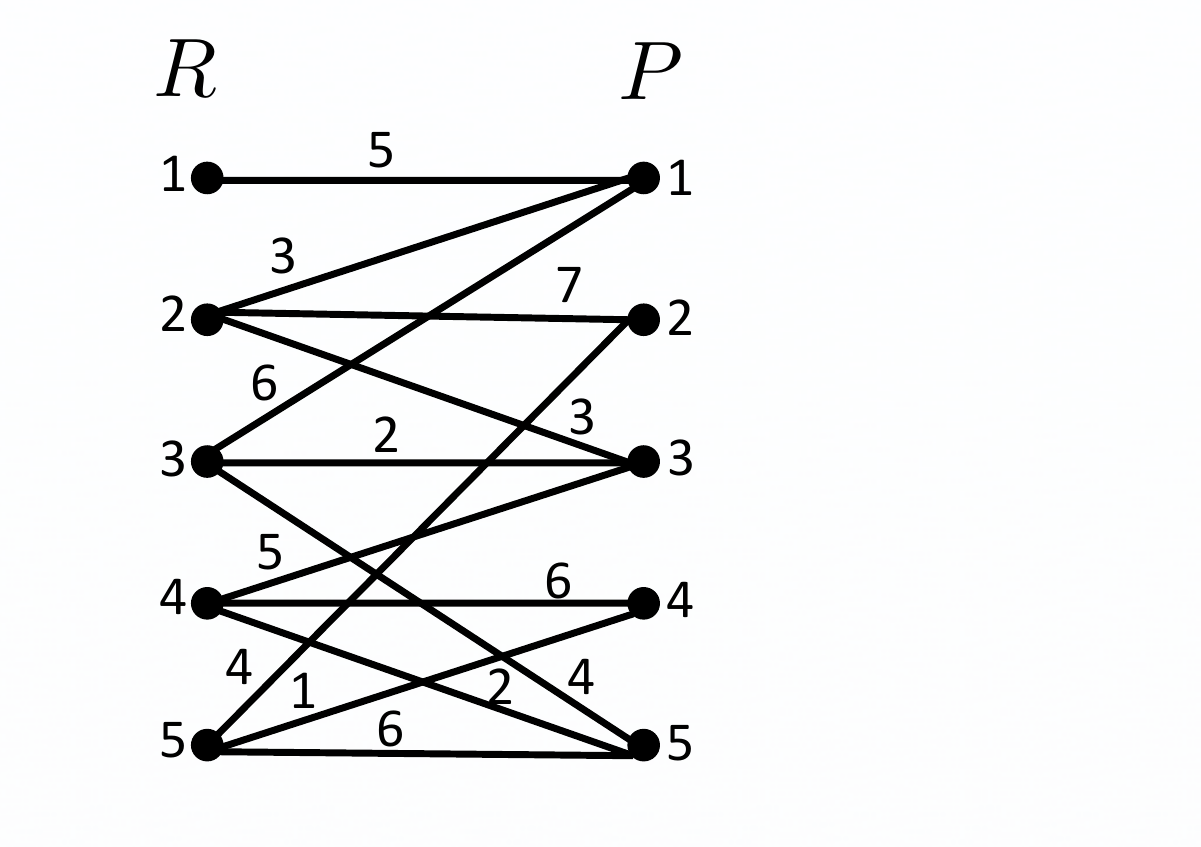}
\label{subfig_hung1}
}
\hfill
\subfloat[Initial feasible vertex labeling function\newline $y:V\rightarrow \mathbb{R}$ (highlighted in yellow).]{
\includegraphics[scale=\scalefactorone]{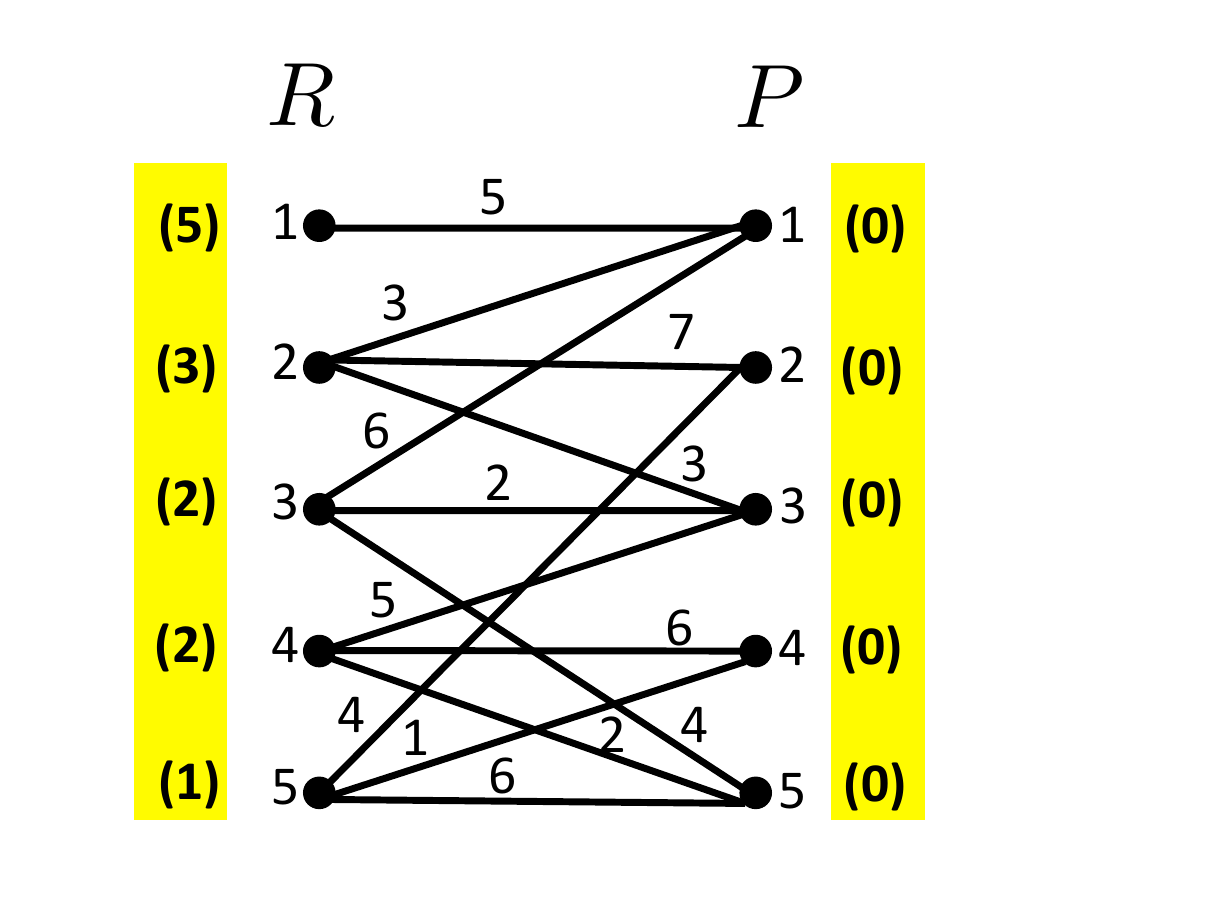}
\label{subfig_hung2}
}
\hfill
\subfloat[Given $y$, the corresponding set of equality subgraph edges $E_y$.]{
\includegraphics[scale=\scalefactorone]{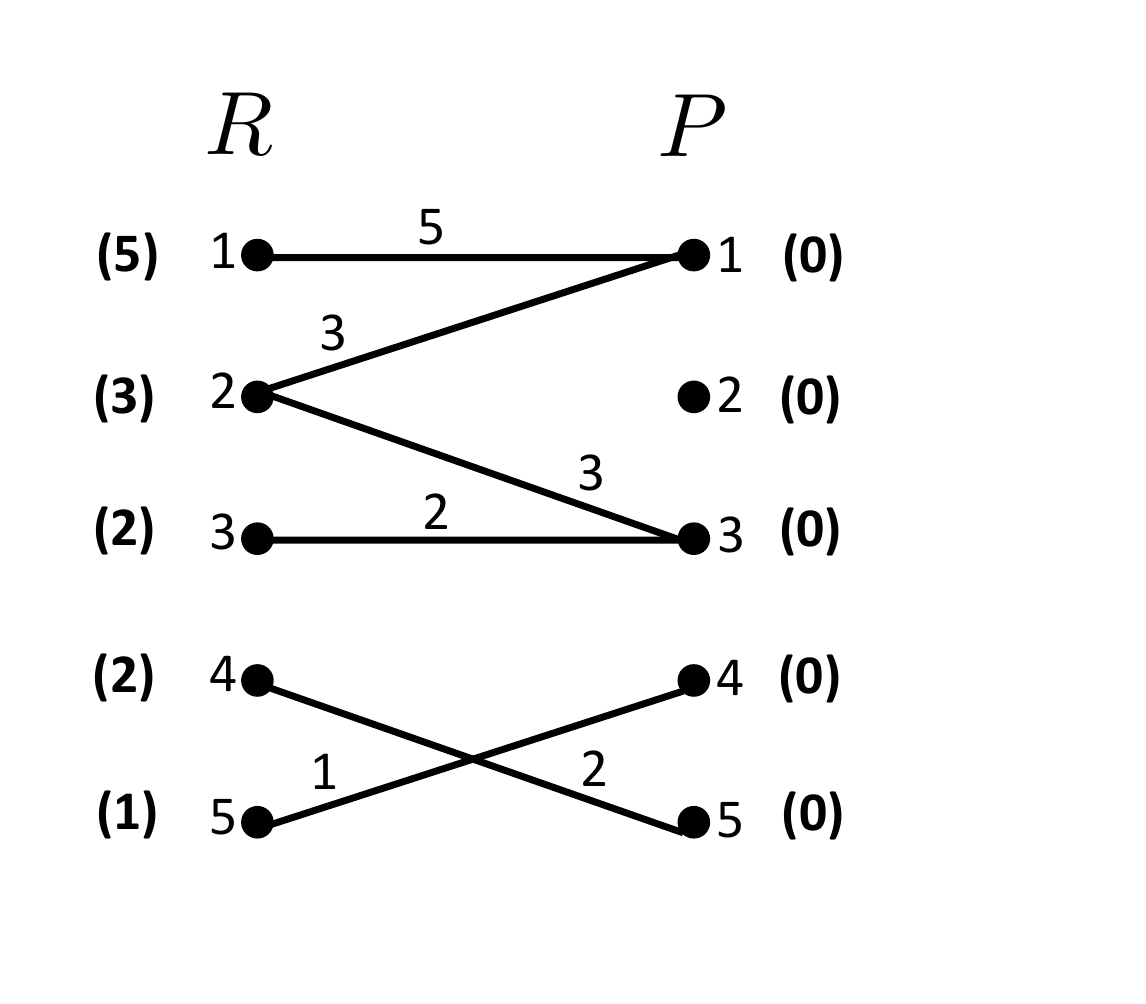}
\label{subfig_hung3}
}
\hfill
\subfloat[Given $E_y$, a maximum cardinality matching $M$ (red edges), and a corresponding minimum vertex cover $V_{c} = (R_{c}, P_{c})$ (red vertices).]{
\includegraphics[scale=\scalefactorone]{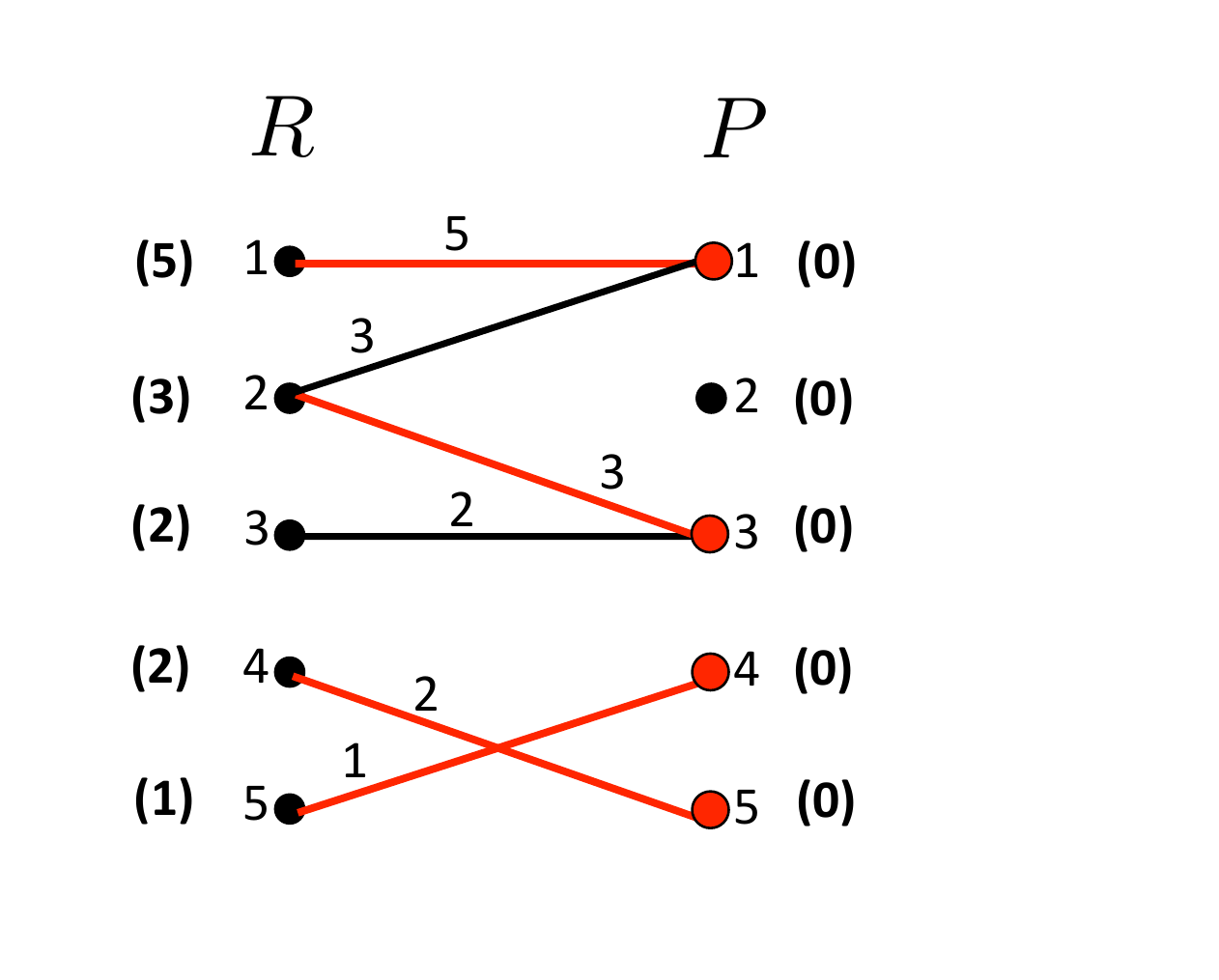}
\label{subfig_hung4}
}
\caption{An instance of the Initialization step of the $\mathrm{Hungarian\_Method}$.}
\label{fig_hunginit}
\end{figure}

\begin{figure}
\subfloat[ \ref{labelstep1}(a): For the given minimum vertex cover $V_{c}$, the isolated set of candidate edges (green edges). \ref{labelstep1}(b): The edge with minimum slack ($\delta$) is identified.]{
\includegraphics[scale=\scalefactorone]{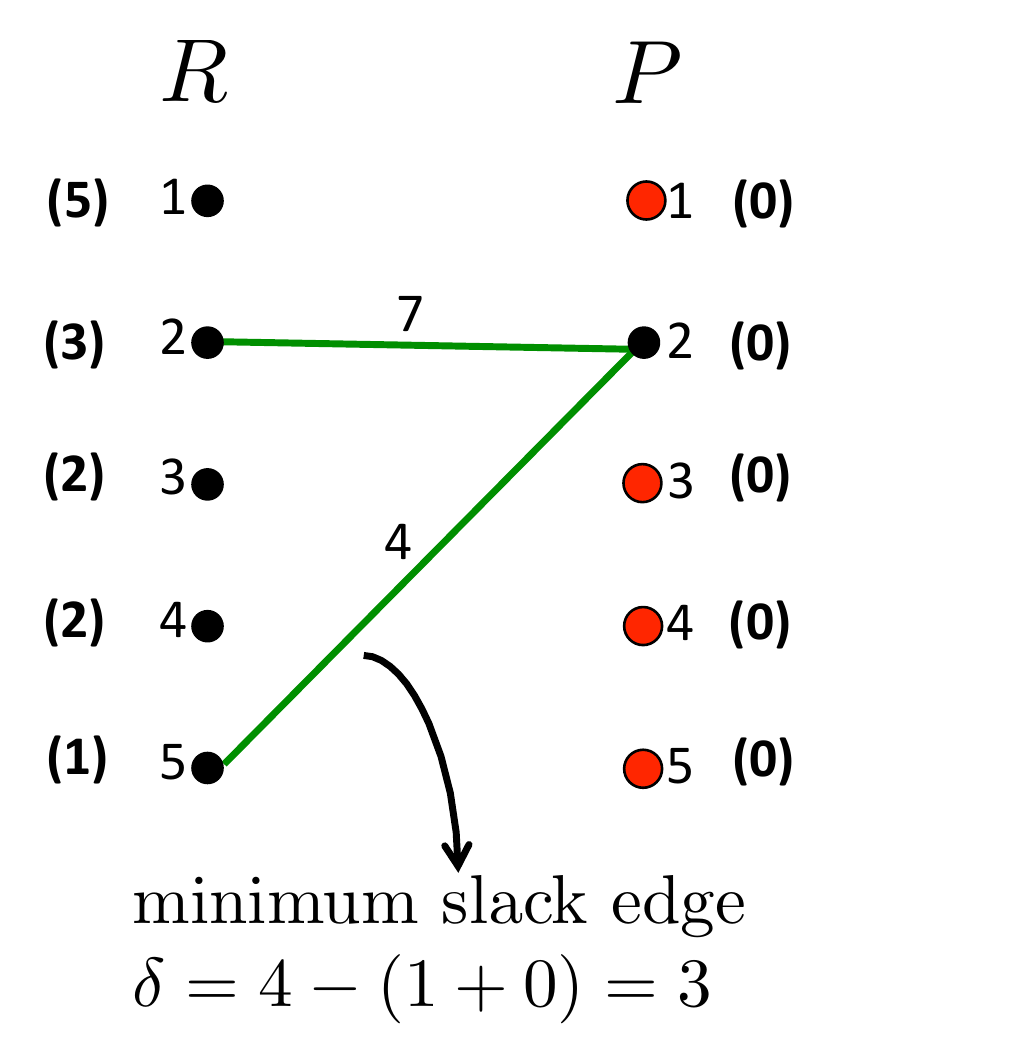}
\label{subfig_hungst11}
}
\hfill
\subfloat[ \ref{labelstep1}(b) contd.: The updated feasible vertex labeling function $y$ (highlighted in yellow), using the minimum slack $\delta$.]{
\includegraphics[scale=\scalefactorone]{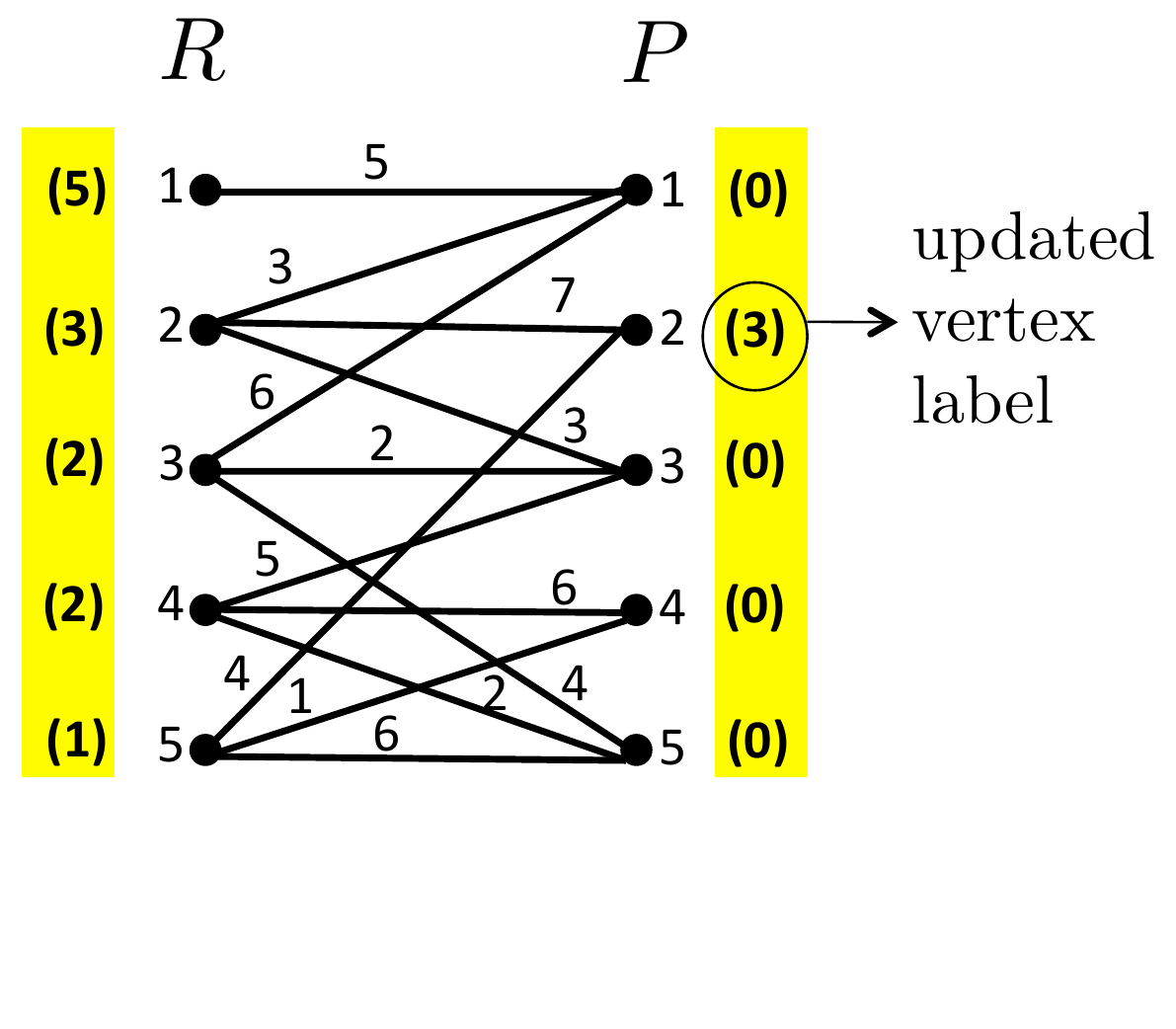}
\label{subfig_hungst12}
}
\hfill
\subfloat[ \ref{labelstep2}: For the updated $y$, the corresponding set of equality subgraph edges $E_y$ (with the new, added edge highlighted in yellow).]{
\includegraphics[scale=\scalefactorone]{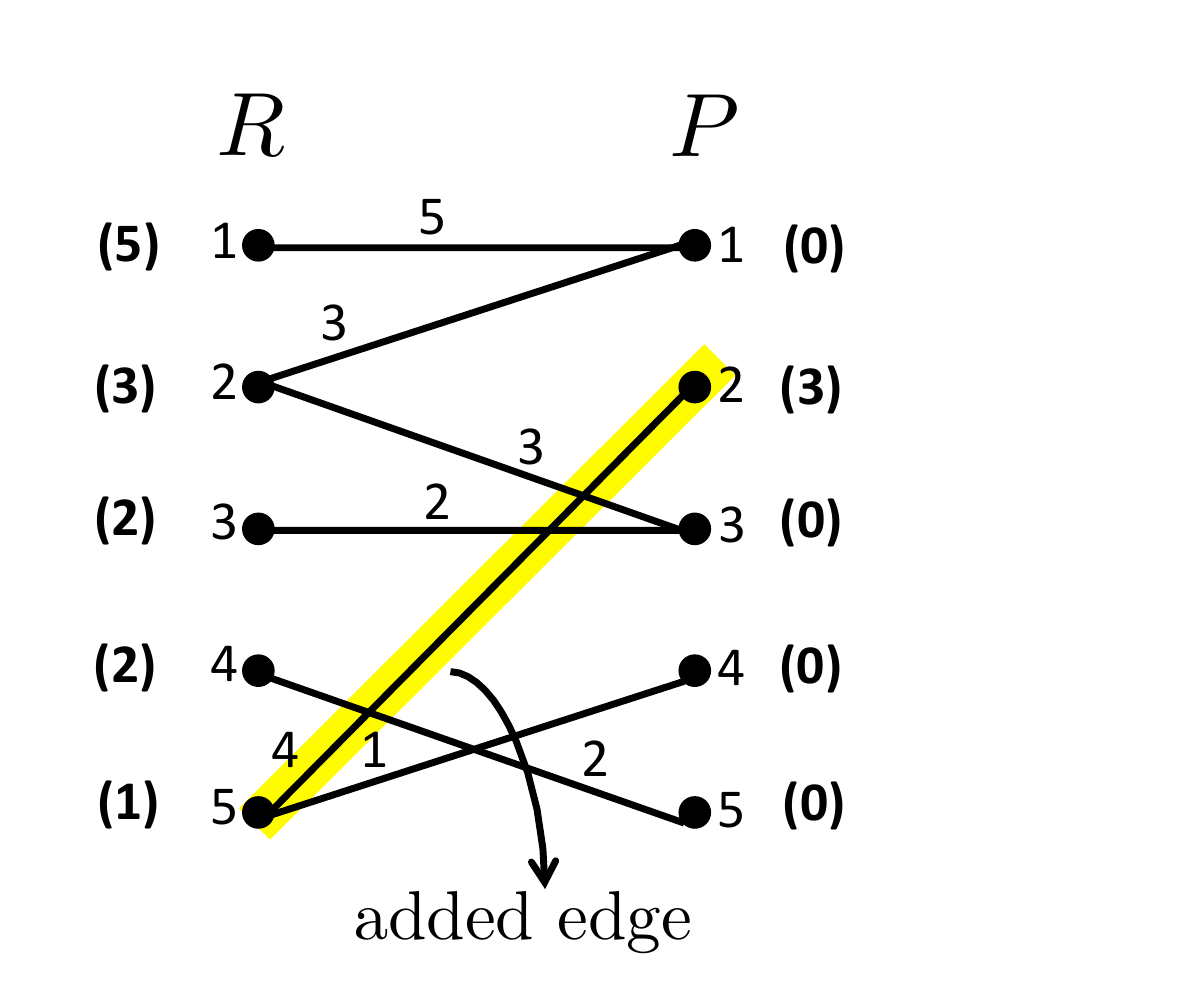}
\label{subfig_hungst21}
}
\hfill
\subfloat[ \ref{labelstep2}: For the updated $E_y$, a maximum cardinality matching $M$ (red edges), and a corresponding minimum vertex cover $V_{c} = (R_{c}, P_{c})$ (red vertices).]{
\includegraphics[scale=\scalefactorone]{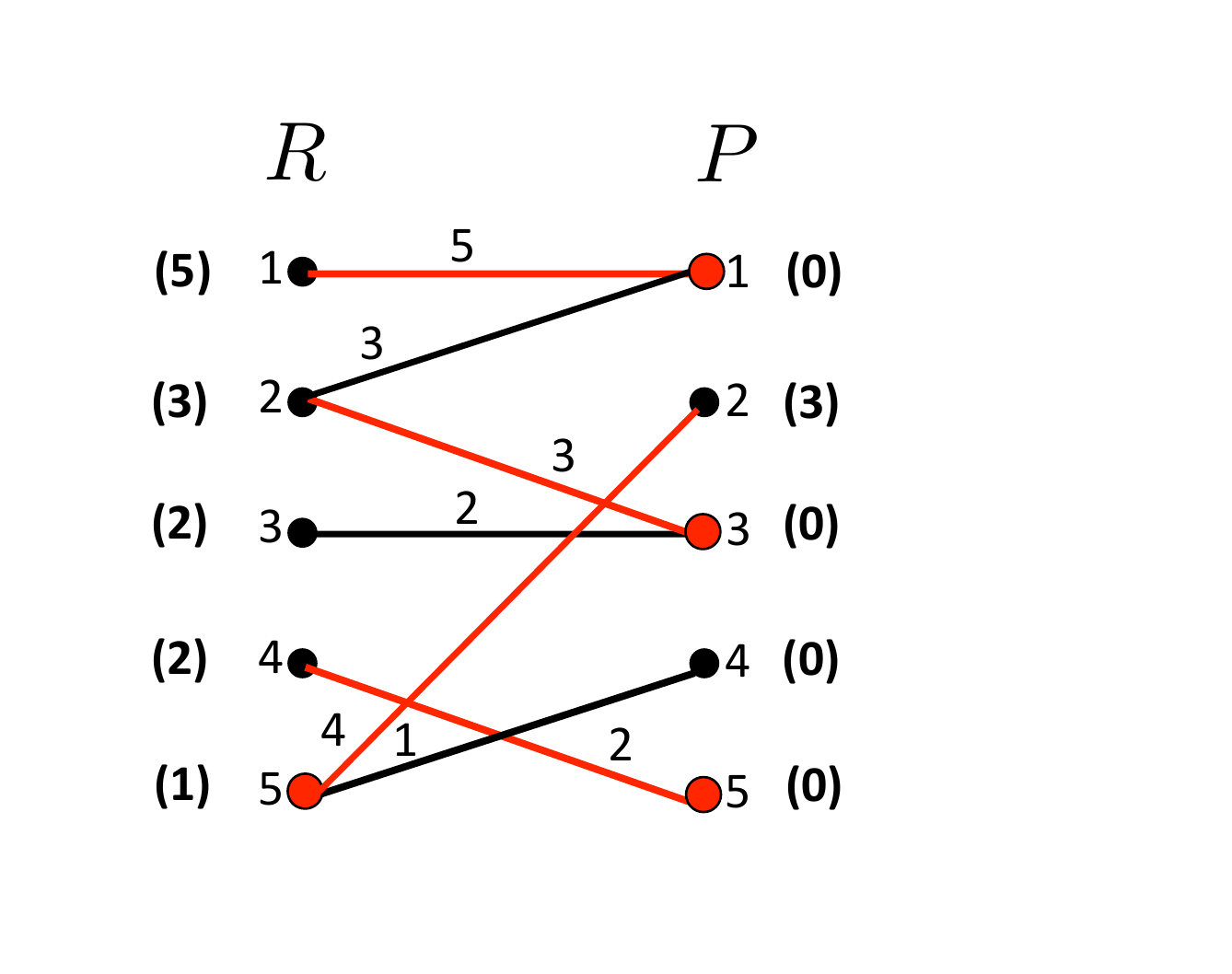}
\label{subfig_hungst22}
}
\caption{An instance of a two-step iteration (\ref{labelstep1} and \ref{labelstep2}) of the $\mathrm{Hungarian\_Method}$.}
\label{fig_hung2step}
\end{figure}

We proceed to provide a brief description of the Hungarian Method\footnote{There are more ways than one to implement the primal-dual Hungarian Method - we describe the implementation that forms the basis of our proposed algorithm.} that 
will assist us in explaining our proposed distributed algorithm in later
sections of this paper (see Figures \ref{fig_hunginit} and \ref{fig_hung2step} for corresponding instances).
\floatname{algorithm}{function}
\begin{algorithm}
\caption{$\mathrm{Hungarian\_Method}$ ($G$)} 
\label{algo_initialize}
\algsetup{indent = 1em}
\begin{algorithmic}[]

\vspace{0.3em}
\INDSTATE[0.5] \textbf{\textit{\% Initialization Step}} 
\vspace{0.3em}

\STATE $y = $ arbitrary feasible labeling, example:
\STATE $y(i \in R) = \min_{j \in P} w(i,j)$ and $y(j \in P) = 0$
\vspace{0.2em}

\STATE $E_y = $ equality subgraph edges using (\ref{eqn_equality_graph})
\vspace{0.2em}

\STATE $(M, V_c) =$ maximum cardinality matching and corresponding minimum vertex cover, given $(V, E_y)$ (see Remark \ref{note_1})
\vspace{0.2em}

\WHILE{$M$ is not a perfect matching}

\vspace{0.3em}
\INDSTATE \textbf{\textit{\% \setword{Step 1}{labelstep1}(a)}} 
\vspace{0.3em}

\FOR{$i \in R\setminus R_c$}

\STATE Choose any $j^\star \in \arg\min_{j\in P\setminus P_c} \mathrm{slack} (w,y, i,j)$, and set $e^i_{cand} = (i,j^\star)$ using (\ref{eqn_slack})

\ENDFOR

\STATE $E_{cand} = \cup_{i\in R\setminus R_c} \, \{e^i_{cand}\}$

\vspace{0.3em}
\INDSTATE \textbf{\textit{\% Step 1(b)}}
\vspace{0.3em}

\STATE $\delta = \min_{(i,j)\in E_{cand}} \mathrm{slack} (w,y, i,j)$
\vspace{0.2em}

\STATE $y(i) = y(i) - \delta, \,\, \forall i \in R_{c}$
\vspace{0.2em}

\STATE $y(j) = y(j) + \delta, \,\, \forall j \in P\setminus P_{c}$

\vspace{0.3em}
\INDSTATE \textbf{\textit{\% \setword{Step 2}{labelstep2}}}
\vspace{0.3em}

\STATE $E_y = $ equality subgraph edges
\vspace{0.2em}

\STATE $(M, V_c) =$ maximum cardinality matching and corresponding minimum vertex cover, given $(V, E_y)$ 
\vspace{0.2em}

\ENDWHILE
\vspace{0.5em}

\end{algorithmic}
\end{algorithm}

\begin{remark}
\label{note_2}
As mentioned in \ref{labelstep1}, the \textit{selection} of the candidate edges
is done based on the minimum vertex cover $V_{c} = (R_{c}, P_{c})$. In
particular, the set of candidate edges $E_{cand}$ represents the edges between
vertices in $R\setminus R_{c}$ and vertices in $P\setminus P_{c}$, i.e. edges
between the so-called \textit{uncovered} vertices in $R$ and \textit{uncovered} vertices in
$P$ (see Figure \ref{subfig_hungst11} for an example). \oprocend
\end{remark}

Without delving into details, we provide an auxiliary lemma, followed by a quick proof sketch that shows the Hungarian Method converges to an optimal solution (see \cite{{burkard:2009a}, {frank:nrl2005a}} for details). We will rely on these fundamental results in later sections of the paper, where we discuss the convergence properties of our proposed distributed algorithm. 

\begin{lemma}
\label{lem_it} 
Given a weighted, bipartite graph $G = (V,E,w)$, with bipartition $(R,P)$, a feasible vertex labeling function $y$, and a corresponding maximal matching $M$, every two-step iteration (\ref{labelstep1} and \ref{labelstep2}) of the $\mathrm{Hungarian\_Method}$ results in the following:
(i) an updated $y$ that remains feasible, and
(ii) an increase in the matching size $|M|$, or no change in the matching $M$, but an increase in $|R_{c}|$ (and corresponding decrease in $|P_{c}|$, such that $|R_{c}| + |P_{c}| = |M|$).\footnote{Either $|R_{c}|$ increases and $|P_{c}|$ decreases, or $|P_{c}|$ increases and $|R_{c}|$ decreases, depending on the particular implementation of the algorithm employed for finding $M$ and $V_{c}$.} \oprocend
\end{lemma}

\begin{remark}[Proof sketch of the $\mathrm{Hungarian\_Method}$]
\label{rem:hungarian_complexity}
The above stated Lemma \ref{lem_it} ensures that the size of a matching $M$
increases after a finite number of two-step iterations (worst-case
$r$, where $r = |R| = |V|/2$). Since the algorithm converges when $M$ is perfect, i.e. $|M| = r$,
Lemma \ref{lem_it} in conjunction with Theorem \ref{thm_hung} proves that the
$\mathrm{Hungarian\_Method}$ converges to an optimal solution (perfect matching with minimum
cost), after $\mathcal{O}(r^2)$ two-step iterations. Each two-step iteration
requires $\mathcal{O}(r^2)$ time, yielding a total running time of
$\mathcal{O}(r^4)$ (through certain modifications, this running time can be
reduced to $\mathcal{O}(r^3)$). \oprocend
\end{remark}

Now that we have reviewed the {LSAP}, as well as the Hungarian Method used for solving it, we proceed to setup the distributed problem central to this paper.

\section{Distributed Problem Setup}
\label{sec_prob_def}
Similar to the previous section, let $R = \{1,2,...,r\}$ denote a set of $r$ robots, and $P = \{1,2,...,p\}$ denote a set of $p$ targets, where $r\geq p$. Let $P^i \subseteq P$ be the set of targets that robot $i\in R$ can be assigned to, with the associated cost function $c^i : P^i \rightarrow \mathbb{R}$. We assume that each robot $i\in R$ knows the sets $R$ and $P$. Moreover, each robot knows the cost function $c^i$, associated with the subset of targets that it can be assigned to.

\begin{remark}
\label{rem_centr}
We can generate the problem data for the corresponding centralized assignment problem (Section \ref{sec_centr_assign}), as the weighted bipartite graph $G = (V, E, w)$, where,
\begin{enumerate}[-]
\item{The edge set $E$ is given by $E = \{(i,j)\,|\,j \in P^i\}$ $\forall i \in R$},
\item{The weight function $w:E \rightarrow \mathbb{R}$ is given by $w((i,j)) = c^i(j)$, $\forall (i,j) \in E$}.
\end{enumerate}
As mentioned before, we can modify $G$ to ensure it is balanced and
complete. For now, assume $|R| = |P| = |V|/2$, and include high-weight edges as
per the big-M method, to make $G$ complete. \oprocend
\end{remark}

Recall that the optimal solution to such an assignment problem is a minimum weight perfect matching. However, due to the inherent degeneracy in assignment problems, there can be multiple minimum weight perfect matchings. Let ${\mathcal{M}}$ denote the set of such minimum weight perfect matchings. Then, for any $M \in \mathcal{M}$, the corresponding \textit{unique} optimal cost $c^\star$ is given by $c^\star = c(M) = \sum_{e\in M} w(e)$.

\textit{Communication network:} We model the communication between the robots by
a time-varying directed graph $\mathcal{G}_c(t) = (R,E_c(t))$,
$t \in \mathbb{R}_{\geq 0}$. In such a graph, an edge from robot $i$ to
robot $j$ at some time $t$ implies that robot $i$ can communicate with robot $j$ at that time instant. Moreover, for robot $i$, we let
$\mathcal{N}_O(i,t)$ denote the set of outgoing neighbors, and
$\mathcal{N}_I(i,t)$ denote the set of the incoming neighbors
respectively. Based on the above discussion, we assume the following:

\begin{assumption}
\label{ass_comm}
For every time instant $t \in \mathbb{R}_{\geq 0}$, the directed graph
$\mathcal{G}_c(t)$ is strongly connected, i.e., there exists a directed path
from every robot, to every other robot in $\mathcal{G}_c(t)$.\oprocend
\end{assumption}

We are interested in the problem of assigning
robots to targets with minimum total cost, where each robot $i\in R$ initially 
knows $(R,P,c^i)$, and can communicate with other robots \textit{only} via the
time-varying communication graph $\mathcal{G}_c(t)$, as per Assumption
\ref{ass_comm}.

Before proceeding to the algorithm central to this paper, we would like to make three remarks
on the proposed set-up.

\begin{itemize}
\item\label{item_1} We have introduced two graphs which have completely different roles. The first
  (fixed, wighted, bipartite) graph $G = (V,E, w)$ models the assignment problem (by
  relating the set of robots $R$ to the set of targets $P$). The second graph
  $\mathcal{G}_c(t)$ describes the peer-to-peer communication network among the
  robots.
\item\label{item_2} Assumption~\ref{ass_comm} can be relaxed by requiring the communication graph to be only
  \textit{jointly} strongly connected over some time period\footnote{There exists a positive and bounded duration $T_c$, such that for every time instant $t \in \mathbb{R}_{\geq 0}$, the directed graph $\mathcal{G}^{t + T_c}_c (t) :=  \bigcup_{\tau = t}^{t + T_c} \mathcal{G}_c (\tau)$}. Note that the relaxation to a jointly strongly connected communication network ties neatly into the framework of an \textit{asynchronous} implementation of any distributed algorithm (e.g. if a robot is still computing, it would have no edges in the underlying communication graph for that time duration, allowing every robot to communicate and compute at its own speed, without any synchronization within the network). However, in order to not overweight the proofs, we prefer to stay with the more stringent condition (strongly connected at \textit{all} times), and assume our algorithm runs synchronously (explained later in detail). We briefly discuss our proposed algorithm in an asynchronous setting in Remark \ref{rem:async}, following the synchronous convergence analysis in Section \ref{sec:convergence}.
\item\label{item_3} Degeneracy of assignment problems\footnote{An assignment
    problem is degenerate when there exists more than one assignment with
    minimum cost.} is of particular concern in a distributed framework, since
  all robots must converge not only to an optimal solution, but to the
  \textit{same} optimal solution. We denote such a solution by
  $\hat{M} \in \mathcal{M}$ (note that $c(\hat{M})= c^\star$).
\end{itemize}

Thus, we define the distributed version of the assignment problem as follows:

\textit{\textbf{Distributed Assignment Problem:}}
\label{subsec_distprob}
Given a set of robots $R$, a set of targets $P$, and a communication graph
$\mathcal{G}_c(t)$ as per Assumption \ref{ass_comm}. Every robot $i\in R$ knows
($R,P,c^i$), i.e. the sets $R$ and $P$, and the cost function associated with
itself and targets that it can be assigned to. Then, the distributed assignment
problem requires all robots to converge to a common assignment,
$\hat{M}$, that is optimal to the centralized assignment
problem, i.e., $\hat{M} \in \mathcal{M}$.

\section{A Distributed Version of the Hungarian Method}
\label{sec_dist_algo}

Drawing from the description of the $\mathrm{Hungarian\_Method}$
(\ref{subsec_hungmethsteps}), we propose the \textit{Distributed-Hungarian}
algorithm for solving the Distributed Assignment Problem
(\ref{sec_prob_def}), where $\mathbb{G}^i$ denotes the state of robot $i$.
Specifically, $\mathbb{G}^i$ contains the following three objects:
\begin{enumerate}[(i)]
\item{$G^i_{lean} = (V, E^i_{lean}, w^i_{lean})$: A bipartite graph with vertex
    partitioning $V = (R,P)$, two disjoint sets of edges $E^i_y$ and $E^i_{cand}$, denoted by the edge partitioning
    $E^i_{lean} = (E^i_y, E^i_{cand})$, and a corresponding edge weight function
    $w^i_{lean}$\footnote{The subscript \textit{lean} refers to the sparseness
      of the graph, with significantly less number of edges than a complete
      graph.}}.
\item{$y^i$: A vertex labeling function for $G^i_{lean}$}
\item{$\gamma^i \in \mathbb{Z}$: A counter variable}
\end{enumerate}

Let $T_s = \{t_0, t_1, t_2,\, ...\, \}$ be the set of discrete time instants
over which the robots \textit{synchronously} run the algorithm. In other words,
at every time instant $t\in T_s$, each robot performs the following two actions
repeatedly: (i) it sends a message $msg^i = \mathbb{G}^i$ to each of its
outgoing neighbors, and (ii) upon receiving messages from its incoming
neighbors, it computes its new state $\mathbb{G}^i$. In this manner, each time
instant represents an ``iteration'' or ``communication-round'' of the
\textit{Distributed-Hungarian} algorithm.

To provide more context, recall that in the Distributed Assignment Problem (\ref{sec_prob_def}), each robot $i$ has access to $(R,P,c^i)$. Thus, before beginning the algorithm, every robot creates its so-called ``original information'' in the form of a weighted, bipartite graph $G^i_{orig} = (V, E^i_{orig}, w^i_{orig})$, where,
\begin{enumerate}[-]
\item{The edge set $E^i_{orig}$ is given by $E^i_{orig} = \{(i,j)\,|\,j \in P\}$},
\item{The weight function $w^i_{orig}:E^i_{orig} \rightarrow \mathbb{R}$ is given by, 
\small
\begin{align*}
w^i_{orig}((i,j)) = 
\begin{cases}
c^i(j) &j\in\ P^i,\\
\mathfrak{M} &j\in\ P\setminus P^i\\
\end{cases}
\end{align*}
\normalsize}
\end{enumerate}

\begin{figure}
\subfloat[Robot 3's original information $(R,P,c^3)$.]{
\fbox{
\includegraphics[scale=0.35]{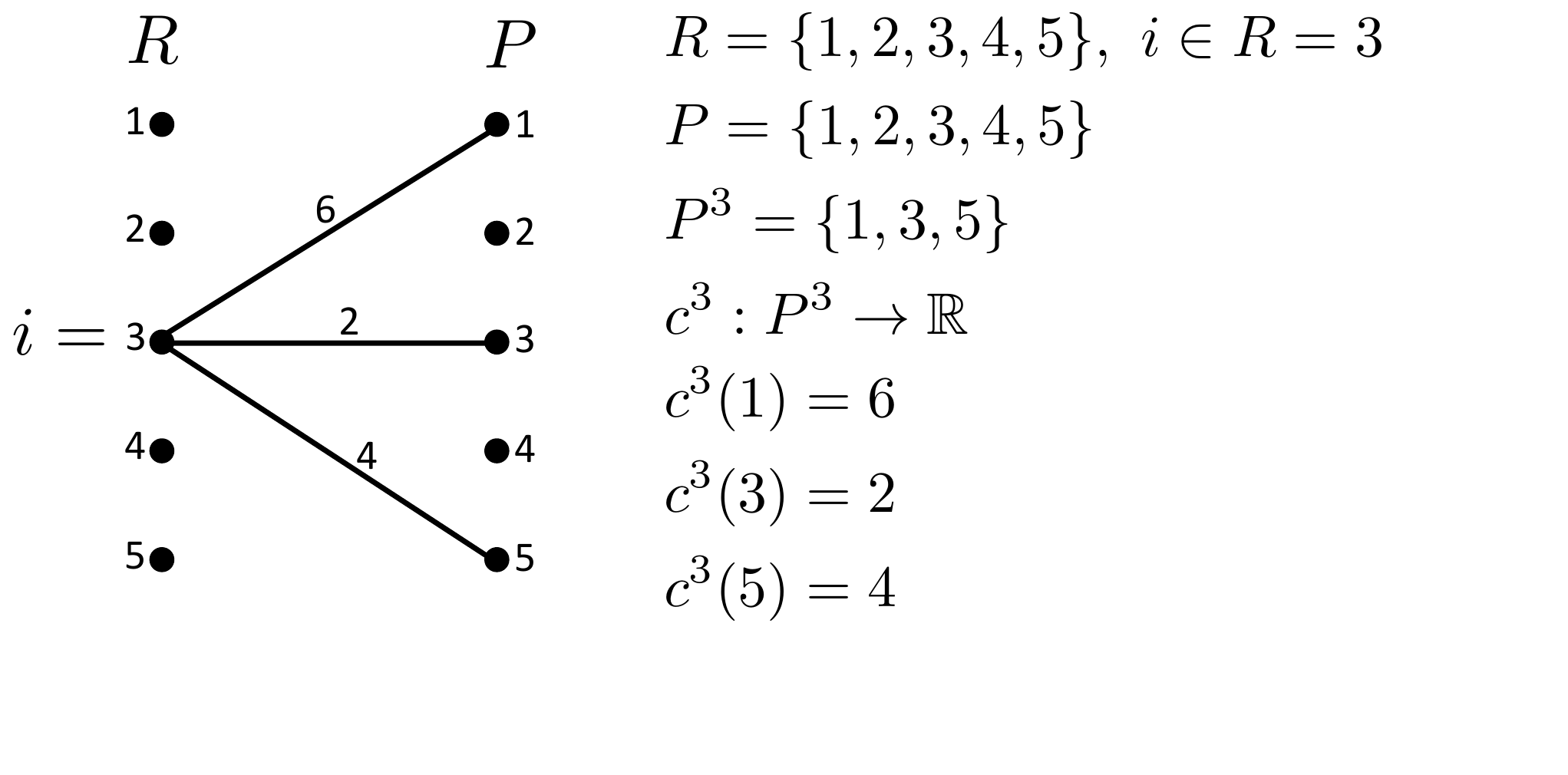}}
\label{fig_ega}}\hfill
\subfloat[Robot 3's original information in the form of the weighted, bipartite graph graph $G^3_{orig} = (V, E^3_{orig}, w^3_{orig}$).]{
\fbox{
\includegraphics[scale=0.35]{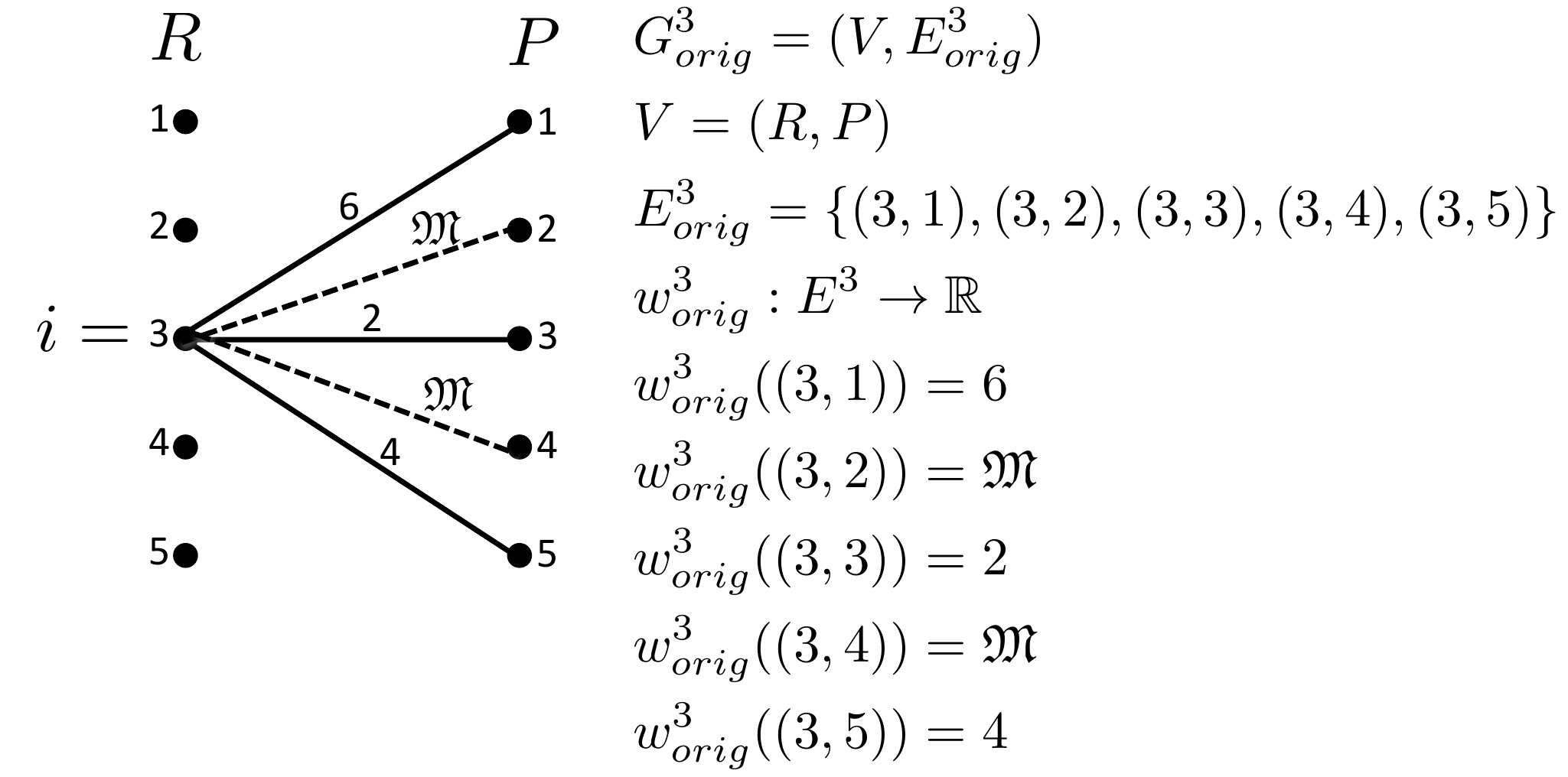}}
\label{fig_egb}}\hfill
\subfloat[Robot 3's initial state (at $t=t_0$), $\mathbb{G}^3 = (G^3_{lean}, y^3, \gamma^3)$.]{
\fbox{
\includegraphics[scale=\scalefactorone]{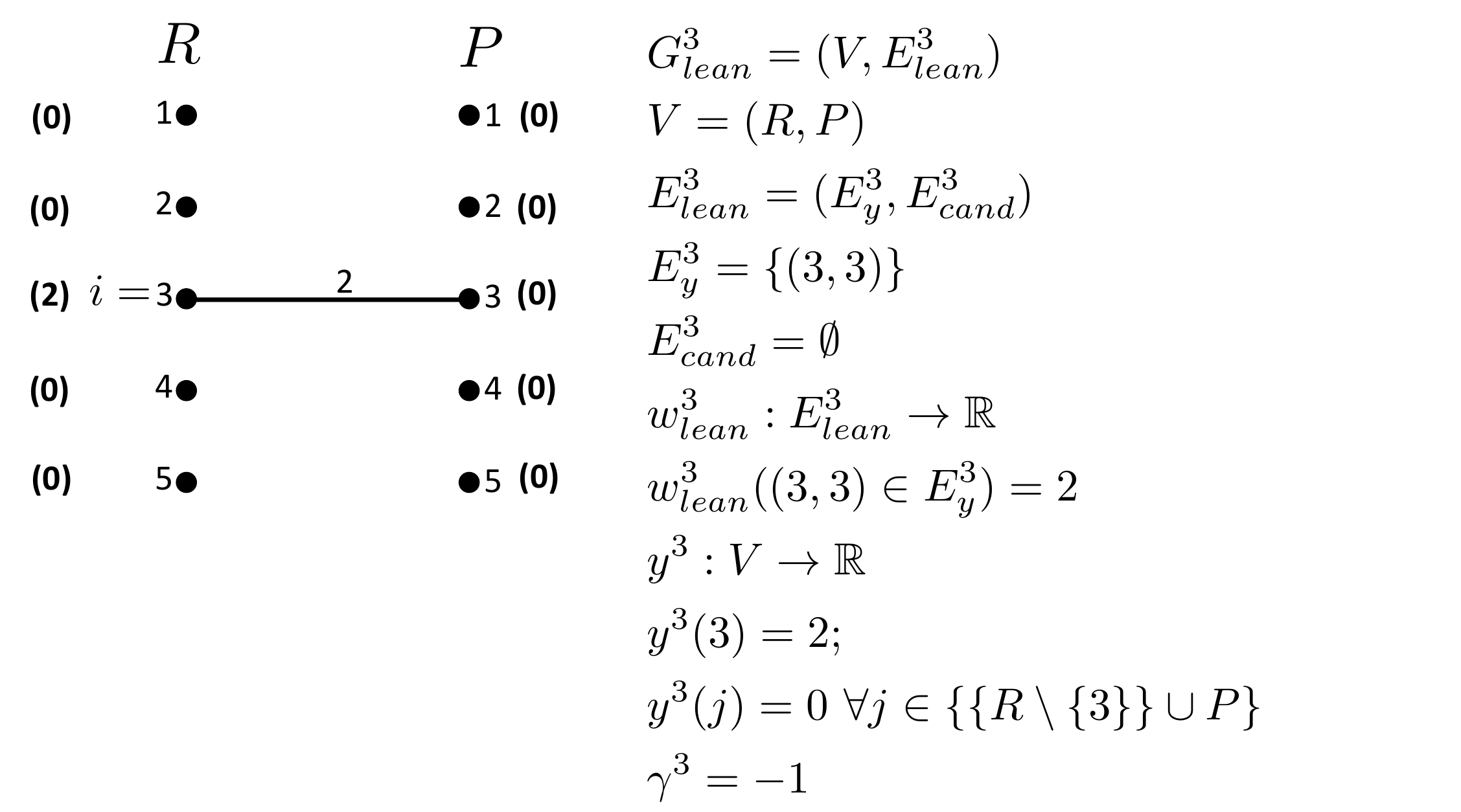}}
\label{fig_egc}}
\caption[An example of a robot's original information, and initial state in the \textit{Distributed-Hungarian} algorithm.]{With the weighted, bipartite graph from Figure \ref{subfig_hung1} as the centralized graph (Remark \ref{rem_centr}), this figure depicts a single robot's (robot $3$) original information, and corresponding initial state, in the \textit{Distributed-Hungarian} algorithm.}  
\label{fig_eg}
\end{figure}

The algorithm is then initialized as follows: 
At $t=t_0$, each robot $i$ selects an edge $(i,j^\star)$ with minimum weight from its original information $G^i_{orig}$. Using
this edge, the robot initializes its state
$\mathbb{G}^i = (G^i_{lean}, y^i, \gamma^i)$, where,
\begin{align}
& G^i_{lean} = (V, (\{(i,j^\star)\}, \emptyset), w^i_{orig}(i,j^\star))\label{initial_state1}\\
& y^i(i) = w^i_{orig}((i,j^\star)); \,y^i(j) = 0, \forall j \in \{\{R\setminus \{i\}\}  \cup P\}\label{initial_state2}\\
& \gamma^i = -1\label{initial_state3}
\end{align}
See Figure \ref{fig_eg} for an example of the above. 

Upon receiving the messages (states) of all incoming neighbors, robot $i$ performs the following steps:
\begin{enumerate}[-]
\item{it calls the $\mathrm{Build\_Latest\_Graph}$ function on all the states in its
    memory, i.e. $\{\mathbb{G}^i\} \cup (\cup_{j\in \mathcal{N}_I(i,t)} \{\mathbb{G}^j\})$, to
    obtain a temporary, most-updated state $\mathbb{G}_{tmp}$.}
\item{using $\mathbb{G}_{tmp}$ and its original information $G^i_{orig}$, robot
    $i$ calls the $\mathrm{Local\_Hungarian}$ function to compute its new state
    $\mathbb{G}^i$.}
\end{enumerate}

We proceed to formally state the $\mathrm{Build\_Latest\_Graph}$ and $\mathrm{Local\_Hungarian}$ functions. 

\subsection{$\mathrm{Build\_Latest\_Graph}$}
Given a set of robots $R' \in R$, and a set of corresponding states
$S = \cup_{j\in R'} \{\mathbb{G}^j\}$, the $\mathrm{Build\_Latest\_Graph}$ function
returns a resultant, most-updated state $\mathbb{G}_{tmp} = (G_{lean}, y,\gamma)$ that contains the
information of \textit{only} those robots that have the highest counter value.
We denote such a subset of robots by $R_{lead}$. If the highest counter
value is positive, the function chooses any one robot $j^\star$ in $R_{lead}$,
and sets $\gamma, y$ and $E_y$ equal to $j^\star$'s corresponding
information. However, the function combines the candidate edges of \textit{all}
robots in $R_{lead}$, i.e. $E_{cand} = \bigcup_{j\in R_{lead}} E^j_{cand}$, and
sets $G_{lean} = (V, (E_y, E_{cand}), w_{lean})$, where $w_{lean}$ is the
corresponding edge weight function.

\floatname{algorithm}{function}
\begin{algorithm}
\caption{$\mathrm{Build\_Latest\_Graph}\, (R', S)$} 
\label{algo_parse_info}
\algsetup{indent = 1em}
\begin{algorithmic}[]

\IF{(all $\gamma^j = -1$)}
\vspace{0.2em} 

\STATE $\gamma = -1$

\STATE$E_{y} = \bigcup_{j\in R'} E^j_{y}$
\vspace{0.3em}

\STATE$E_{cand} = \emptyset$
\vspace{0.3em}

\STATE $G_{lean} = (V, (E_y, E_{cand}), w_{lean})$
\vspace{0.3em}

\STATE\label{gen_y} 
\vspace{0.3em}
$y(j) = y^j(j), \, \forall j \in R'$; $y(j) = 0,  \, \forall j \in \{\{R\setminus R'\}  \cup P\}$
\vspace{0.05em}

\IF{$|E_y| = |R|$} 
\vspace{0.3em}

\STATE $\gamma = 0$ 
\vspace{0.3em}

\ENDIF

\ELSE

\STATE $R_{lead} = \arg\max_{j\in R'} \gamma^j $  
\vspace{0.3em}

\STATE\label{linemaxinfo} choose any $j \in R_{lead}$ and set $\gamma = \gamma^j$; $y = y^j$;  $E_y = E^j_{y}$

\STATE $E_{cand} = \bigcup_{j\in R_{lead}} E^j_{cand}$ 
\vspace{0.3em}

\STATE $G_{lean} = (V, (E_y, E_{cand}), w_{lean})$
\vspace{0.3em}

\ENDIF

\STATE \textbf{return} $\mathbb{G} = (G_{lean}, y, \gamma)$

\end{algorithmic}
\end{algorithm}

A special instance of the function occurs when all counter values are $-1$. In
this case, the function simply sets $\gamma = -1$, $E_{cand} = \emptyset$, and
combines the equality subgraph edges of all robots in $R'$, i.e.
$E_{y} = \bigcup_{j\in R'} E^j_{y}$. With
$G_{lean} = (V, (E_y, E_{cand}), w_{lean})$, the function sets the vertex labels
$y(i)$ of every robot $i$ that has an edge in $E_y$, to the corresponding weight
of that edge. If \textit{all} robots in $R$ have an edge in $E_y$, the function
sets $\gamma$ to $0$.

\begin{remark}
To provide context to the $\mathrm{Build\_Latest\_Graph}$ with respect to the centralized $\mathrm{Hungarian\_Method}$, we note here that robot $i$'s most updated state $\mathbb{G}_{tmp}$ contains $y$, a \textit{globally} feasible vertex labeling function with respect to the centralized graph $G = (V, E, w)$. In other words, robot $i$'s information is a sparse (lean) version of the centralized method's information at the beginning of every two-step iteration. We prove this fact in later sections of the paper (Lemma 2 and extensions). \oprocend
\end{remark}

\subsection{$\mathrm{Local\_Hungarian}$}
Given robot $i$'s temporary state $\mathbb{G}_{tmp} = (G_{lean}, y, \gamma)$, and its original information $G^i_{orig} = (V, E^i_{orig}, w^i_{orig})$, the $\mathrm{Local\_Hungarian}$ function computes robot $i$'s new state $\mathbb{G}^i$ as follows:

For the bipartite graph $G_{lean} = (V, (E_y, E_{cand}),w_{lean})$, and the vertex labeling function $y$, the $\mathrm{Local\_Hungarian}$ function computes the maximum cardinality matching $M$, and the corresponding minimum vertex cover $V_c$, as per Remark \ref{note_1}.
If $M$ is not a perfect matching, the function chooses a single candidate edge from robot $i$'s original information using the $\mathrm{Get\_Best\_Edge}$ sub-function (formally stated below), and adds it to $E_{cand}$. Note that robot $i$ can contribute a candidate edge \textit{only} if it is uncovered.

\begin{algorithm}
\caption{$\mathrm{Get\_Best\_Edge}$ ($G^i_{orig}, y, V_c$)} 
\label{algo_get_cand}
\algsetup{indent = 1em}
\begin{algorithmic}[1]

\IF{$i \in R\setminus{R}_c$}

\STATE\label{choiceecand}  Choose any $j^\star \in \arg\min_{j\in P\setminus P_c} \mathrm{slack} (w^i_{orig},y, i,j)$, and set $e_{cand} = (i,j^\star)$
\vspace{0.2em}

\ENDIF

\STATE \textbf{return} $e_{cand}$

\end{algorithmic}
\end{algorithm}

Up until this point, the $\mathrm{Local\_Hungarian}$ function mimics the $\mathrm{Hungarian\_Method}$ (\ref{subsec_hungmethsteps}). However, following this, the function can have one of two outcomes:
\begin{enumerate}[-]
\item{\textit{There exists an uncovered robot with no edge in $E_{cand}$}\\
The above case corresponds to an \textit{incomplete} \ref{labelstep1}(a) of the $\mathrm{Hungarian\_Method}$, and as such, the $\mathrm{Local\_Hungarian}$ function simply sets robot $i$'s new state $\mathbb{G}^i = (G_{lean}, y, \gamma)$, where $G_{lean}$ contains the (possibly) updated $E_{cand}$.}
\vspace{0.3em}
\item{\textit{Every uncovered robot has an edge in $E_{cand}$}\\
Such a case corresponds to a \textit{completed} \ref{labelstep1}(a) of the $\mathrm{Hungarian\_Method}$, and thus, the $\mathrm{Local\_Hungarian}$ function continues with \ref{labelstep1}(b) and \ref{labelstep2}, updating $y$, $E_y$, $M$ and $V_c$, and incrementing the counter value $\gamma$ by $1$. Using the updated information, the function resets $E_{cand}$ to a new candidate edge from robot $i$'s original information (if none exists, $E_{cand} = \emptyset$). Moreover, the function calls a $\mathrm{Reduce\_Edge\_Set}$ sub-function that essentially prunes $E_y$ to contain the minimum number of equality subgraph edges, such that (i) $M$ and $V_c$, when calculated with respect to the pruned $E_y$, remain unchanged from their previous state, and (ii) the total number of edges in $E_{lean}$ is at most $(2r-1)$, i.e. $|E_y| + |E_{cand}| \leq 2r-1$.}
\end{enumerate}
The $\mathrm{Local\_Hungarian}$ function then sets robot $i$'s new state $\mathbb{G} = (G_{lean}, y, \gamma)$.
\begin{algorithm}
\caption{$\mathrm{Local\_Hungarian}$ ($\mathbb{G}_{tmp}, G^i_{orig}$)} 
\label{algo_initialize}
\algsetup{indent = 1em}
\begin{algorithmic}[]

\STATE $(M, V_c) =$ maximum cardinality matching and corresponding minimum vertex cover, given $(V, E_y)$
\vspace{0.2em}

\IF{$M$ is not a perfect matching}
\vspace{0.2em}

\STATE\label{choiceecand}  $e_{cand} = \mathrm{Get\_Best\_Edge} (G_{orig}^i, y, V_c)$
\vspace{0.2em}

\STATE $E_{cand} =  E_{cand} \cup \{e_{cand}\}$
\vspace{0.3em}

\IF[\ref{labelstep1}(a) completed] {$|R\setminus R_{c}| = |E_{cand}|$}  
\vspace{0.3em}

\STATE perform \ref{labelstep1}(b) and \ref{labelstep2}
\vspace{0.3em}

\STATE $\gamma = \gamma + 1$
\vspace{0.3em}

\STATE \textit{\% Prune the number of edges}
\vspace{0.3em}

\STATE\label{choiceecand}  $e_{cand} = \mathrm{Get\_Best\_Edge}$ ($G_{orig}^i, y, V_c$)
\vspace{0.2em}

\STATE $E_{cand} = \{e_{cand}\}$
\vspace{0.2em}

\STATE $E_y = \mathrm{Reduce\_Edge\_Set}$ ($V, E_y, M, V_c$)
\vspace{0.2em}

\ENDIF

\ENDIF

\STATE $G_{lean} = (V,(E_y, E_{cand}), w_{lean})$
\vspace{0.2em}

\STATE\label{line_msg} \textbf{return} $\mathbb{G} = (G_{lean}, y, \gamma)$ 

\end{algorithmic}
\end{algorithm}

Now that we have stated the functions used by the \textit{Distributed-Hungarian} algorithm, we proceed to provide a formal description of the algorithm (see Figures \ref{fig_disthunginit} - \ref{fig_disthungstep12(3)} for corresponding instances). 

\floatname{algorithm}{Algorithm}
\begin{algorithm}[H]
\caption{{\textit{Distributed-Hungarian}} $(G_{orig}^{i})$} 
\label{algo_dist_assign}
\algsetup{indent = 1em}
\begin{algorithmic}[]

\INDSTATE[2] \textbf{\textit{Initialization}} 
\vspace{0.3em} 

\STATE choose any $j^\star \in \arg\min_{j\in P} w^i_{orig}((i,j))$
\vspace{0.3em}

\STATE $E^i_y = \{(i,j^\star)\}; \,\, E^{i}_{cand} = \emptyset; \,\, w^i_{lean} = w^i_{orig}((i,j^\star))$
\vspace{0.3em}

\STATE $G^i_{lean} = (V, (E^i_y, E^i_{cand}), w^i_{lean})$
\vspace{0.3em}

\STATE $y^i(i) = w^i_{orig} ((i,j^\star))$; $y^{i}(j) = 0$, $\forall j\in \{\{R\setminus \{i\}\}  \cup P\}$ 
\vspace{0.3em}

\STATE $\gamma^{i} = -1$
\vspace{0.3em}

\STATE $\mathbb{G}^i = (G^i_{lean}, y^i, \gamma^i)$ 

\vspace{0.3em}
\INDSTATE[2] \textbf{\textit{Evolution}} 
\vspace{0.3em}

\WHILE[See Corollary \ref{cor_stop}.]{$\neg$ stopping criterion}

\vspace{0.3em} 
\INDSTATE {\textit{\% Receive and Parse:}}
\vspace{0.3em}

\STATE $R' = \mathcal{N}_I(i,t) \cup \{i\}$, $t\in T_s$
\vspace{0.3em}

\STATE $S = \{\mathbb{G}^i\} \cup (\cup_{j\in R'} \{msg^j\})$
\vspace{0.3em}

\STATE\label{lineparseinfo} $\mathbb{G}_{tmp} = \mathrm{Build\_Latest\_Graph} \,(R', S)$
\vspace{0.2em}

\IF {$\gamma \geq 0$} 

\vspace{0.3em} 
\INDSTATE[2.3] {\textit{\% Compute:}} 
\vspace{0.3em} 

\STATE $\mathbb{G}^{i} = \mathrm{Local\_Hungarian} \, (\mathbb{G}_{tmp}, G_{orig}^{i})$
\vspace{0.2em}

\ELSE
\vspace{0.2em}

\STATE $\mathbb{G}^{i} = \mathbb{G}_{tmp}$
\vspace{0.2em}

\ENDIF
\vspace{0.2em}

\ENDWHILE
\end{algorithmic}
\end{algorithm}
\section{Convergence Analysis}
\label{sec:convergence}
In this section we prove that the \textit{Distributed-Hungarian} algorithm
converges to an optimal solution in finite time. In other words, all the robots agree on a common assignment that minimizes the total cost, in finite time.
We begin by proving an auxiliary lemma.

\begin{lemma}
\label{lemma:unique_and_feasible_y}
For any counter value $\gamma \in \mathbb{N}_0$, there exists a (unique) vertex
labeling function $y_{\gamma}$, and a corresponding set of equality subgraph
edges, $E_{y_\gamma}$, such that for robot $i$'s state $\mathbb{G}^i$, if
$\gamma^i=\gamma$, then $y^i= y_\gamma$ and
$E^i_y = E_{y_\gamma}$. Moreover, for the centralized graph $G = (V, E, w)$ (Remark \ref{rem_centr}), $y_\gamma$ is a \textit{feasible
  labeling} (as per Problem D - the dual of the Minimum Weight Bipartite
Matching Problem).
\end{lemma}

\begin{IEEEproof}[Proof by induction (Prove for counter value $0$)]
Recall that every robot $i$ starts running the \textit{Distributed-Hungarian}
algorithm at time $t = t_0$, with its state $\mathbb{G}^i$ initialized as per Equations (\ref{initial_state1}) - (\ref{initial_state3}), to a bipartite graph
$G^i_{lean}$, a vertex labeling function $y^i$ and a counter value
$\gamma^i = -1$. First, notice that $G^i_{lean}$, or more precisely
$E^i_y$, contains exactly one, minimum weight (equality subgraph) edge from
its original information $G^i_{orig}$. Thus, it is clear that $y^i$
(generated using the minimum weight edge) is a feasible labeling with respect to
the centralized graph $G = (V, E, w)$. Moreover, there exists a unique set of
equality subgraph edges, $E_{y_0} = \cup_{i\in R} E^i_y$, that contains
exactly one such edge from \textit{every} robot, and a corresponding vertex
labeling function denoted by $y_0$ (generated using $r$ minimum weight edges). It is clear that 
$y_0$ is a feasible labeling with respect to the centralized graph $G$.

With the evolution of the algorithm, every robot $i$ repeatedly receives the
states of its incoming neighbors, and runs the $\mathrm{Build\_Latest\_Graph}$
function. 
If all its neighbors $j$ have $\gamma^j=-1$, then it populates its set of
equality subgraph edges $E_y$ in its ``most-updated'' state $\mathbb{G}_{tmp}$, by simply merging the received equality subgraph edges, i.e.,
$E_{y} = \cup_{j\in \mathcal{N}_I(i,t) \cup \{i\}} E^j_y$. 

Thus, according to the $\mathrm{Build\_Latest\_Graph}$ function, the only way
the \textit{first} robot, say $i$, sets its counter value $\gamma^i = 0$ is by building $\mathbb{G}_{tmp}$ with $E_{y} = E_{y_0}$. Moreover, another robot $k$ sets $\gamma^k=0$ either
by building the set $E_{y_0}$ on its own (as above), or by inheriting it
from robot $i$ (by virtue of robot $i$ being its in-neighbor with highest counter value $\gamma^i=0$).
Iterating this argument, it follows immediately that all robots $i$ with counter value
$\gamma^i=0$ have identical vertex
labeling functions $y_0$, and corresponding equality subgraph edges $E_{y_0}$.  

\paragraph*{Assume true for counter value $n$}
Assume that all robots with counter value $n$ have identical vertex labeling
functions, and identical equality subgraph edges, denoted by $y_n$ and $E_{y_n}$
respectively. Moreover, assume that $y_n$ is a \textit{feasible labeling} with
respect to the centralized graph $G$.
 
\paragraph*{Prove for counter value $n+1$}
Given $y_n$, $E_{y_n}$, and consequently a maximum cardinality matching
$M_n$ and corresponding minimum vertex cover $V_{c_n}$, there exists a unique
set $E_{{cand}_n}$ that comprises of \textit{exactly one edge} from each uncovered robot in $R\setminus R_{c_n}$
(going to an uncovered target in $P\setminus P_{c_n}$). 
Such an edge is uniquely determined by the
$\mathrm{Get\_Best\_Edge}$ function, identical to the process of selecting candidate edges in the centralized  $\mathrm{Hungarian\_Method}$. Thus, $E_{{cand}_n}$ corresponds to a \textit{completed} \ref{labelstep1}(a) of the $\mathrm{Hungarian\_Method}$.
As such, if any robot $i$ constructs a ``most-updated'' state $\mathbb{G}_{tmp}$
with the unique set $E_{{cand}_n}$, then by construction, its
$\mathrm{Local\_Hungarian}$ function proceeds to perform \ref{labelstep1}(a),
\ref{labelstep1}(b) and \ref{labelstep2} of the $\mathrm{Hungarian\_Method}$ on $\mathbb{G}_{tmp}$, resulting in a counter value of $n+1$, and an updated vertex labeling $y_{n+1}$ (with corresponding set of
equality subgraph edges $E_{y_{n+1}}$). 
Most importantly, the two-step iteration of the $\mathrm{Local\_Hungarian}$ function satisfies
the conditions in Lemma~\ref{lem_it}, proving that the updated $y_{n+1}$ is \textit{still} a feasible labeling with respect
to the centralized graph $G$. 

As the algorithm evolves, every robot $i$ with counter value $n$ includes in its
state, a candidate edge $e_{cand} \in E_{{cand}_n}$ from its original information $G^i_{orig}$, using the $\mathrm{Get\_Best\_Edge}$ function. Upon receiving the states of incoming neighbors, if all its neighbors $j$ have
$\gamma^j\leq n$, then robot $i$ populates its set of candidate edges in $\mathbb{G}_{tmp}$, by simply
merging the received candidate edges from \textit{only} those robots, with highest counter
value, i.e., $n$.

Thus, according to the $\mathrm{Build\_Latest\_Graph}$ and the $\mathrm{Local\_Hungarian}$ function,
the only way the \textit{first} robot, say $i$, sets $\gamma^i=n+1$ is
by building $\mathbb{G}_{tmp}$ with $E_{cand} = E_{cand_n}$. Similar to the argument for counter
value 0, another robot $k$ sets $\gamma^k=n+1$ either by building the
set $E_{cand_n}$ on its own (as above), or inheriting the latest information
directly from robot $i$ (by virtue of robot $i$ being its in-neighbor with highest counter value $\gamma^i=n+1$), thereby concluding the proof.
\end{IEEEproof}

\begin{corollary}
\label{corollary:equal_gammas}
If two robots $i$ and $j$ have identical counter values, then with respect to
the graphs $(V, E^i_y)=(V, E^j_y)$, they have identical maximal matchings,
i.e.,
\begin{align*}
\gamma^i = \gamma^j \,\Rightarrow \, M^i = M^j,  \forall i,j \in R.          
\end{align*}
\end{corollary}
\begin{IEEEproof}
Let $\gamma^i  = \gamma^j = n$, for some $n \in \mathbb{N}_0$. Since the counter value $n$ corresponds to a unique set of equality subgraph edges $E_{y_{n}}$, the maximal matching found from within the set of such edges is also unique. Thus, robots $i$ and $j$ have identical maximal matchings, denoted by $M_n$.
\end{IEEEproof}

\begin{corollary}
\label{corollary:unequal_gammas}
If robot $i$'s counter value is higher than robot $j$'s counter value,
i.e. $\gamma^i > \gamma^j$, then with respect to the graphs $(V, E^i_y)$ and
$(V, E^j_y)$, one of the following is true,
\begin{enumerate}[-]
\item{robot $i$'s maximal matching, $M^i$, is greater in size than robot $j$'s maximal
    matching, $M^j$, i.e. $|M^i| > |M^j|$};
\item{robots $i$ and $j$ have the same maximal matching, i.e. $M^i = M^j$,
    but in the context of their corresponding minimum vertex covers $V^i_c$ and $V^j_c$, robot $i$ has more covered vertices in $R$ than robot $j$, i.e., $|R^i_{c}| > |R^j_{c}|$.}
\end{enumerate}
\end{corollary}

\begin{IEEEproof}
  Let $\gamma^i = p$, and $\gamma^j = q$, where $p,q \in \mathbb{N}_0$,
  $q<p$. From the proof of Lemma \ref{lemma:unique_and_feasible_y}, we know that
  for a counter value to increment to say $n+1$, there must have existed at
  least one robot with counter value $n$, at some previous iteration of the
  \textit{Distributed-Hungarian} algorithm. In other words, for every
  $n \in \{q,...,p\}$, there \textit{existed} a robot with counter value $n$, a
  corresponding feasible vertex labeling function $y_n$, a set of equality
  subgraph edges $E_{y_{n}}$, and consequently, a maximal matching $M_n$, during
  some previous iteration of the algorithm. By construction, since every counter value
  update that occurs from $n$ to $n+1$, $n \in \{q,...,p\}$ corresponds to
  a two-step iteration of the $\mathrm{Hungarian\_Method}$, then (statement (ii) of)
  Lemma~\ref{lem_it} holds. Iteratively, the proof follows.
\end{IEEEproof}

\begin{figure}
\subfloat[Robot 2's original information as the weighted, bipartite graph $G^2_{orig} = (V, E^2_{orig}, w^2_{orig})$.]{
\includegraphics[scale=\scalefactortwo]{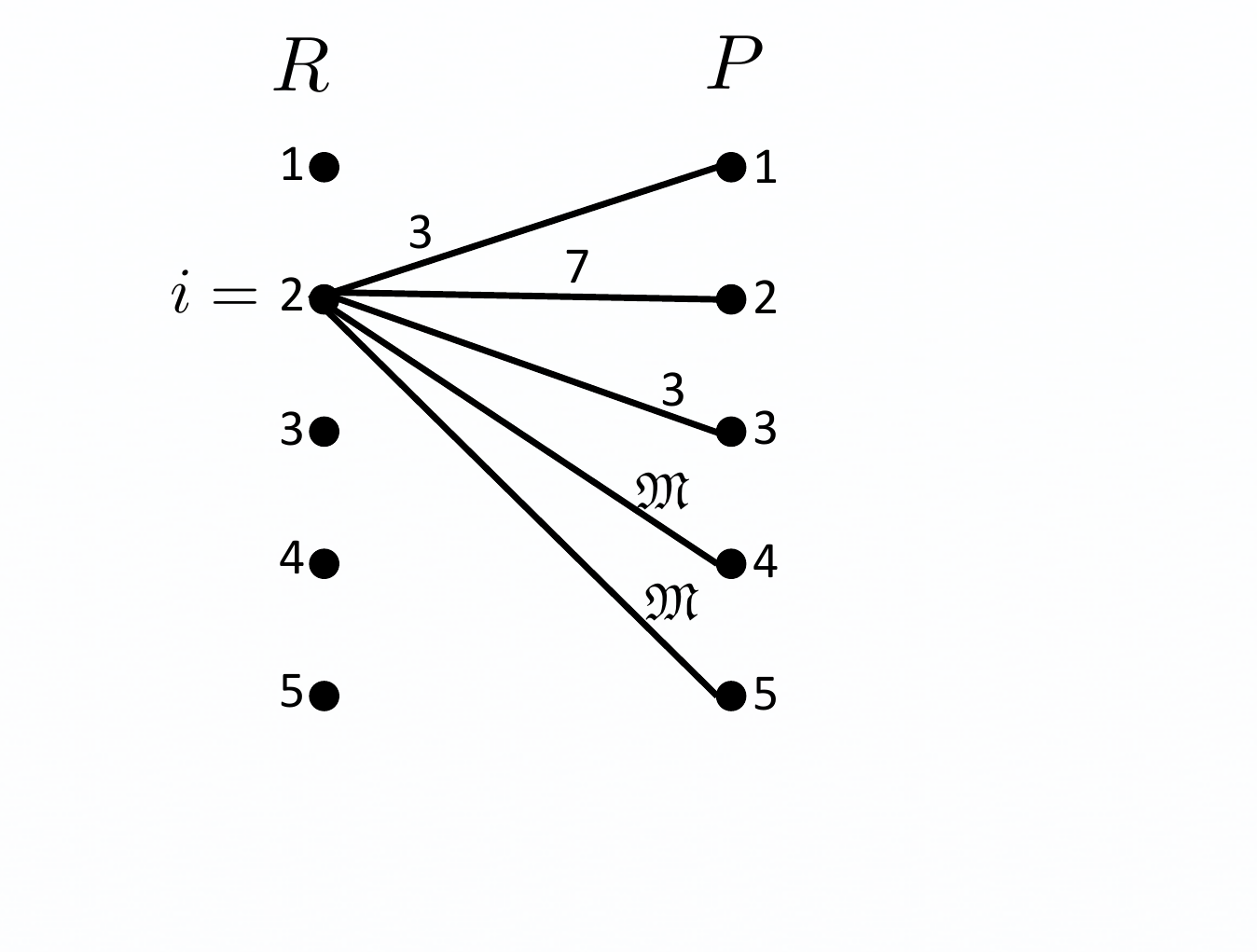}
\label{subfig_disthung1}
}
\hfill
\subfloat[Robot 2's temporary, most-updated state $\mathbb{G}_{tmp} = (G_{lean}, y, \gamma)$ with $\gamma = 0$. $E_{lean}$ contains the equality subgraph edges $E_y$ (black edges) and the candidate edges $E_{cand}$ (green edges, none in this case).]{
\includegraphics[scale=\scalefactortwo]{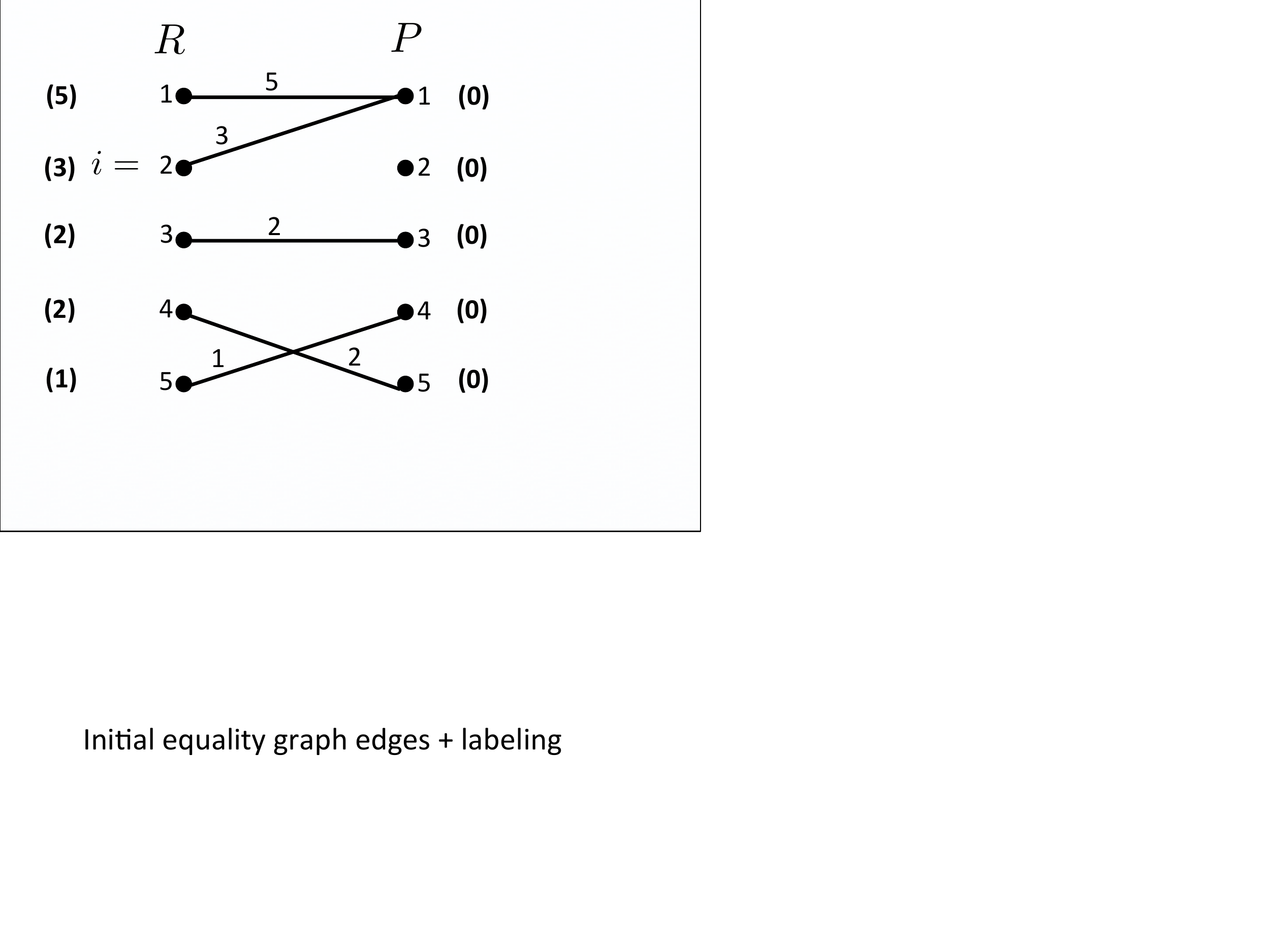}
\label{subfig_disthung2}
}
\hfill
\subfloat[The isolated set of equality subgraph edges $E_y$ in $\mathbb{G}_{tmp}$.]{
\includegraphics[scale=\scalefactortwo]{Dist_Hung_depiction_01_receivedinfo.pdf}
\label{subfig_disthung3}
}
\hfill
\subfloat[Given $E_y$, a maximum cardinality matching $M_y$ (red edges), and a corresponding minimum vertex cover $V_{c} = (R_{c}, P_{c})$ (red vertices).]{
\includegraphics[scale=\scalefactortwo]{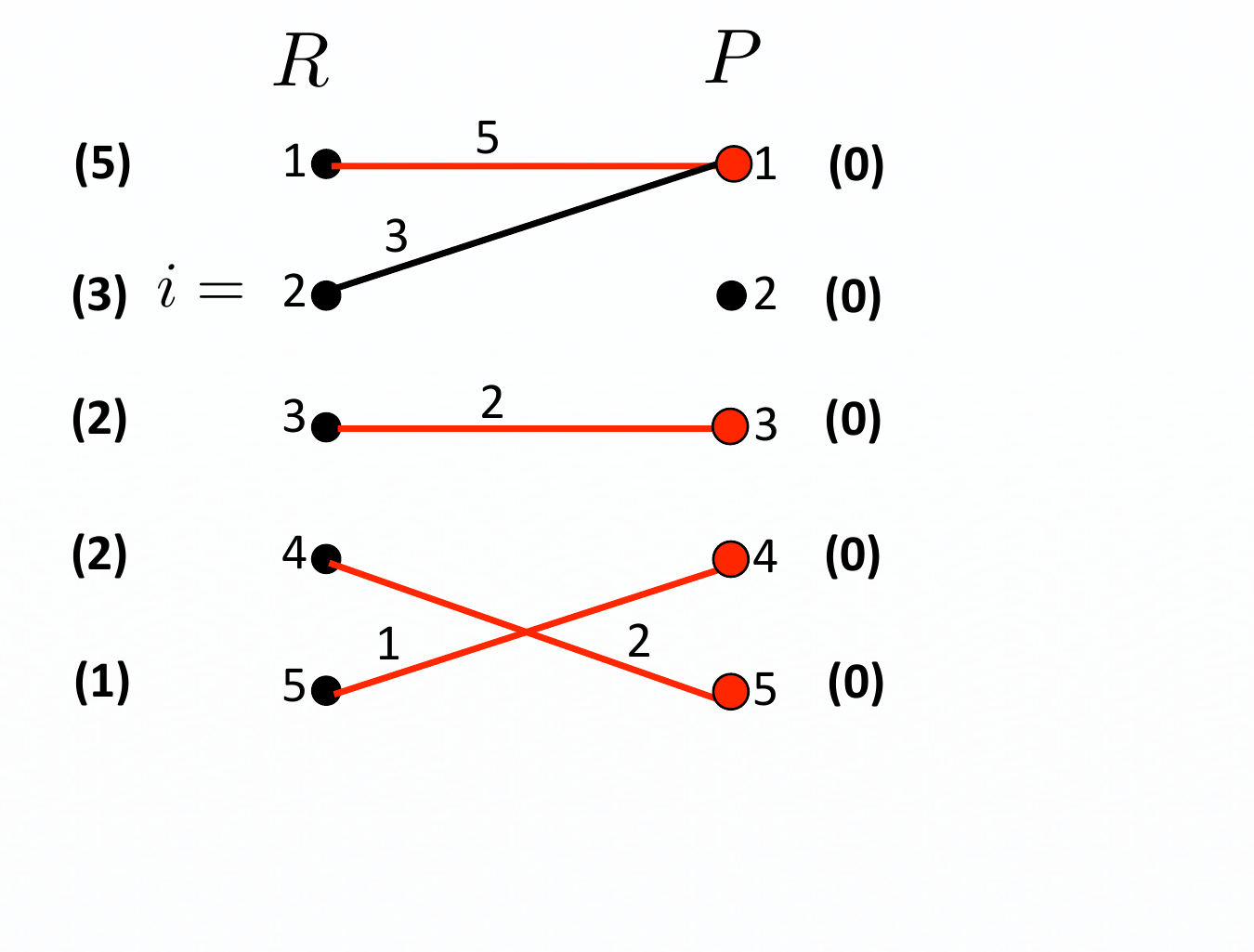}
\label{subfig_disthung4}
}
\caption{An instance of the \textit{Distributed-Hungarian} algorithm. (a) Robot $2$'s original information, (b) Example output for the $\mathrm{Build\_Latest\_Graph}$ function, (c) \& (d) Within the $\mathrm{Local\_Hungarian}$ function: isolated set $E_y$, and corresponding $M_y$ and $V_c$ respectively.}
\label{fig_disthunginit}
\end{figure}
\begin{figure}
\subfloat[The isolated set of candidate edges $E_{cand}$ in $\mathbb{G}_{tmp}$ (none in this case).]{
\includegraphics[scale=\scalefactortwo]{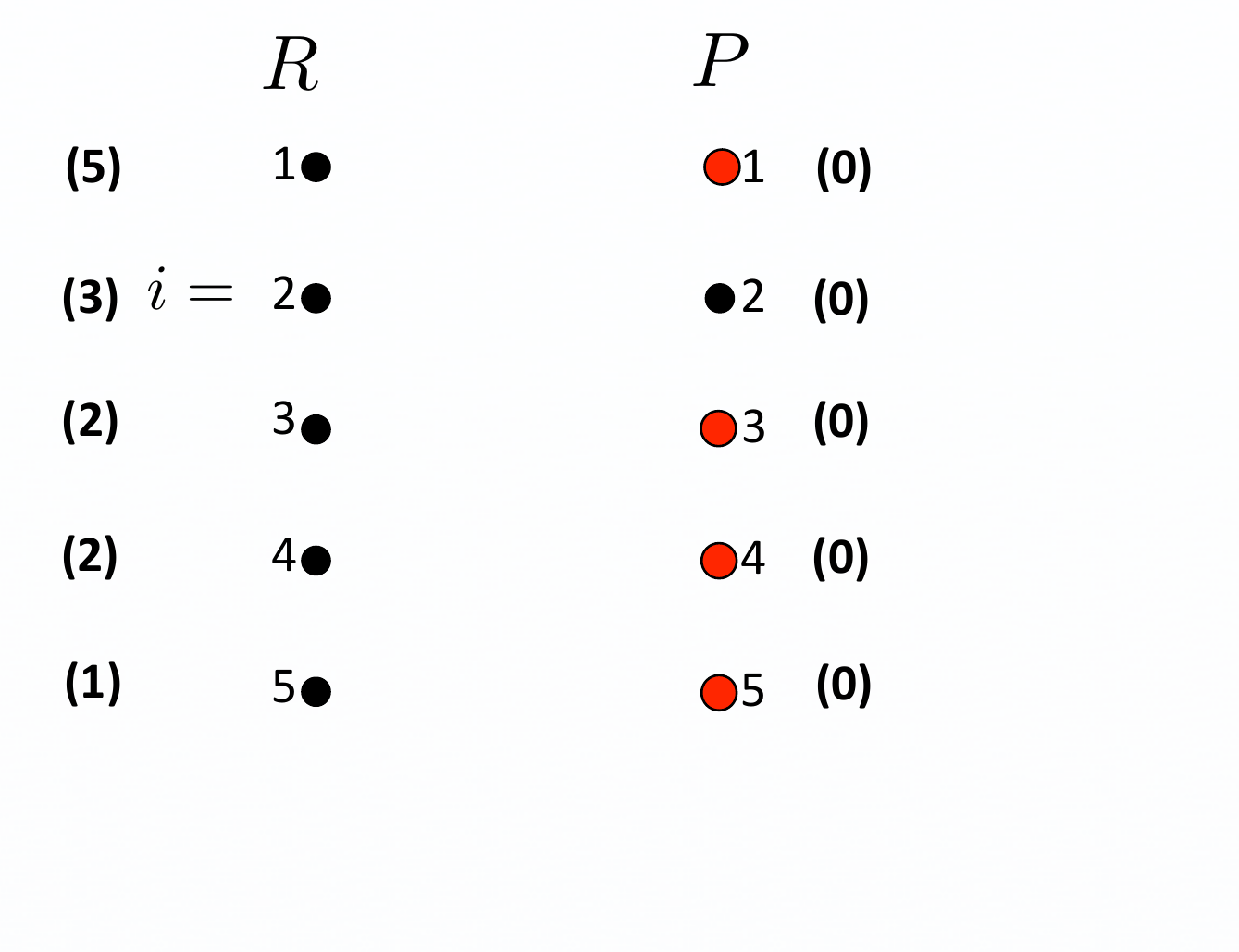}
\label{subfig_disthung6}
}
\hfill
\subfloat[Given $V_{c}$, Robot $2$'s set of candidate edges (green edges, only one in this case) using its original information $G^2_{orig}$. Exactly one edge with minimum slack $\delta$ is chosen for inclusion in $E_{cand}$.]{
\includegraphics[scale=\scalefactortwo]{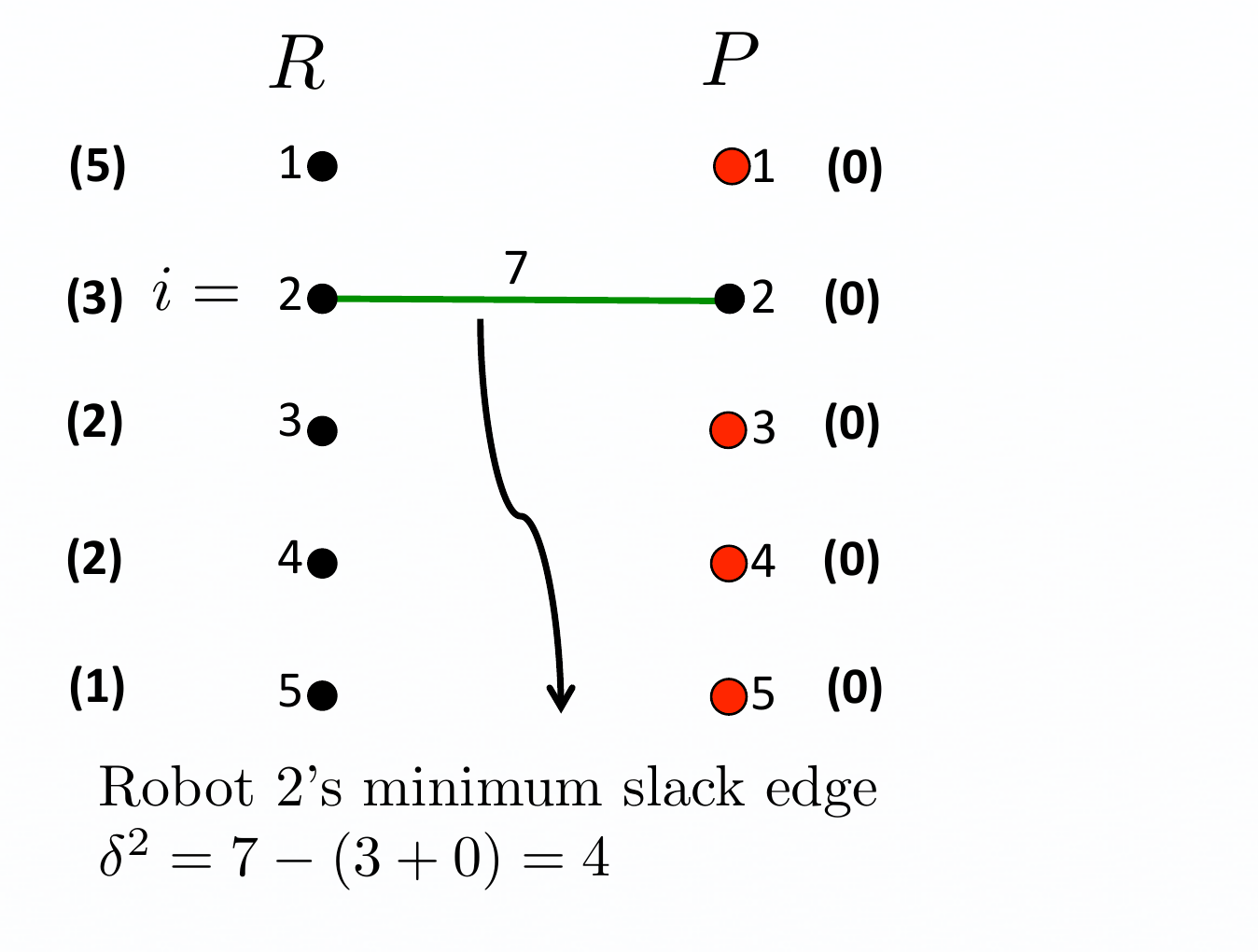}
\label{subfig_disthung5}
}
\hfill
\subfloat[The updated set of candidate edges $E_{cand}$ (green edges), combining the edges from Figures \ref{subfig_disthung5} and \ref{subfig_disthung6}. Note that the number of edges is \textit{not sufficient} to proceed.]{
\includegraphics[scale=\scalefactortwo]{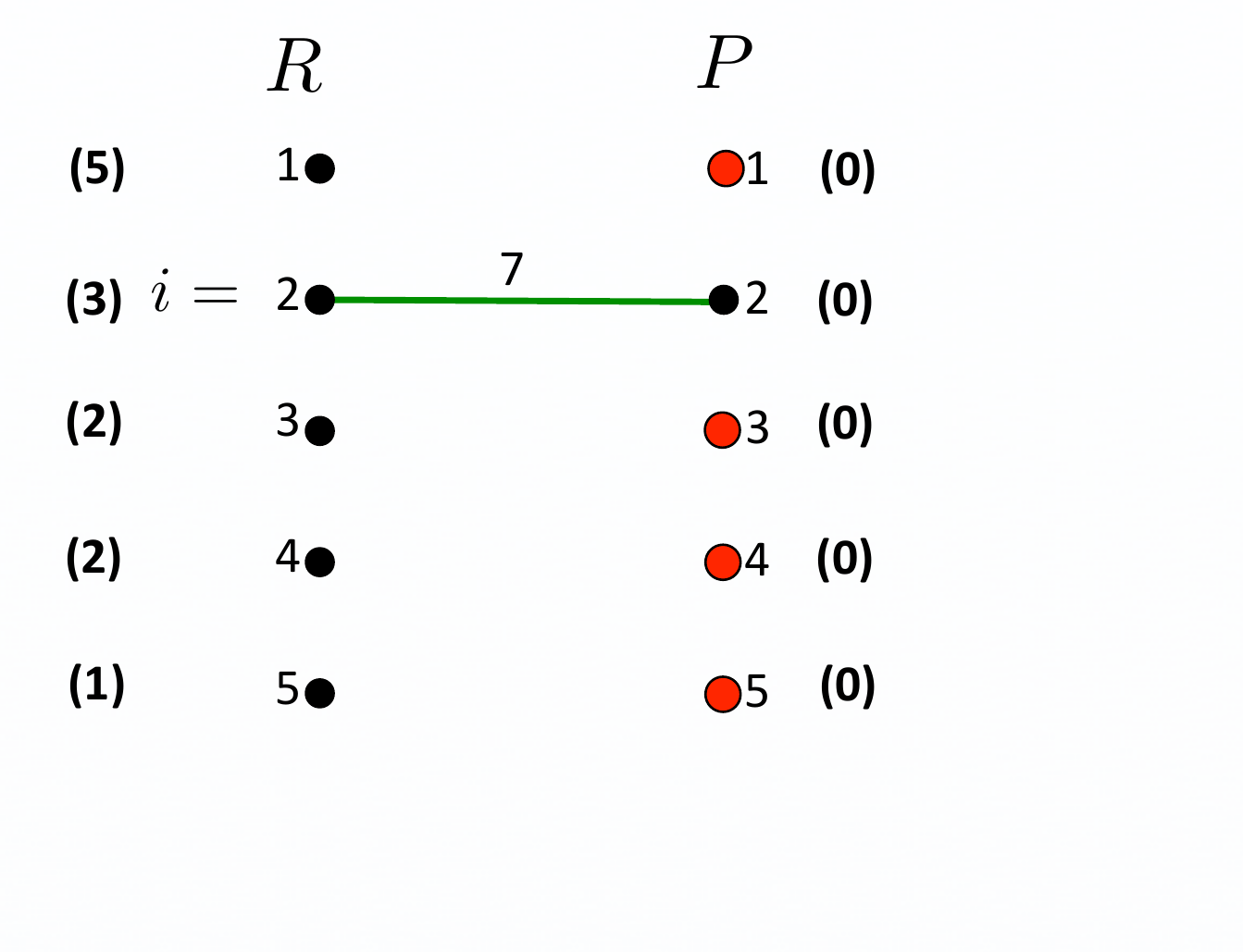}
\label{subfig_disthung7}
}
\hfill
\subfloat[Robot $2$'s outgoing message $msg^2 = \mathbb{G}^2$, comprising of the graph $G^2_{lean} = (V, E^2_{lean}, w^2_{lean})$, the vertex labeling function $y^2$, and the counter value $\gamma^2 = 0$.]{
\includegraphics[scale=\scalefactortwo]{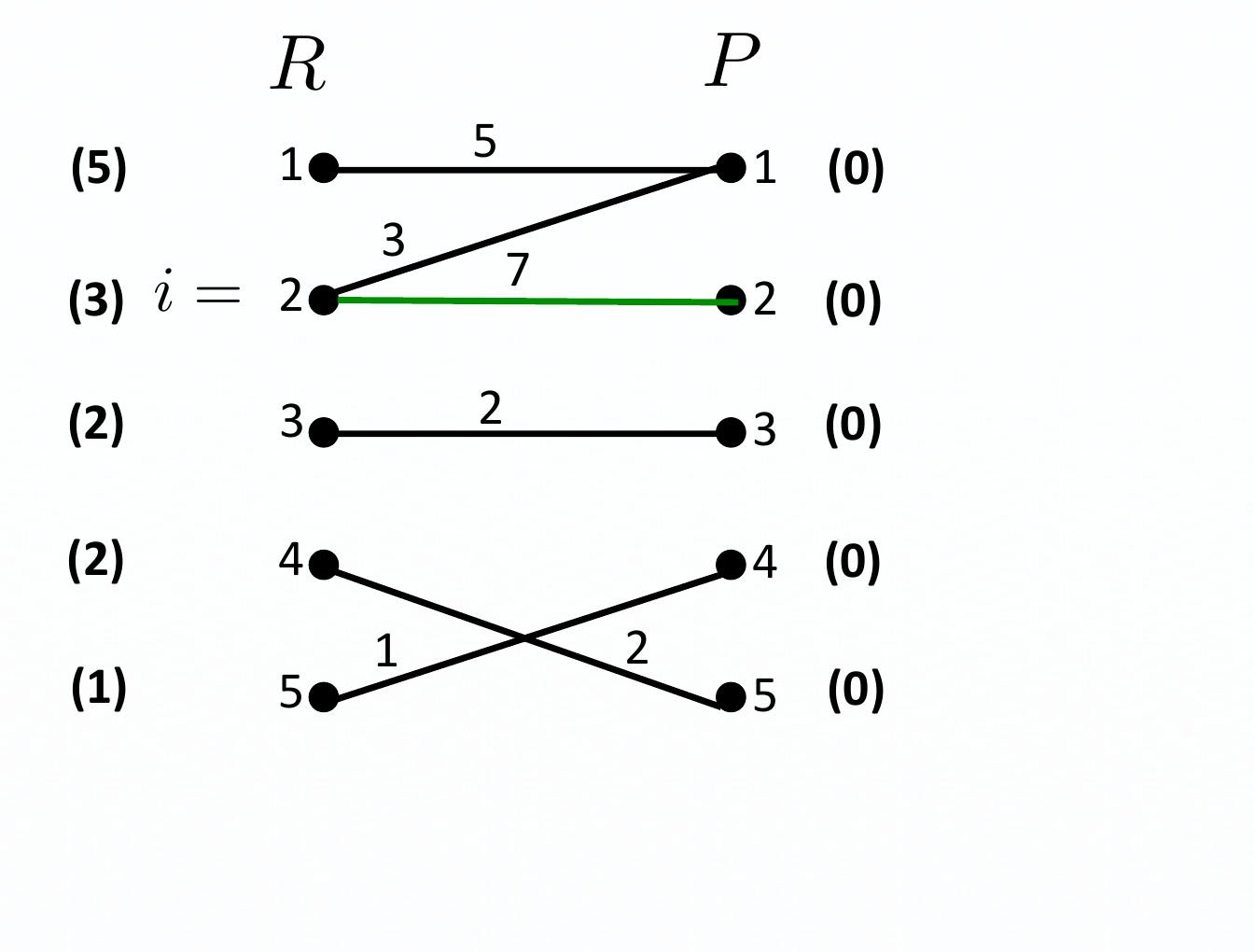}
\label{subfig_disthung8}
}
\caption{Contd. from Figure \ref{fig_disthunginit}. (a), (b) \& (c) Within the $\mathrm{Local\_Hungarian}$ function: isolated set $E_{cand}$, output of the $\mathrm{Get\_Best\_Edge}$ function, updated set $E_{cand}$ with \textit{insufficient} edges, respectively and (d) Output of the $\mathrm{Local\_Hungarian}$ function, i.e. Robot $2$'s outgoing message $msg^2 = \mathbb{G}^2$.}
\label{fig_disthungnotenufinfo}
\end{figure}
\begin{figure}
\subfloat[Robot 4's original information as the weighted, bipartite graph $G^4_{orig} = (V, E^4_{orig}, w^4_{orig})$.]{
\includegraphics[scale=\scalefactortwo]{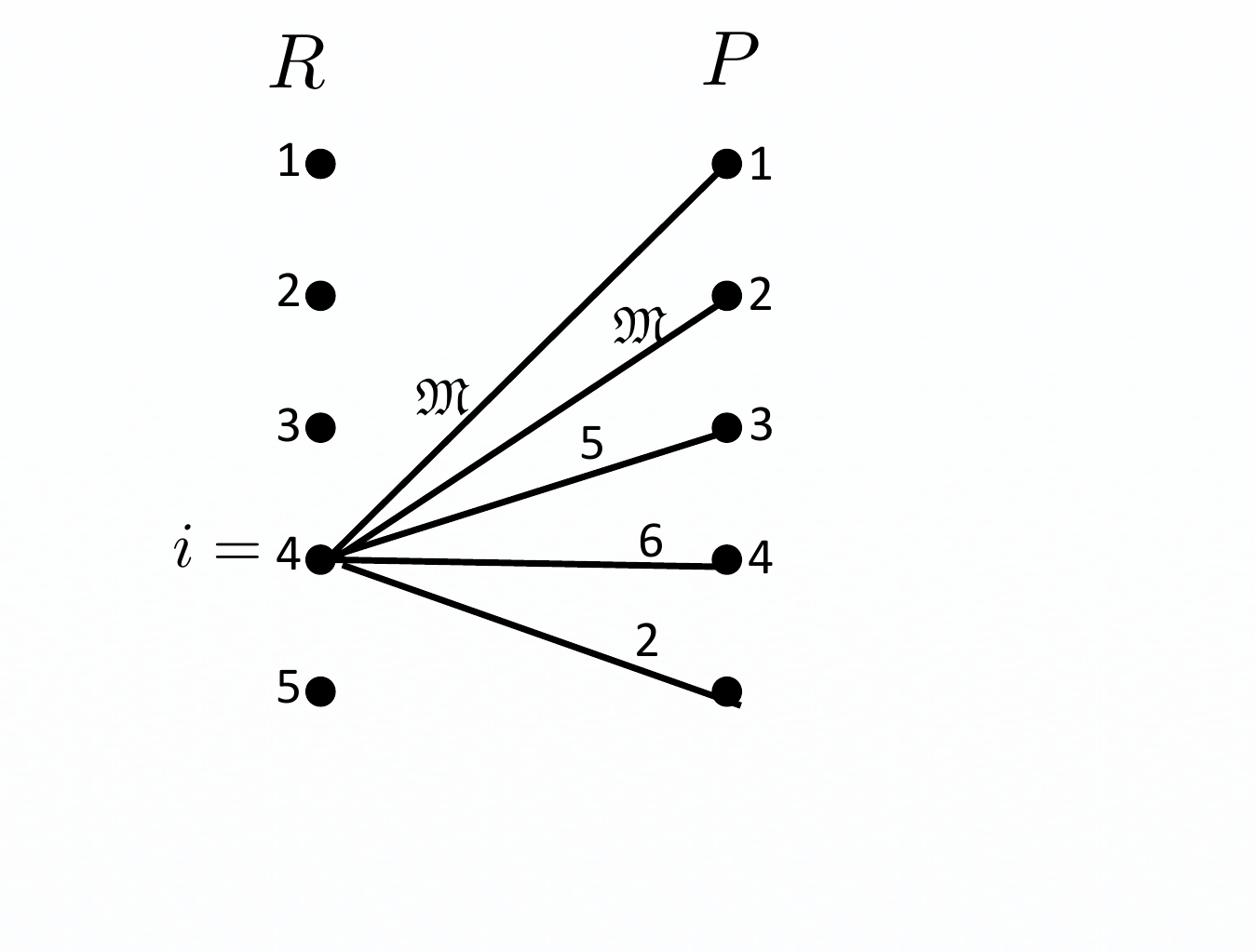}
\label{subfig_disthungstep1}
}
\hfill
\subfloat[Robot 4's temporary, most-updated state $\mathbb{G}_{tmp} = (G_{lean}, y, \gamma)$, with $\gamma = 0$. $E_{lean}$ contains the equality subgraph edges $E_y$ (black edges) and the candidate edges $E_{cand}$ (green edges).]{
\includegraphics[scale=\scalefactortwo]{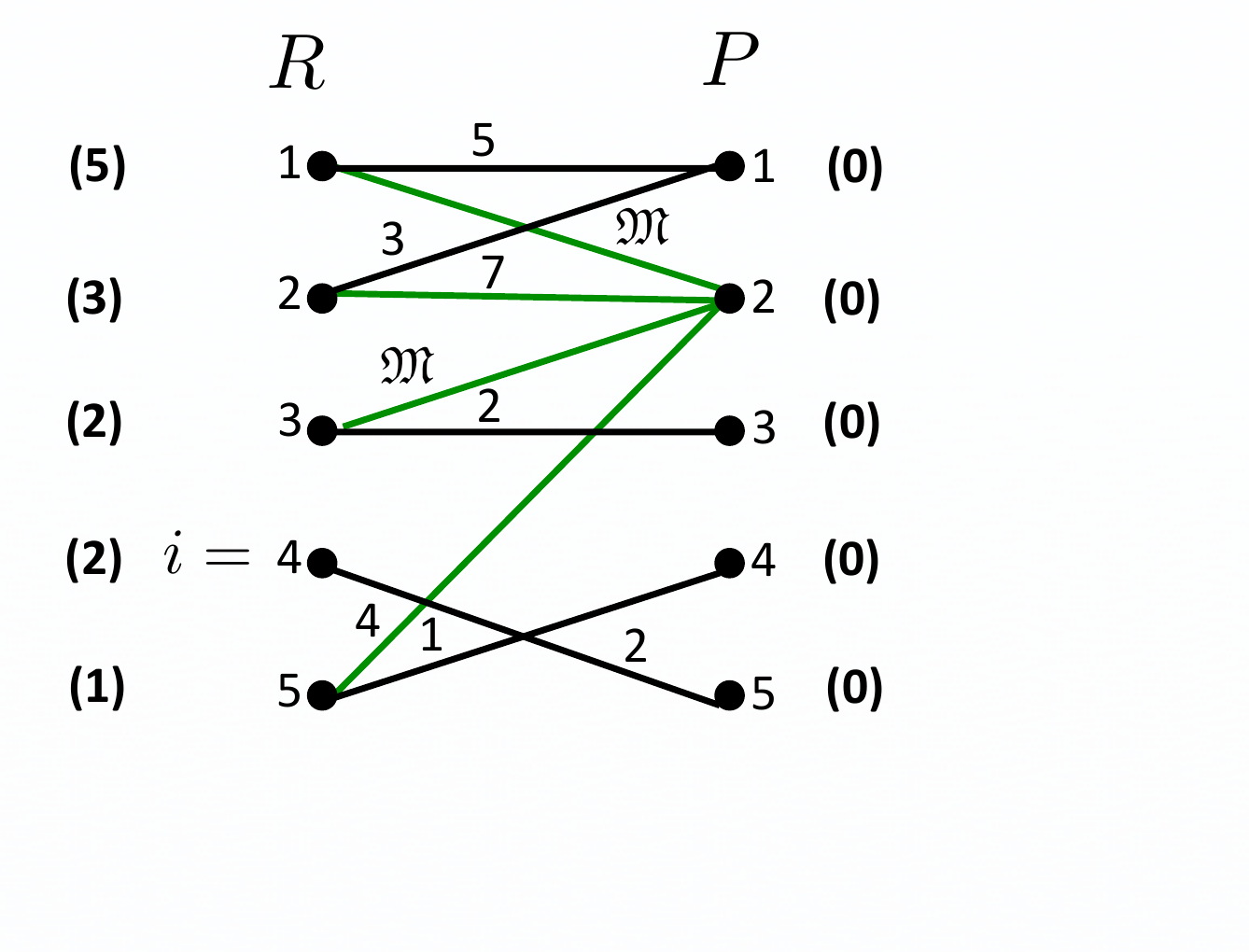}
\label{subfig_disthungstep2}
}
\hfill
\subfloat[The isolated set of equality subgraph edges $E_y$ in $\mathbb{G}_{tmp}$.]{
\includegraphics[scale=\scalefactortwo]{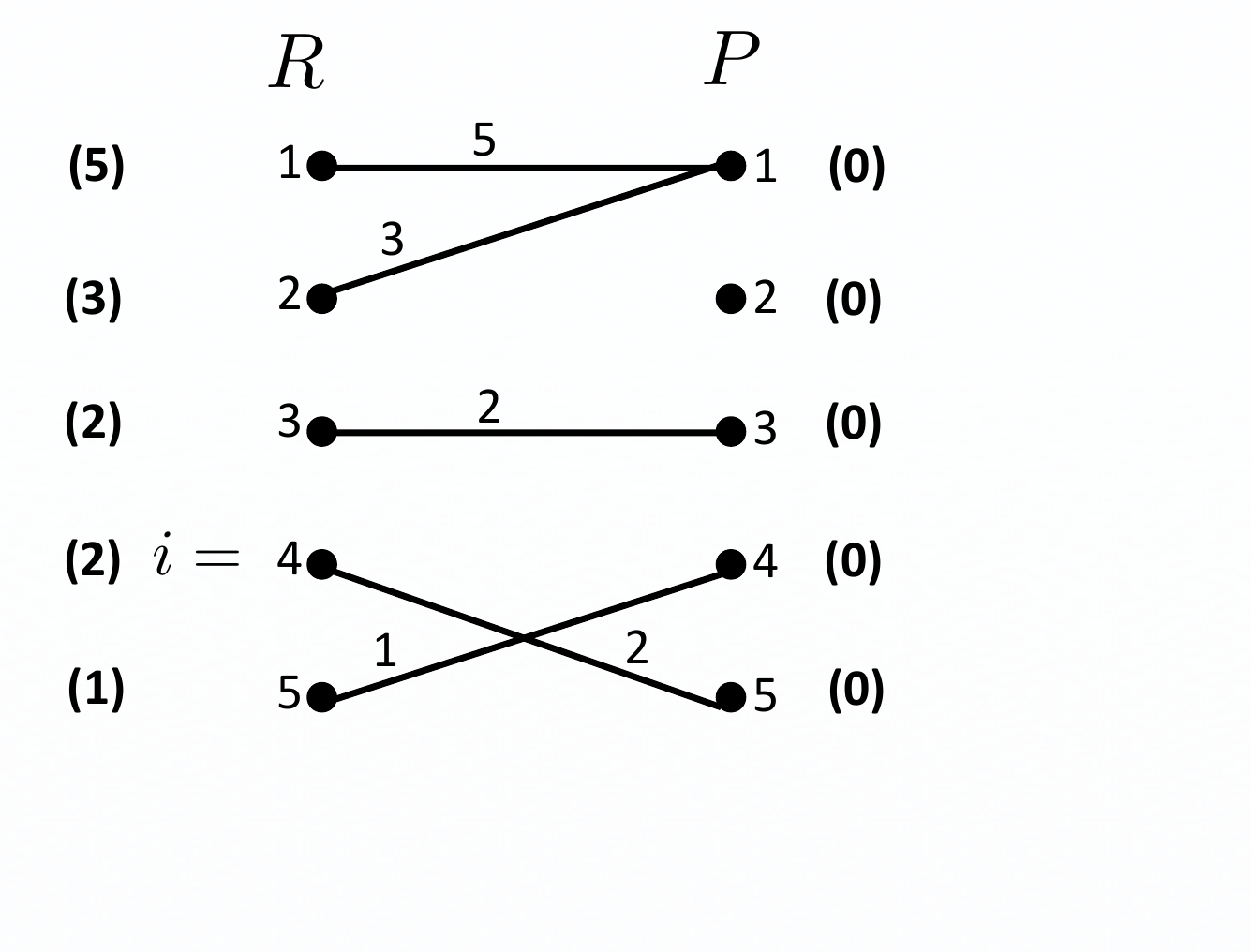}
\label{subfig_disthungstep3}
}
\hfill
\subfloat[Given $E_y$, a maximum cardinality matching $M_y$ (red edges), and a corresponding minimum vertex cover $V_{c} = (R_{c}, P_{c})$ (red vertices).]{
\includegraphics[scale=\scalefactortwo]{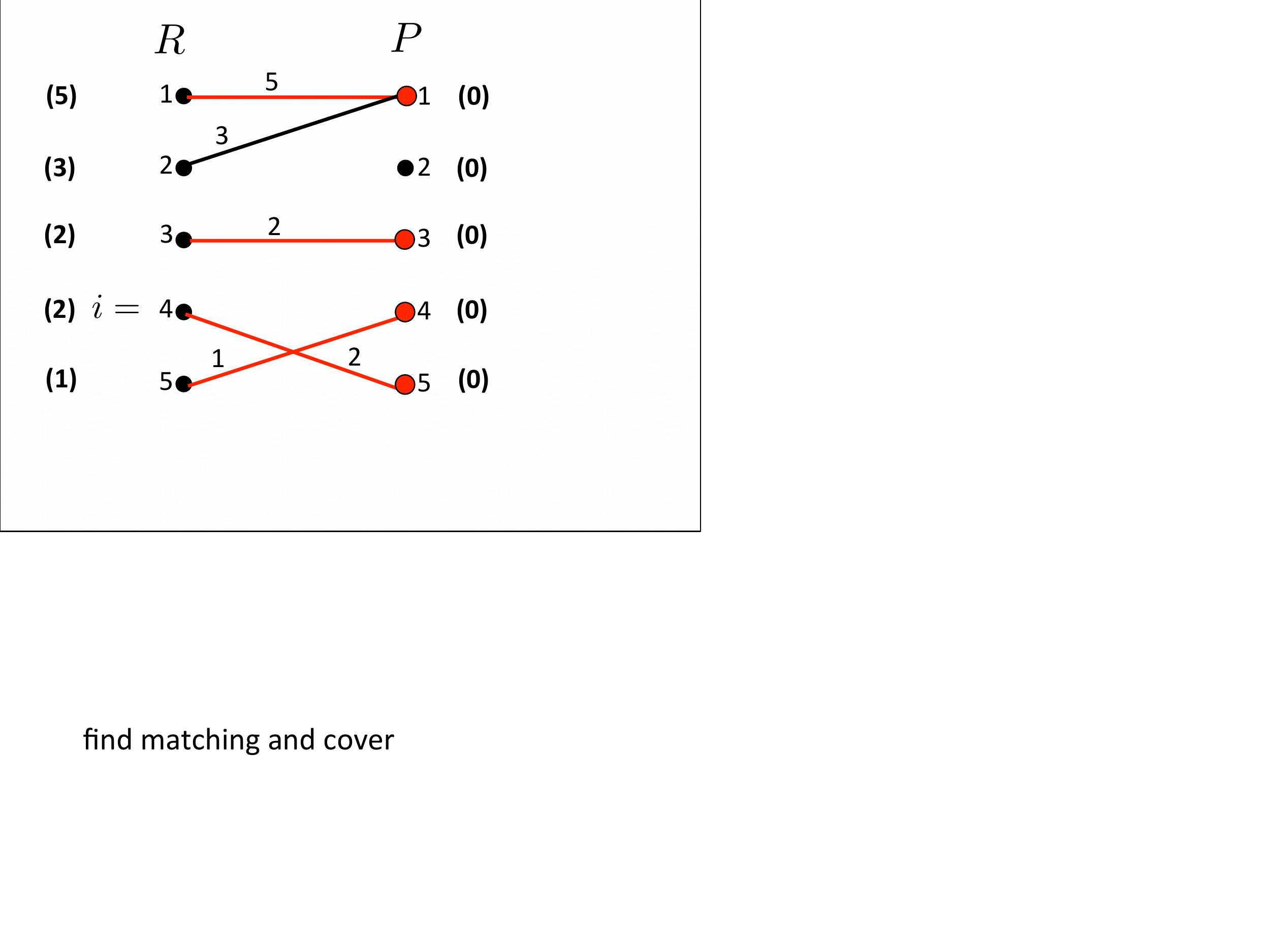}
\label{subfig_disthungstep4}
}
\caption{Another instance of the \textit{Distributed-Hungarian} algorithm. (a) Robot $4$'s original information, (b) Example output for the $\mathrm{Build\_Latest\_Graph}$ function, (c) \& (d) Within the $\mathrm{Local\_Hungarian}$ function: isolated set $E_y$, and corresponding $M_y$ and $V_c$ respectively.}
\label{fig_disthungstep12(1)}
\end{figure}
\begin{figure}
\hfill
\subfloat[The isolated set of candidate edges $E_{cand}$ in $\mathbb{G}_{tmp}$.]{
\includegraphics[scale=\scalefactortwo]{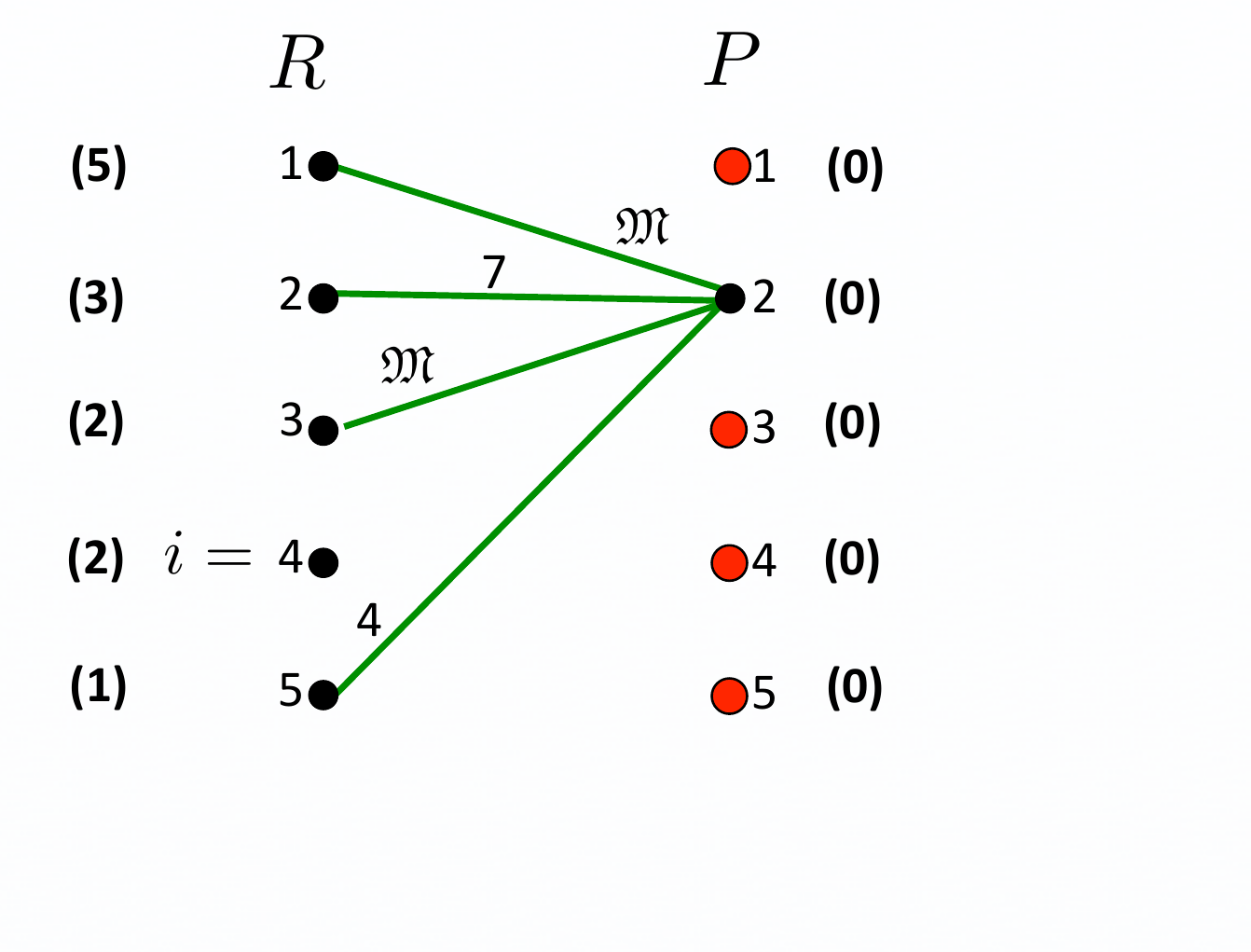}
\label{subfig_disthungstep6}
}
\hfill
\subfloat[Given $V_{c_y}$, Robot $4$'s set of candidate edges (green edges) using its original information $G^4_{orig}$. Exactly one edge (with corresponding minimum slack $\delta$) is chosen for inclusion in $E_{cand}$.]{
\includegraphics[scale=\scalefactortwo]{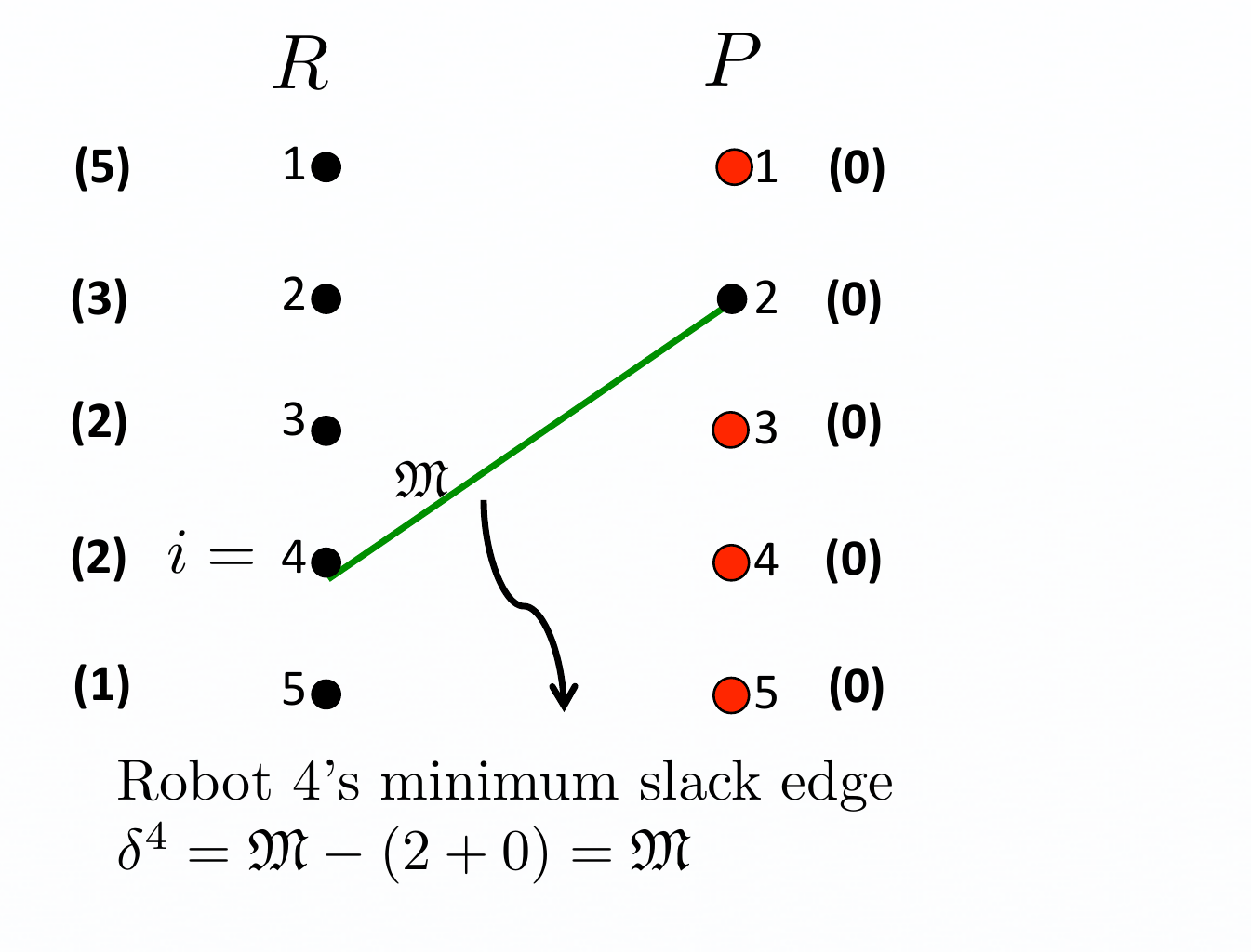}
\label{subfig_disthungstep5}
}
\hfill
\subfloat[ \ref{labelstep1}(a): The updated set of candidate edges $E_{cand}$ (green edges), combining the edges from Figures \ref{subfig_disthungstep5} and \ref{subfig_disthungstep6}. \ref{labelstep1}(b): Since the number of edges is sufficient to proceed, the minimum slack edge is identified.]{
\includegraphics[scale=\scalefactortwo]{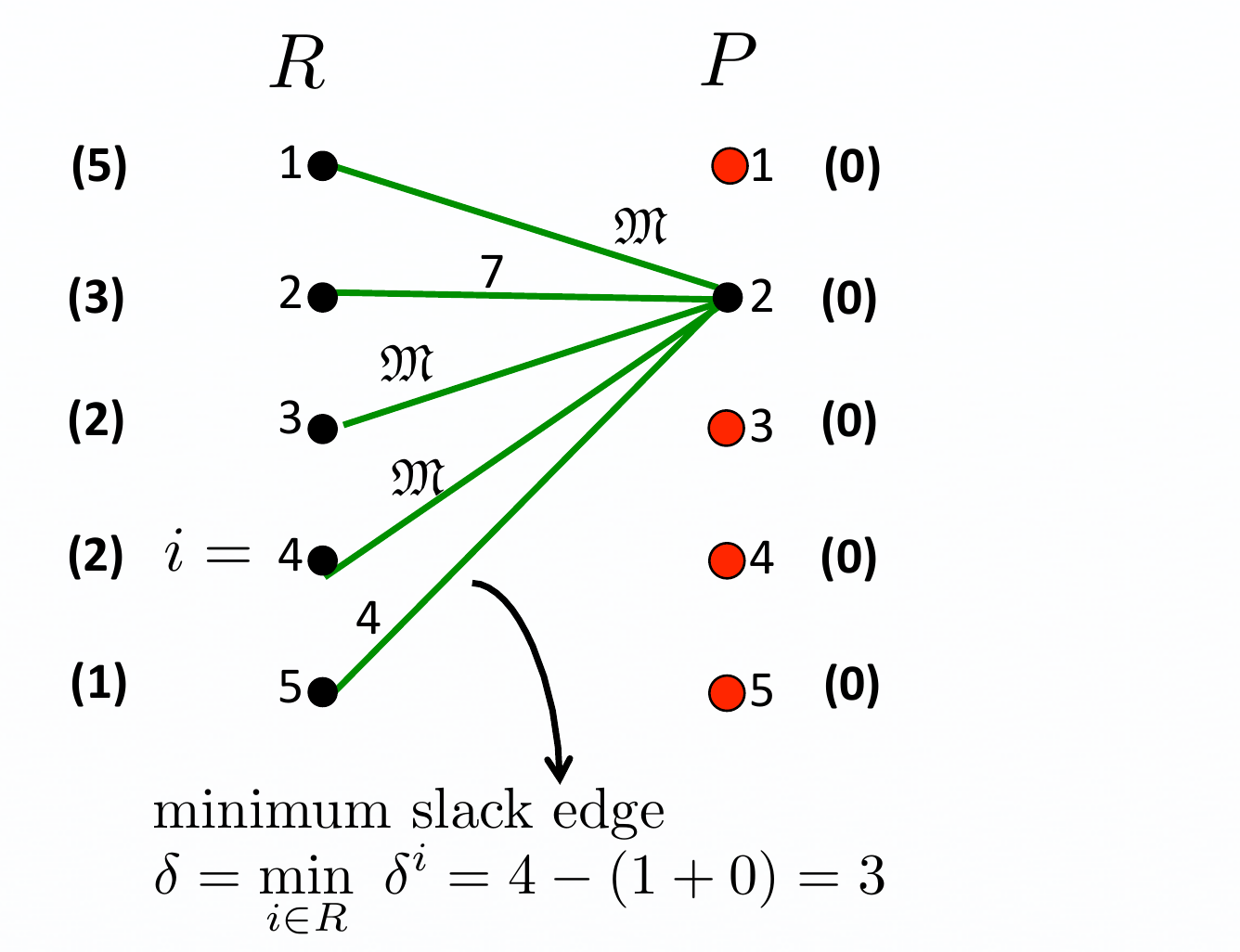}
\label{subfig_disthungstep7}
}
\hfill
\subfloat[ \ref{labelstep1}(b) contd.: The updated feasible vertex labeling function $y$ (highlighted in yellow), using the minimum slack $\delta$.]{
\includegraphics[scale=\scalefactortwo]{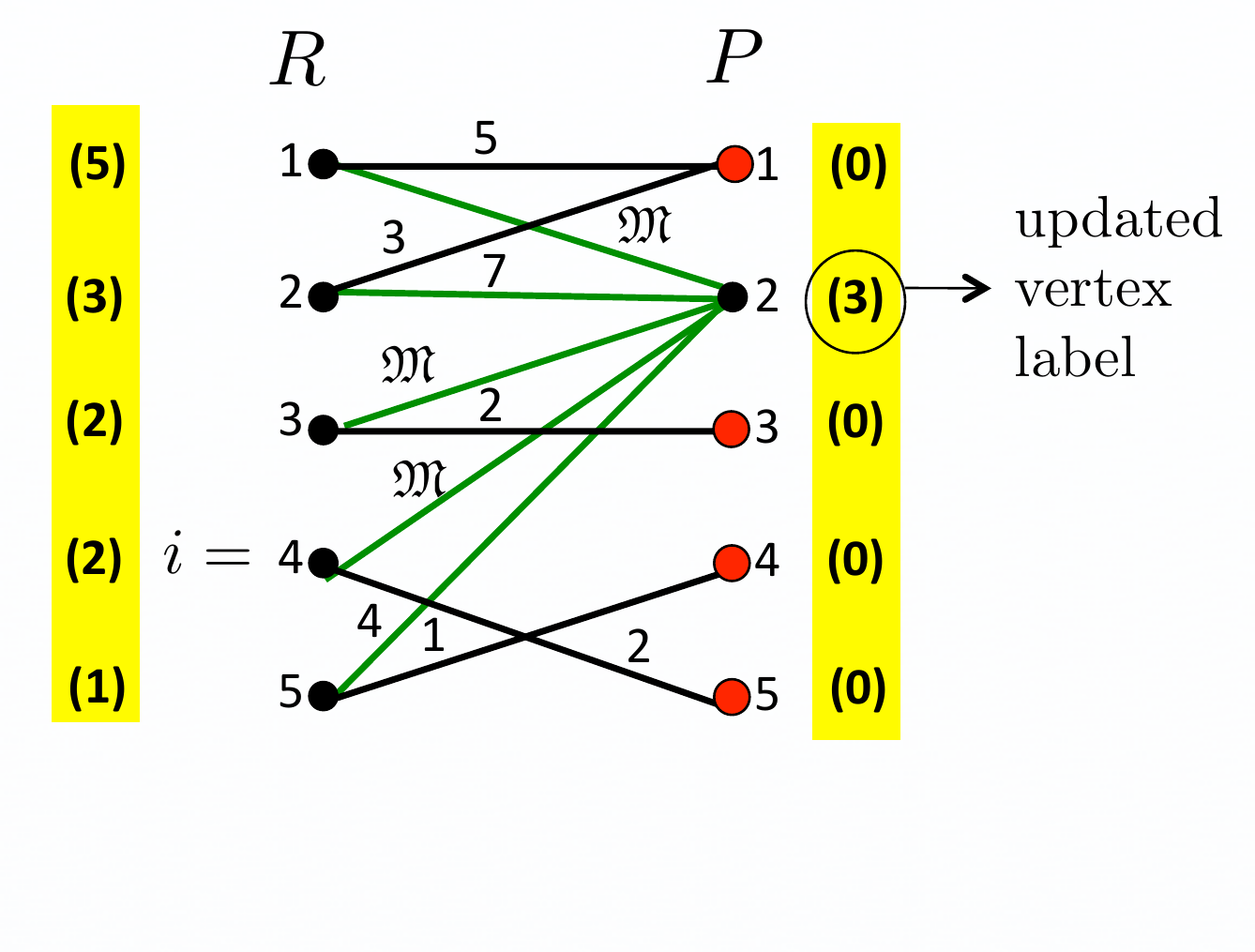}
\label{subfig_disthungstep8}
}
\caption{Contd. from Figure \ref{fig_disthungstep12(1)}. Within the $\mathrm{Local\_Hungarian}$ function: (a) Isolated set $E_{cand}$, (b) Output of the $\mathrm{Get\_Best\_Edge}$ function, (c) Updated set $E_{cand}$ with \textit{sufficient} edges (\ref{labelstep1}(a)), and identified minimum slack edge (\ref{labelstep1}(b)) and (d) Updated $y$ (\ref{labelstep1}(b) contd.).}
\label{fig_disthungstep12(2)}
\end{figure}
\begin{figure}
\subfloat[ \ref{labelstep2}: For the updated $y$, the corresponding set of equality subgraph edges $E_y$ (with the new, added edge highlighted in yellow).]{
\includegraphics[scale=\scalefactortwo]{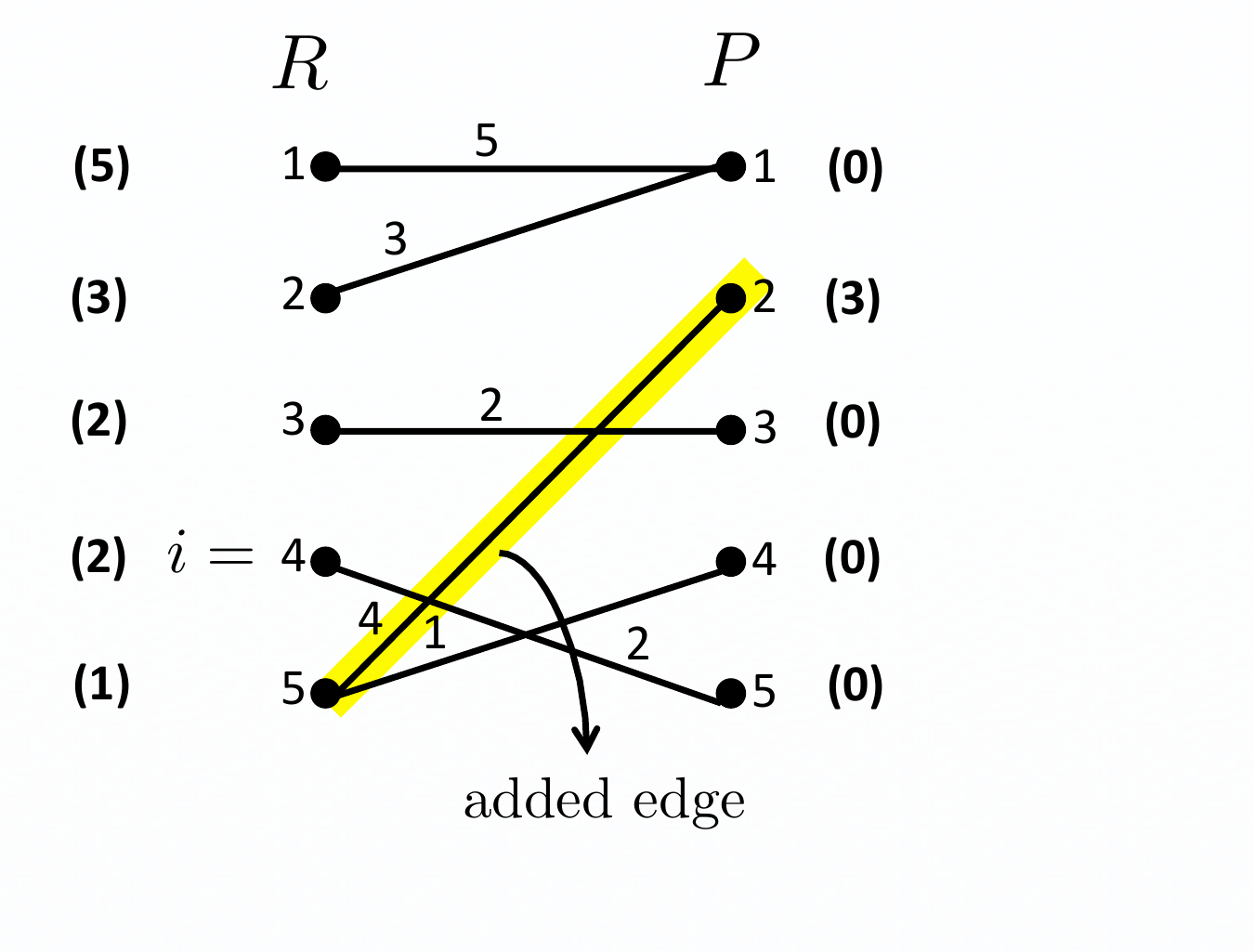}
\label{subfig_disthungstep9}
}
\hfill
\subfloat[ \ref{labelstep2} contd.: For the updated $E_y$, a maximum cardinality matching $M_y$ (red edges), and a corresponding minimum vertex cover $V_{c} = (R_{c}, P_{c})$ (red vertices).]{
\includegraphics[scale=\scalefactortwo]{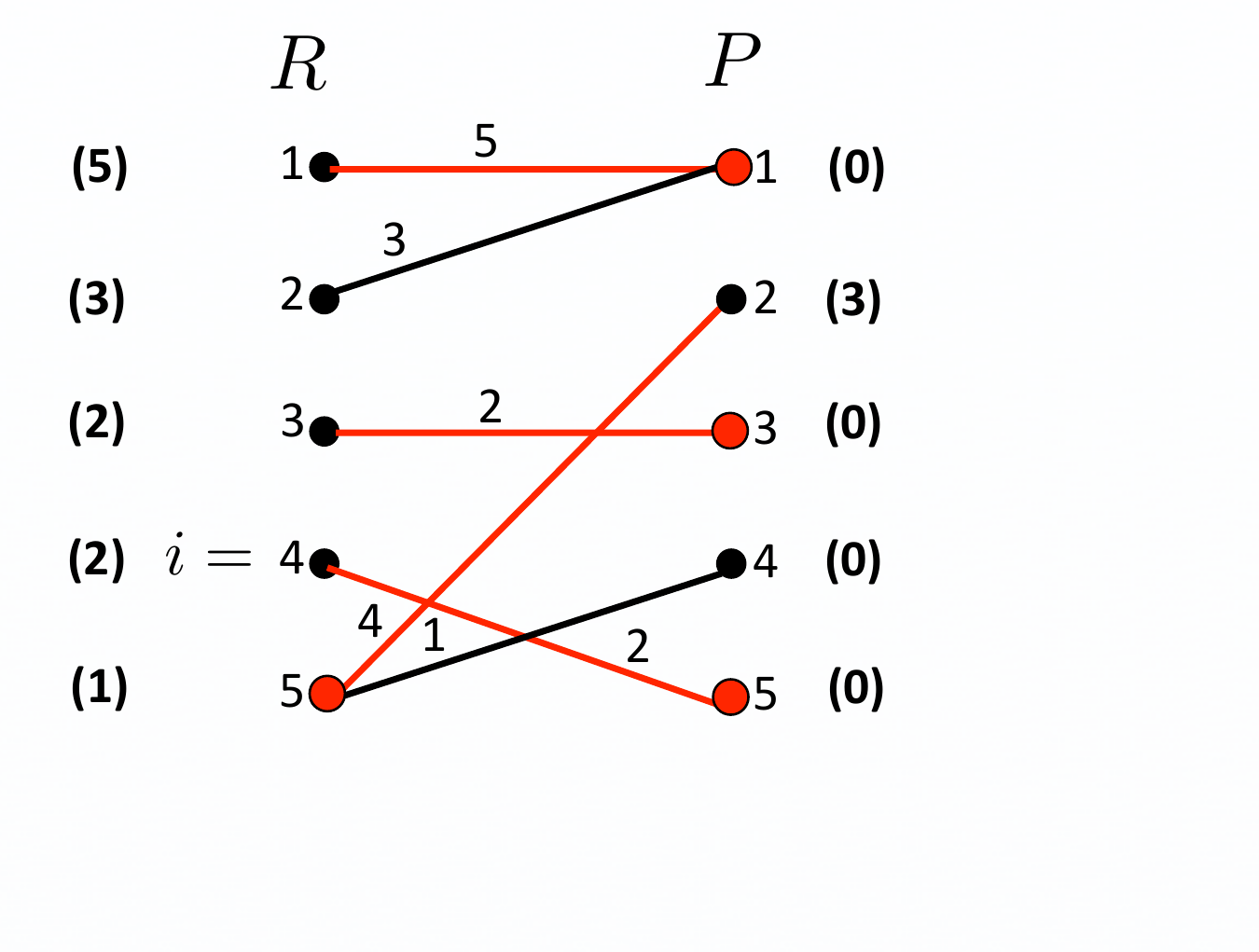}
\label{subfig_disthungstep10}
}
\hfill
\subfloat[For the updated $V_{c}$, Robot $4$'s set of candidate edges (green edges) using its original information $G^4_{orig}$. Exactly one edge with minimum slack $\delta$ is chosen for inclusion in $E_{cand}$]{
\includegraphics[scale=\scalefactortwo]{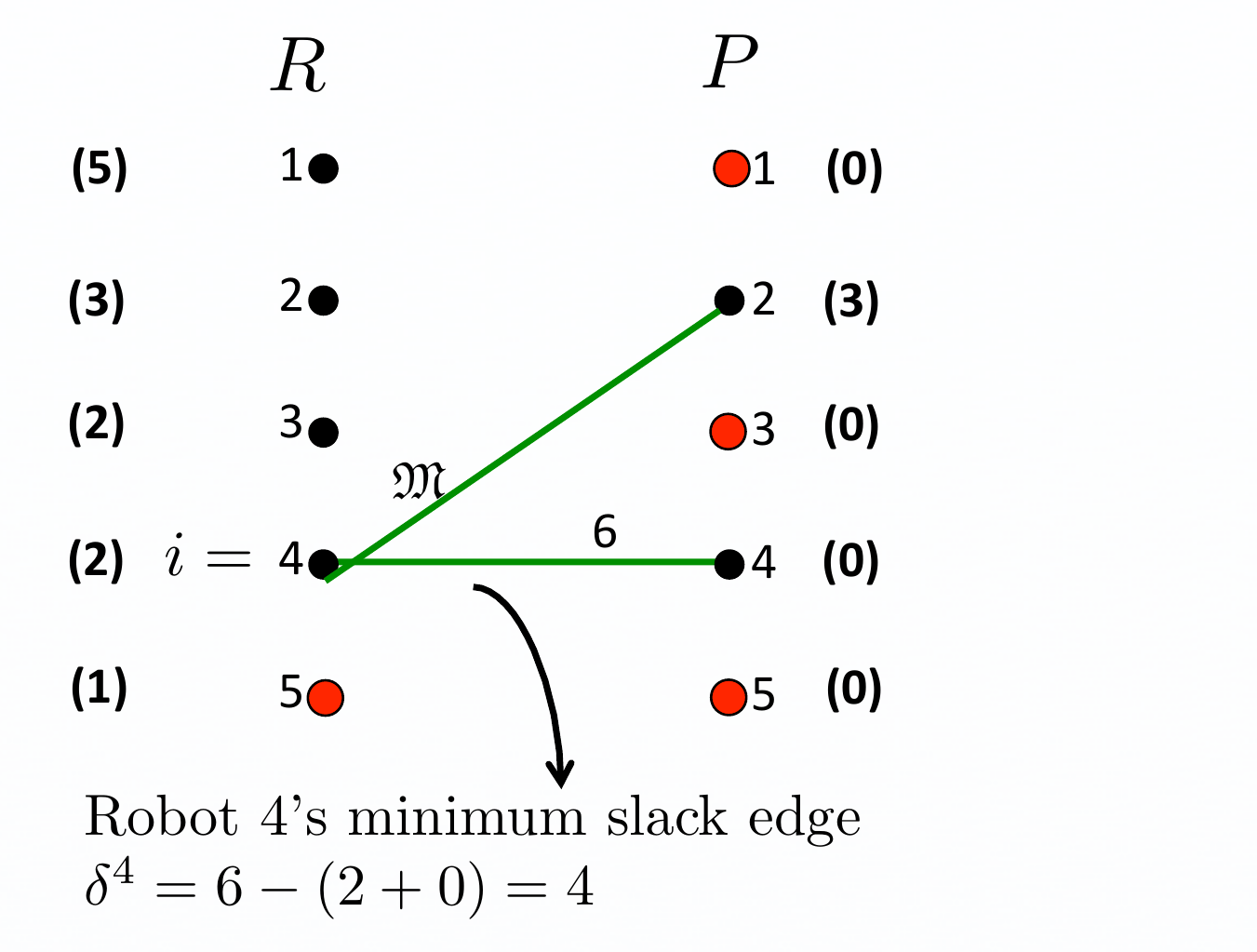}
\label{subfig_disthungstep11}
}
\hfill
\subfloat[Robot $4$'s outgoing message $msg^4 = \mathbb{G}^4$, comprising of the graph $G^4_{lean} = (V, E^4_{lean}, w^4_{lean})$, the vertex labeling function $y^4$, and the counter value $\gamma^4 = 1$.]{
\includegraphics[scale=\scalefactortwo]{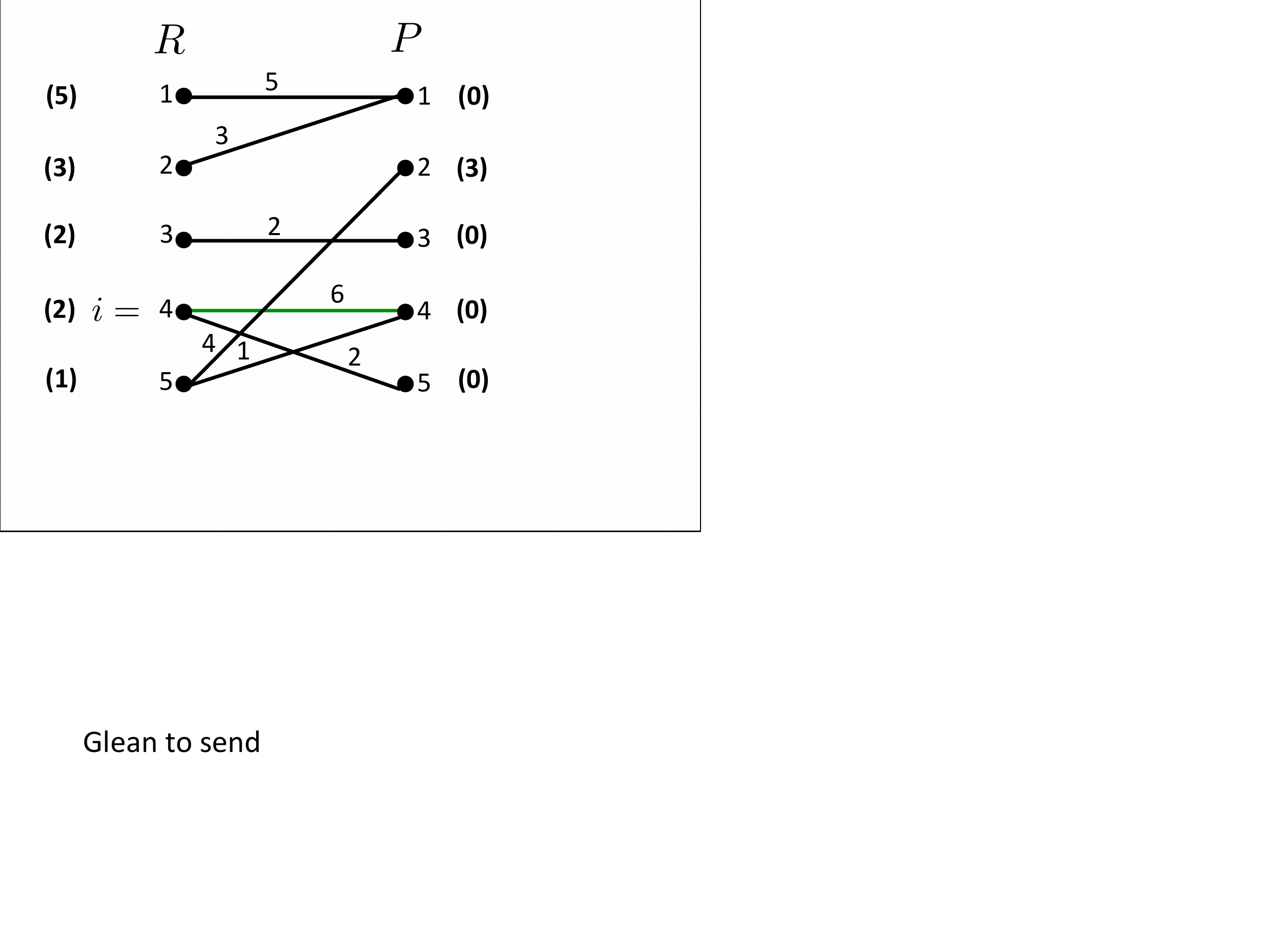}
\label{subfig_disthungstep12}
}
\caption{Contd. from Figure \ref{fig_disthungstep12(2)}. (a), (b) \& (c) Within the $\mathrm{Local\_Hungarian}$ function: updated set $E_y$ (\ref{labelstep2}), corresponding $M_y$ and $V_c$ (\ref{labelstep2} contd.), and output of the $\mathrm{Get\_Best\_Edge}$ function respectively, and (d) Output of the $\mathrm{Local\_Hungarian}$ function, i.e. Robot $4$'s outgoing message $msg^4 = \mathbb{G}^4$.}
\label{fig_disthungstep12(3)}
\end{figure}
\begin{theorem}
\label{thm_main}
Given a set of robots $R$, a set of targets $P$, and a time-varying
communication graph $\mathcal{G}_c(t)$, $t \in \mathbb{R}_{\geq 0}$, satisfying
Assumption~\ref{ass_comm}, assume every robot $i\in R$ knows $(R, P , c^i)$. If
the robots execute the \textit{Distributed-Assignment} algorithm, there exists a
finite time $T_f$ such that all robots converge to a common assignment
$\hat{M}$, which is an optimal solution of the (centralized) assignment problem,
i.e., $\hat{M} \in \mathcal{M}$.
\end{theorem}
\begin{IEEEproof}
  We proceed to prove the theorem in the following three steps: (i) for every robot $i$ in $R$, the 
  counter value $\gamma^i$ evolves as a monotone, non-decreasing sequence that converges in finite time, (ii) once all the counter values have
  converged, they must be at the same value, and (iii) at steady-state, a common,
  perfect matching corresponding to the optimal assignment is computed.
   
  By the connectivity assumption (Assumption \ref{ass_comm}), there exists a finite time
  interval in which at least one robot $i$ is able to construct a
  ``most-updated'' state $\mathbb{G}_{tmp}$ that contains \textit{enough
    candidate edges} to perform \ref{labelstep1} and \ref{labelstep2} in the
  $\mathrm{Local\_Hungarian}$ function, thereby incrementing its counter value from
  $n$ to $n+1$. Moreover, from Lemma \ref{lemma:unique_and_feasible_y}, we know that such a counter value increment  satisfies the conditions in Lemma \ref{lem_it}. As such, we can use the proof-sketch in Remark \ref{rem:hungarian_complexity} to show that there exists only a finite 
  number of such counter value increments (worst case $\mathcal{O}(r^2)$, with
  $r=|R|$) before which the matching found is perfect. Moreover, by construction, once a robot computes a perfect (and hence optimal) matching, the counter value stops increasing. Therefore, each so-called counter-sequence must converge in finite time.

 Once all the counter-sequences have converged, they must be at the same
  value. Indeed, if this is not the case, by Assumption~\ref{ass_comm}, there
  must exist two robots, say $i$ and $j$ at some time $t \in T_s$, such that
  $\gamma^i < \gamma^j$, and $j$ is an in-neighbor to $i$. By construction, robot $i$ would receive a message from robot $j$ and set its counter value $\gamma^i = \gamma^j$, contradicting the fact that all the sequences have converged.

  Next, at some time $t$, let $\gamma^i = \gamma^j = n$, $\forall i,j \in R$. Then by Lemma
  \ref{lemma:unique_and_feasible_y} and Corollary \ref{corollary:equal_gammas},
  we know that all robots have the same vertex labeling
  function $y_n$ that is also a \textit{feasible labeling} for the centralized graph $G$, and the same maximal matching $M_n$ respectively. Thus, if
  $M_n$ is a perfect matching, then by the Kuhn-Munkres Theorem (Theorem
  \ref{thm_hung}), it is an optimal solution to the centralized assignment
  problem.
  Suppose, by contradiction, $M_n$ is not a perfect matching. In that case,
  every robot contributes a candidate edge to its state (if such an edge
  exists), and sends it to its outgoing neighbors. As mentioned previously, there exists a finite time
  interval in which at least one robot $i$ is able to construct a
  ``most-updated'' state $\mathbb{G}_{tmp}$ that contains \textit{enough
    candidate edges} to update its counter value. This contradicts the fact that the counter-sequences of all the robots 
  have converged and concludes the proof.  
\end{IEEEproof}

Note that since we describe a synchronous implementation of the \textit{Distributed-Hungarian} algorithm, we can take the results from Theorem \ref{thm_main} one step further and quantify both, the stopping criterion and convergence time, in terms of \textit{iterations} (communication rounds) of the algorithm. As such, we provide the following auxiliary lemma,

\begin{lemma}[Information Dispersion]
\label{lem_info}
If robot $i$ sends its information to its outgoing neighbors (which in turn propagate the information forward, and so on), then within a maximum of $(r-1)$ iterations (communication rounds) of the \textit{Distributed-Hungarian} algorithm, robot $i$'s information reaches every other robot in the network.
\end{lemma}

\begin{IEEEproof}[Proof sketch]
By Assumption \ref{ass_comm}, at every iteration (time instant $t \in T_s$) of the \textit{Distributed-Hungarian} algorithm, the underlying dynamic, directed communication network $\mathcal{G}_c(t)$ is strongly connected. Such a criterion implies that at every time $t \in T_s$, there exists a \textit{directed path} (sequence of edges) from every robot to every other robot in the network. Thus, on the first iteration of the \textit{Distributed-Hungarian} algorithm, at least one robot $j$ is robot $i$'s out-neighbor and receives its information. It follows that on each consequent iteration, at least one \textit{new} robot is added to the set of robots that have already received robot $i$'s information, by virtue of being an out-neighbor to at least one of them, thus concluding the proof-sketch. 
\end{IEEEproof}

\begin{corollary}[Stopping Criterion]
\label{cor_stop}
Suppose robot $i$ finds a perfect matching on some iteration of the
\textit{Distributed-Hungarian} algorithm. Then robot $i$ can stop sending its
corresponding message $msg^i$ after $(r-1)$ iterations (communication rounds).
\end{corollary}
\begin{IEEEproof}
Let $M^i$ denote the perfect matching found by robot $i$. As discussed in the
proof of Theorem \ref{thm_main}, $M^i$ is also optimal with respect to the
centralized assignment problem, and can be denoted by $\hat{M}$. Thus, using Lemma \ref{lem_info}, within a maximum of
$(r-1)$ iterations, every robot in the network will receive robot $i$'s message,
and update its own information to robot $i$'s solution, at which point, robot
$i$ need not send its message anymore.
\end{IEEEproof}

\vspace{0.5em}

\begin{corollary}[Convergence Time]
A common, optimal solution $\hat{M} \in \mathcal{M}$ is found in $\mathcal{O}(r^3)$ iterations (communication rounds) of the \textit{Distributed-Hungarian} algorithm.
\end{corollary}
\begin{IEEEproof}
From Theorem \ref{thm_main}, we know that the convergence of a counter-sequence implies that a common, optimal solution has been found (where the number of counter value increments cannot exceed $\mathcal{O}(r^2)$). Moreover, from Lemma \ref{lem_info}, we know that within $(r-1)$ iterations of the algorithm, the highest counter value among all robots is incremented (irrespective of the robot it belongs to). Thus, a common, optimal solution $\hat{M}$ is found in $\mathcal{O}(r^3)$ iterations.  \end{IEEEproof}

\vspace{0.5em}

\begin{remark}[Detecting Infeasibility]
  If the centralized assignment problem $(G, w)$ is infeasible, the
  \textit{Distributed-Hungarian} algorithm converges to a matching $\hat{M}$
  that contains edges with infeasible weights (i.e. denoted by $\mathfrak{M}$. \oprocend
\end{remark}

\vspace{0.5em}

\begin{remark}[Message Size]
Recall that a robot $i$'s message, $msg^i$, comprises of a sparse graph $G^i_{lean} = (V, E^i_{lean}, w^i_{lean})$ with at most $(2r-1)$ edges, a vertex labeling function $y^i: V \rightarrow \mathbb{R}$, and a counter value, $\gamma^i\in \mathbb{Z}$. Edges in $E_{lean}$ can be encoded with $\left \lceil{\frac{1}{4}\log_2(r)}\right \rceil$ bytes each, while edge weights and vertex labels can be encoded with 2 bytes each (approximating a real number as a 16-bit floating point value). Moreover, since the counter value represents the number of two-step iterations (maximum $r^2$ as per Remark \ref{rem:hungarian_complexity}), it can be encoded as an integer with $\left \lceil{\frac{1}{4}\log_2(r)}\right \rceil$ bytes. Thus, at each iteration (communication round) of the algorithm, $((2r)\,.\,(4 + \left \lceil{\frac{1}{4}\log_2(r)}\right \rceil) - 2)$ bytes are sent out by each robot. \oprocend
\end{remark}

\begin{remark}\label{rem:async}[Asynchronous Implementation]
As mentioned previously, the connectivity assumption (Assumption~\ref{ass_comm}) can be relaxed by requiring the communication graph to be only \textit{jointly} strongly connected over some time period $T_c$, lending towards an asynchronous framework. Also, in our implementation of the \textit{Distributed-Hungarian} algorithm, time $t$ is universal time, and as such, need not be explicitly known by the robots. Due to this independence, the \textit{Distributed-Hungarian} algorithm can indeed run asynchronously under the joint connectivity assumption. The proof sketch for such an implementation is similar to the synchronous case explained earlier in this section, and relies on the fact that within a maximum time interval $T_c$, there exists a time dependent directed path between every pair of robots, thus ensuring subsequent convergence. \oprocend
\end{remark}

\subsection{Simulation Experiments}
To assess the performance of our proposed \textit{Distributed-Hungarian} algorithm in practice, we performed simulation experiments on multiple instances of the LSAP with varying problem sizes, and plotted the average number of iterations required for convergence. In particular, the simulation experiments were performed in MATLAB, and executed on a PC with an Intel Quad Core i5, 3.3GHz CPU and 16GB RAM. For every $r$ (total number of robots) varying from 5 to 160, we performed 20 runs of the \textit{Distributed-Hungarian} algorithm, over randomly generated $(r \times r)$ cost matrices with cost $c_{i,j} \in (0, 1000)$, and a strongly connected, communication network with \textit{dynamic} incoming and outgoing edges between robots at each time instant of the synchronous implementation. 

\begin{figure}
\centering
\subfloat[Average number of iterations required for convergence, versus the number of robots.]{
\includegraphics[scale=0.3]{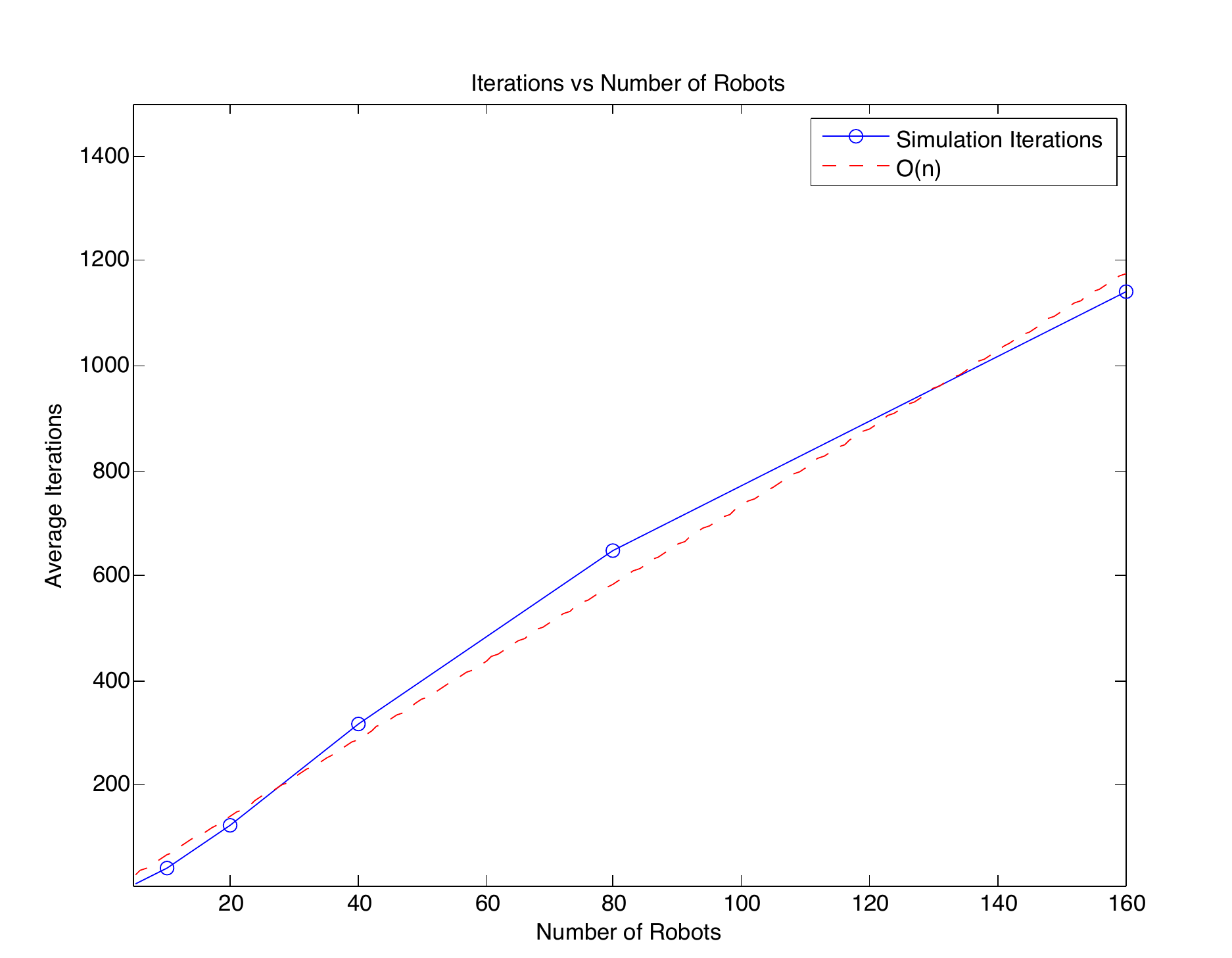}
\label{subfig_1}
}
\hfill
\subfloat[Average, worst-case computational time of an iteration, versus the number of robots.]{
\includegraphics[scale=0.3]{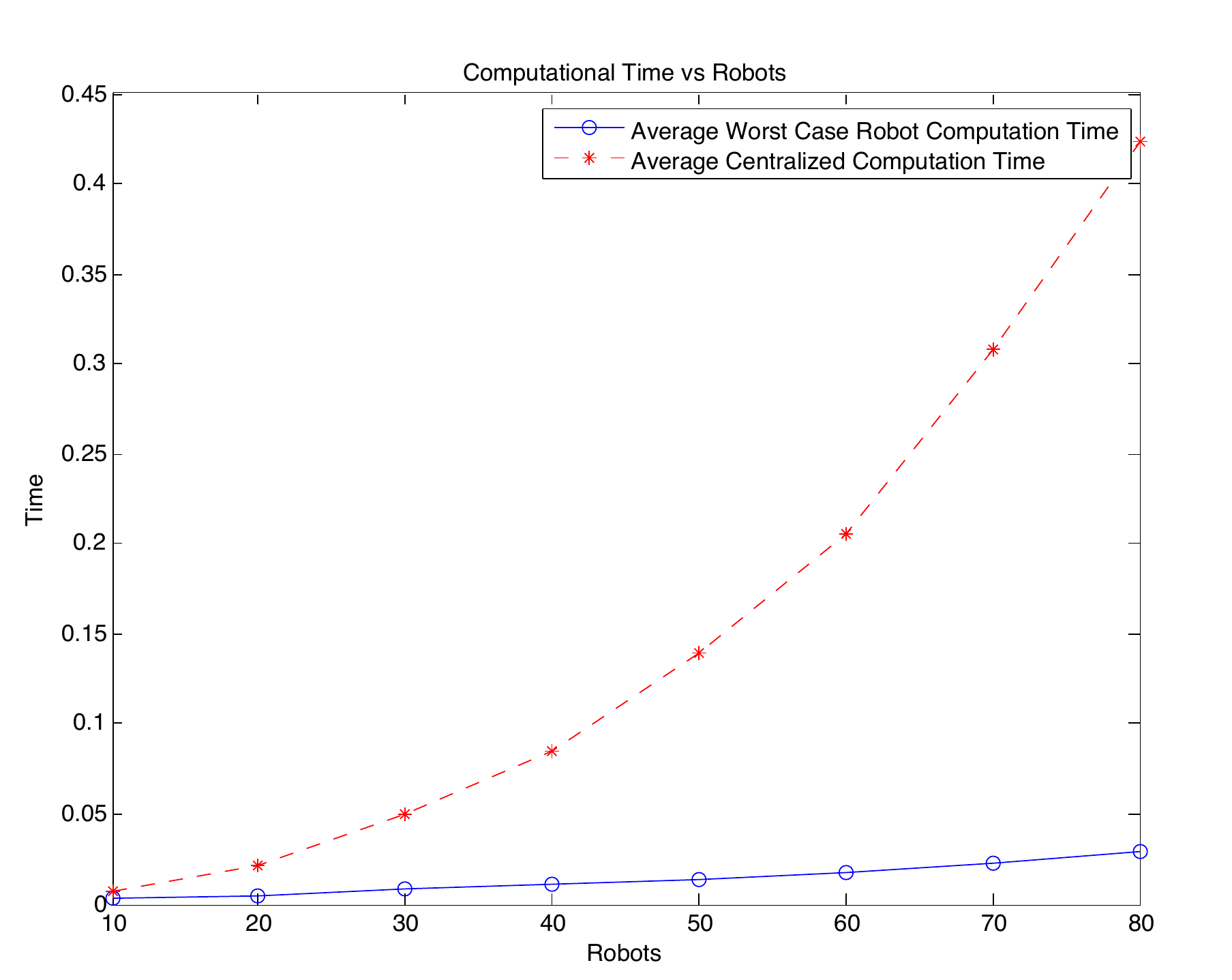}
\label{subfig_2}
}
\caption{Simulation experiments characterizing different aspects of the \textit{Distributed-Hungarian} algorithm.}
\label{fig_disthunginit}
\end{figure}

As seen in Figure \ref{subfig_1}, the average number of iterations required for convergence is well under $\mathcal{O}(r^3)$ (worst-case bound). Moreover, to qualify the computational load on a robot at any given time, on each run of the \textit{Distributed-Hungarian} algorithm, we observed the iteration that took the maximum computational time (across all robots), and plotted the average of all 20 runs, against the computational time of the corresponding centralized $\mathrm{Hungarian\_Method}$ (Figure \ref{subfig_2} ). The experiments support the applicability of the \textit{Distributed-Hungarian} algorithm to varying sized teams of mobile robots in practice. As such, we proceed to describe the motivating application in this paper, which provides an intuitive and immediate testbed for demonstrating our proposed algorithm.

\section{A Motivating Application:  Dynamic Spatio-Temporal Multi-Robot Routing}
\label{sec_motiv_eg}

To demonstrate the distributed algorithm central to this paper, we consider
multi-robot routing as a motivating application. In particular, we consider a
special class of routing problems called ``spatio-temporal routing
problems'', previously introduced in \cite{chopra:pacc2014a}. In such routing
problems, each target is associated with a \textit{specific time instant} at
which it requires servicing. Additionally, each target, as well as each robot,
is associated with one or more skills, and a target is serviceable by a robot
\textit{only if} the robot has a skill in common with the skill set of that
target (or in other words, the robot is authorized to service the target). As
shown in \cite{chopra:pacc2014a}, task assignment is a convenient framework to
attack such spatio-temporal routing problems. %

In this section, we briefly discuss the extension of our previous work in \cite{chopra:pacc2014a} on
spatio-temporal routing from a static, centralized solution to its dynamic and
distributed counterpart. More specifically, for a series of spatio-temporal
requests, the robots cooperatively determine their routes online by using the
\textit{Distributed-Hungarian} algorithm.
The online scheme we propose is based on a simple, widely applied idea, where assignments are solved 
iteratively between consecutive time instants.
This scheme provides an effective framework for incorporating
our distributed algorithm towards dynamic spatio-temporal
routing. We further illustrate this setup in a musical environment through a
novel, ``multi-robot orchestral'' framework. We acknowledge that we are not trying to find the most ``optimal '' or ``efficient'' solution to the routing problem itself. Instead, we are interested in demonstrating the applicability of our proposed algorithm, in a practical, and intuitive setting.

\begin{figure}
\begin{center}
\includegraphics[scale=0.3]{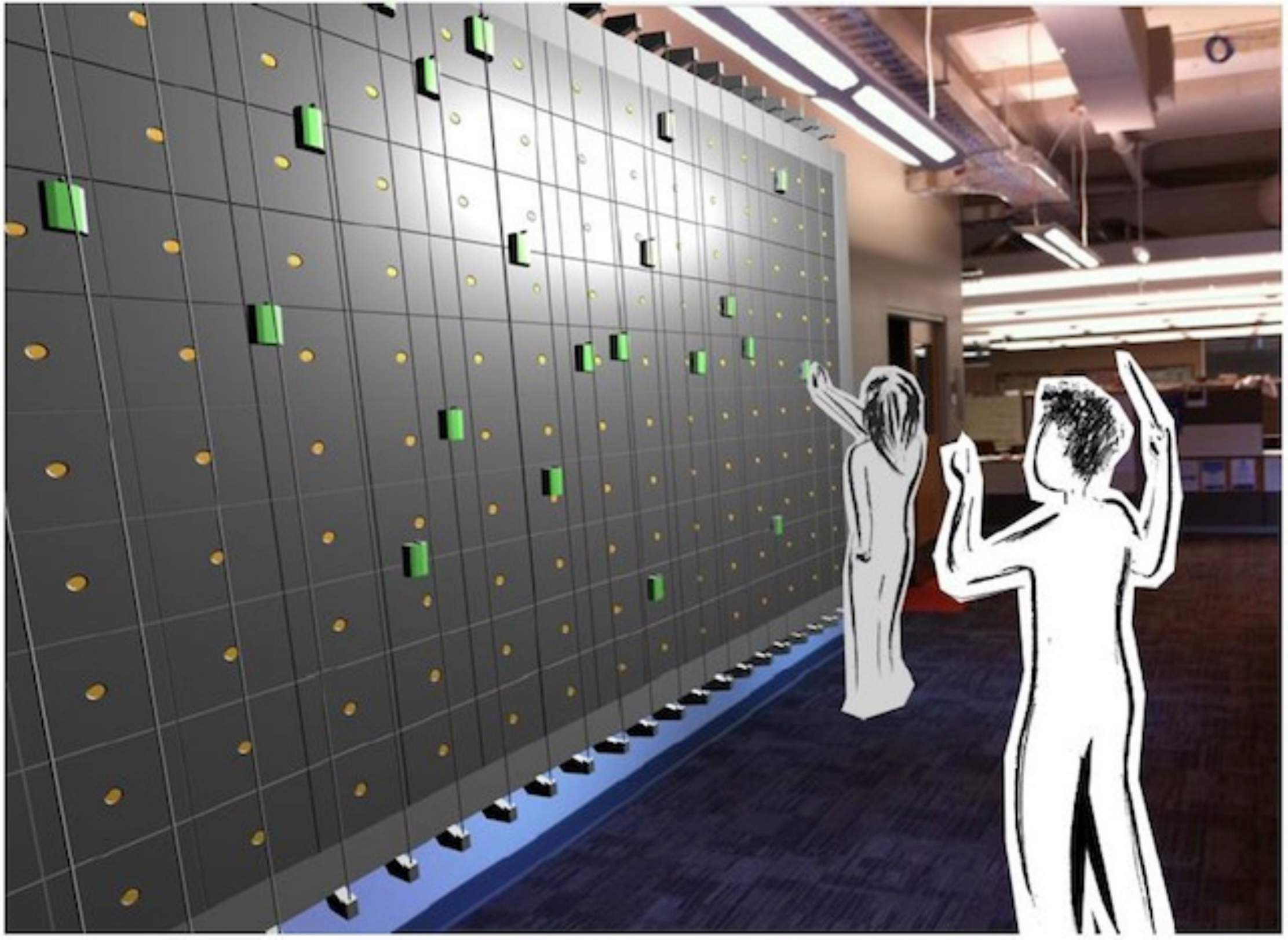}    %
\caption{A rendering of the \textit{Robot Orchestral Floor} concept (viewed in the picture, as a vertical wall instead of a floor).}  %
\label{fig_robotmusicwall}                                %
\end{center}                                 %
\end{figure}

Each robot can play one or more instruments (essentially, a piano, a guitar and
drums), and a piece of music can be interpreted as a series of spatio-temporal
requests on the so-called \textit{Robot Orchestral Floor} - a music surface
where planar positions correspond to distinct notes of different instruments
(see Figure \ref{fig_robotmusicwall} for an illustration).

A user (acting similar to a ``conductor'') can change the piece of
music in real-time, while the robots adapt their routes accordingly, to
incorporate the changes. The user interacts with the team of robots by means of
a tablet interface through which it broadcasts the spatio-temporal
requests.\footnote{This paper considers the generalized version of the routing
  problem that associates a \textit{set of skills or instruments} with each
  spatio-temporal request. However, for the purpose of musical demonstration on
  the orchestral floor, a \textit{single} skill (instrument) is associated with
  each request (w.l.o.g.).}

\subsection{Overview of the Methodology}
For convenience, we assume that the minimum time difference between two timed positions
is always greater than the time needed by the robots to solve an instance of the \textit{Distributed Hungarian} algorithm, and to reach their assigned positions.

\begin{figure}  
    \centering \subfloat[\textit{Future}
  portions of the routes of the robots (between time instants
  $t_{i+2}$ to $t_{i+3}$) are currently being determined, while the robots
  execute the portion of the routes that have already been determined (up until 
  $t_{i+2}$).]{\includegraphics[scale=0.25]{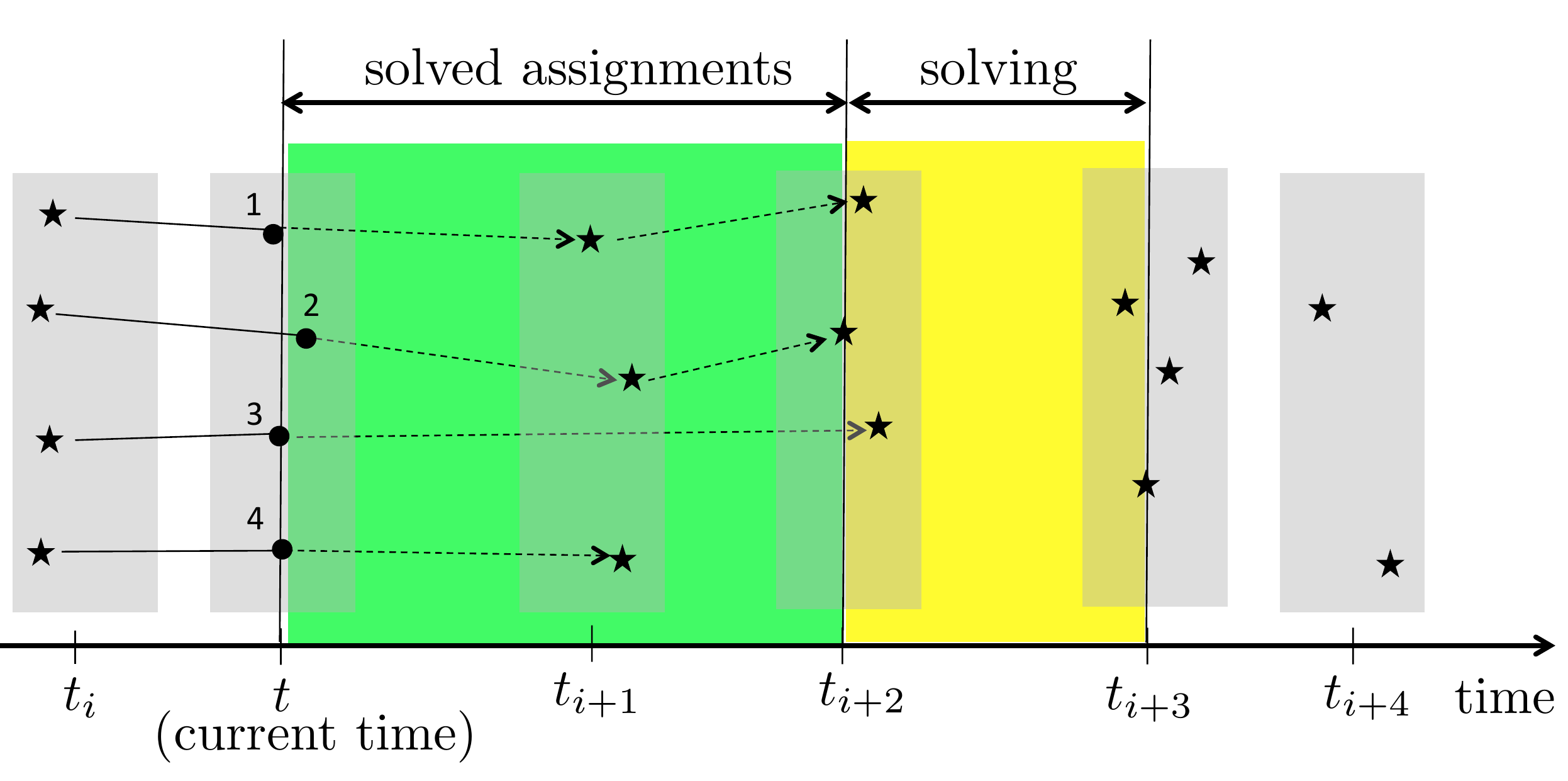}
      \label{fig_dynrob} } \\ \vspace{1em}
      
    \centering \subfloat[Case 1: If at the current time $t$, a modification is
    received, specified at a time instant in the depicted time interval
    (yellow), \textit{all} robots retain their routes up until
    $t_{i+1}$ and begin recalculating their routes, from their positions at
    $t_{i+1}$
    onwards.]{\includegraphics[scale=0.25]{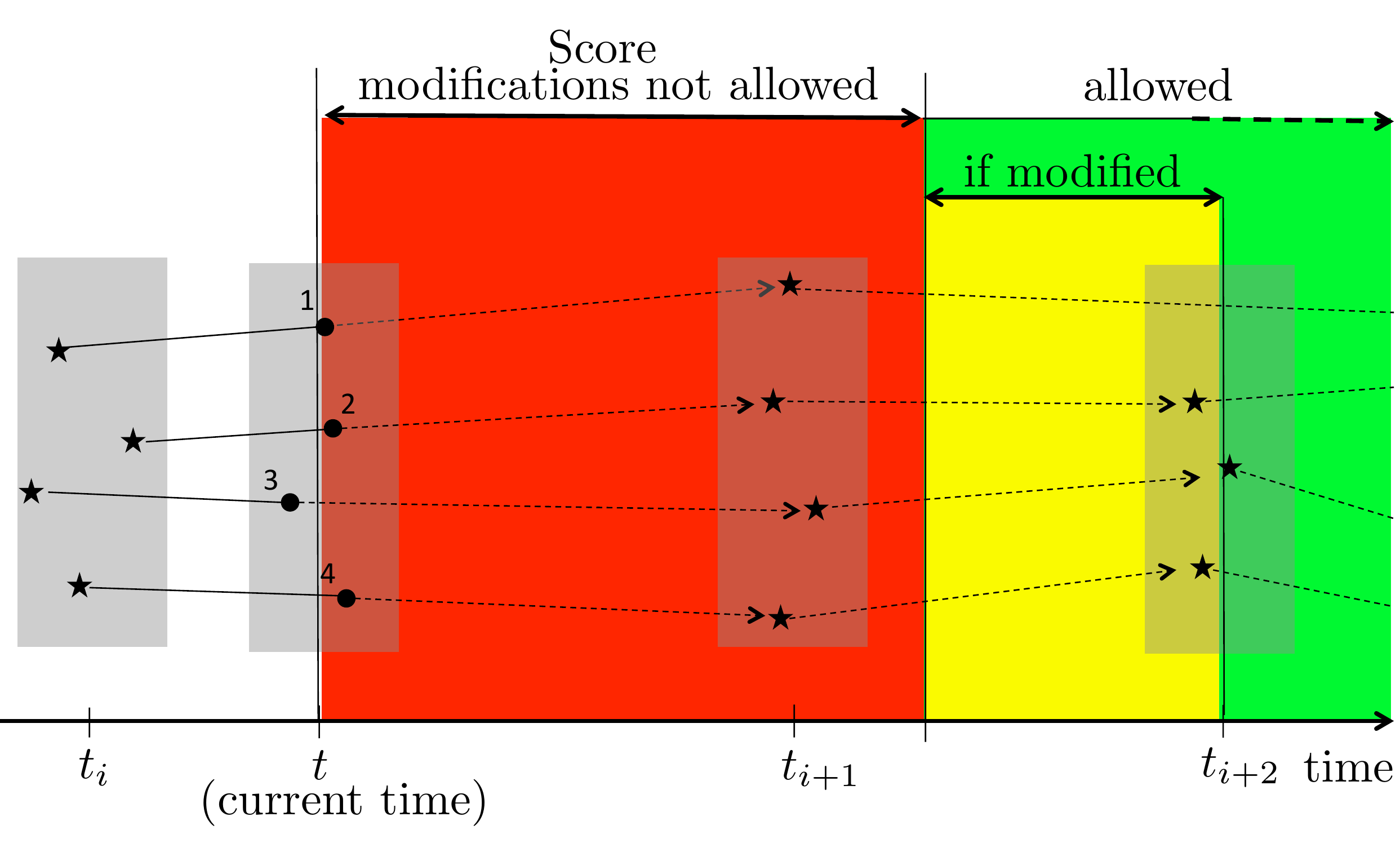}
      \label{fig_dyn1} } \\ \vspace{1em}
     
     \centering \subfloat[Case 2: If at the
    current time $t$, a modification is received, specified at a time instant in
    the depicted time interval (yellow), \textit{all} robots begin recalculating
    their routes, from their current positions at time $t$
    onwards.]{\includegraphics[scale=0.25]{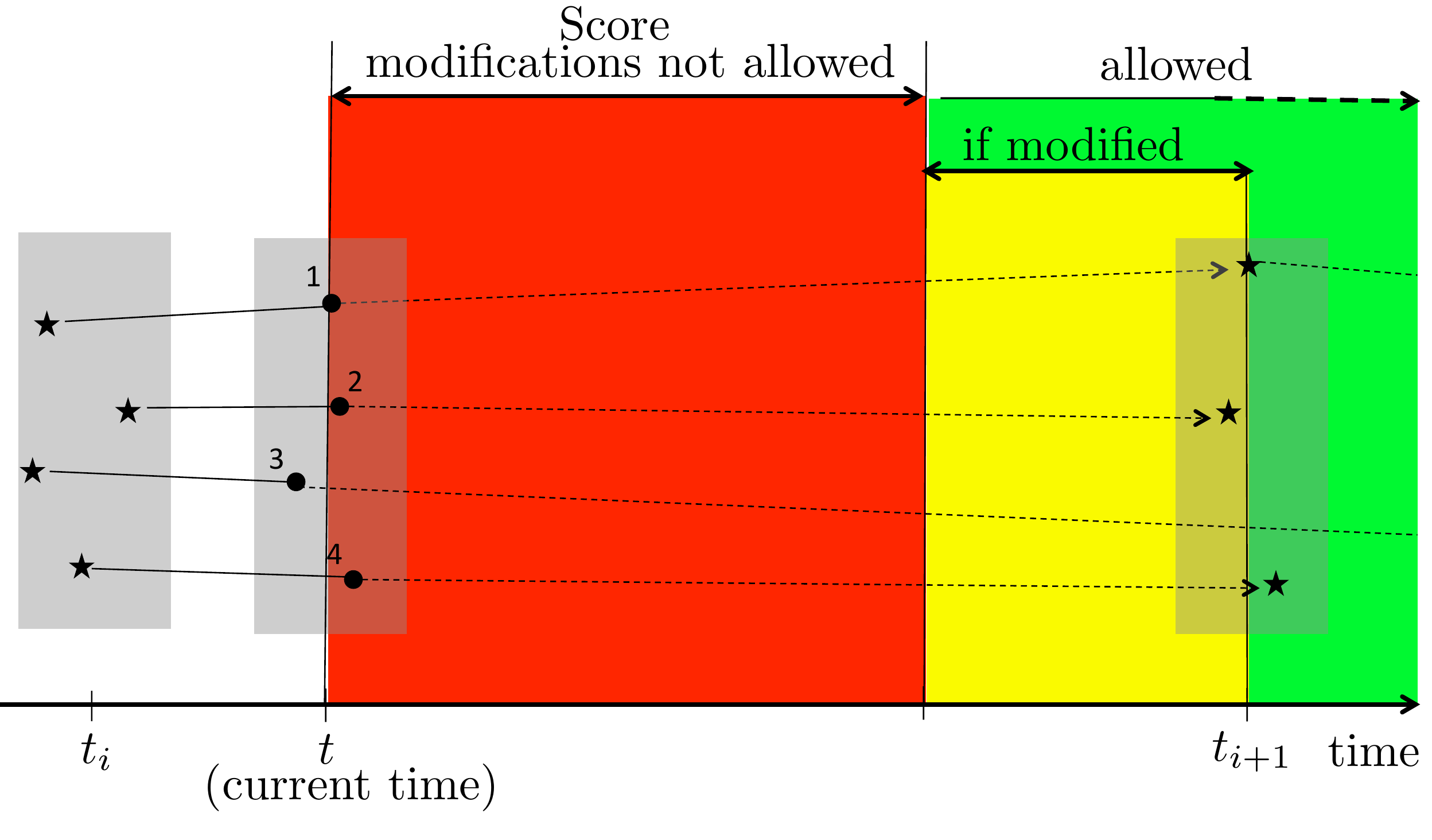}
      \label{fig_dyn2}} \\ \vspace{1em}
     
     \centering \subfloat[Case 3: If at the
    current time $t$, a modification is received, specified at a time instant in
    the depicted time interval (yellow), \textit{only} robots $1,2$ and $3$
    retain their routes up until $t_{i+1}$. The robots begins
    recalculating their routes from their positions at $t_{i+1}$ onwards,
    however robot $4$ begins recalculating its route from its current position
    at time $t$
    onwards.]{\includegraphics[scale=0.25]{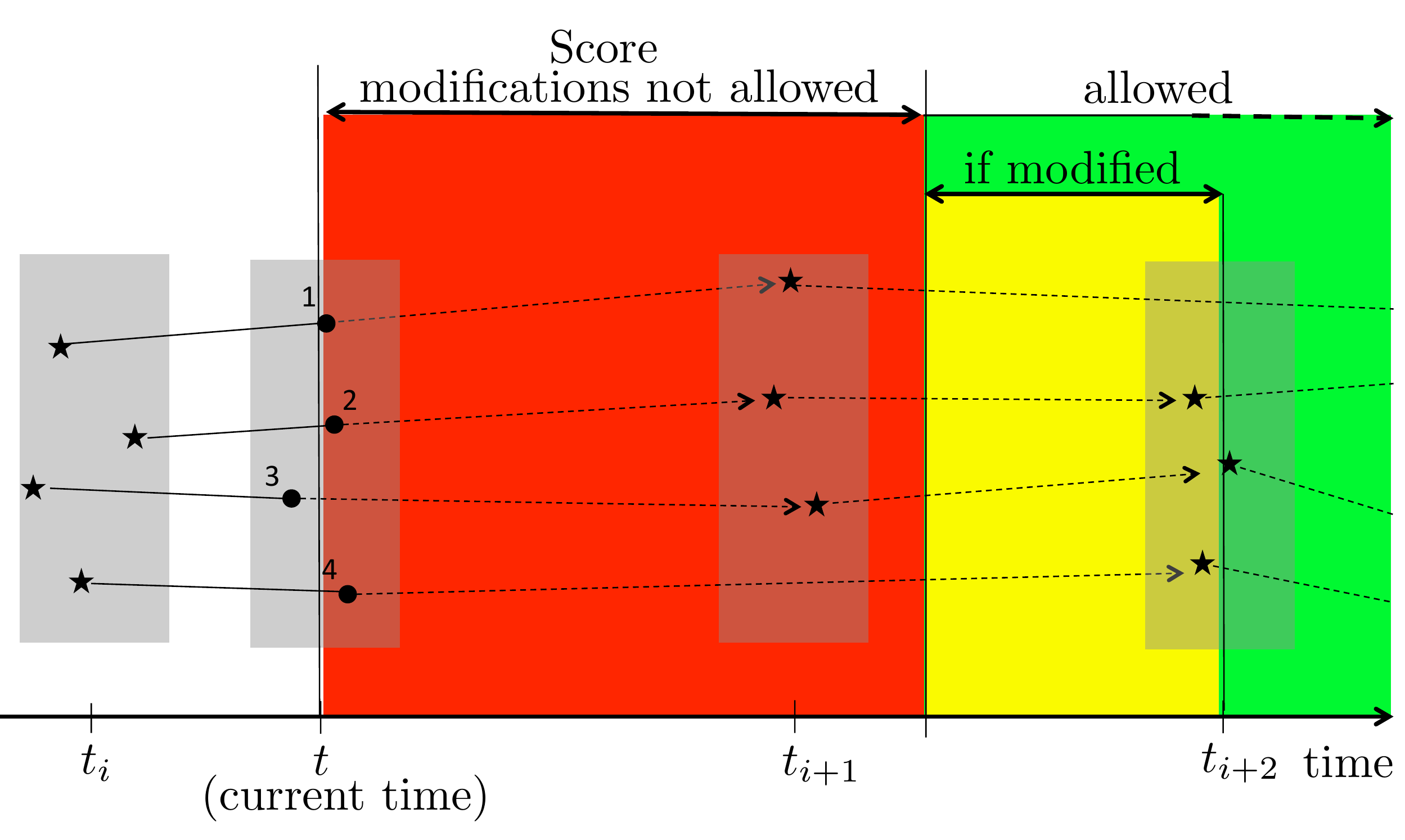}
      \label{fig_dyn3}}
      
    \caption{Robots
      (circles) at time $t$, and timed positions (stars) grouped in grey boxes
      representing specific time instants.}
    \label{fig_dynusr}
  \end{figure}

In the following paragraph, we put forth the two key ideas, central to the
scheme described above. 

\begin{enumerate}[-]
\item{\textit{Distributed Aspect:}} Given a Score, the robots determine routes
  by iteratively solving assignments using the \textit{Distributed-Hungarian} algorithm, between successive time instants. Each instance of an assignment can be formulated as an unbalanced Linear Sum Assignment Problem (0-1 linear program) \cite{burkard:1999a}, using a mapping $l(p,\alpha)$ as follows, 
 \vspace{0.5em} 
  \paragraph*{For the consecutive time instants $t_{i}$ to $t_{i+1}$, given the quintuple ($Sc_i, R, M_{pos},M_{rbt}$), and the function $P_{rbt}: R \rightarrow \mathbb{R}^2$, denoting the planar positions of the robots\footnote{Under the iterative scheme, each robot's planar position is a \textit{previously assigned} timed position at some time instant $t_j \leq t_{i}$ (unless $i=0$).}, find $l$ such that} 
\begin{align}
\min_{l} \sum_{p\in R} \sum\limits_{\alpha \in 1}^{|Sc_i|} ||P_{i,\alpha} - P_{rbt}(p)||\,l(p,\alpha)
\label{eqn_lp}
\end{align}
subject to:
\begin{align}
& l(p,\alpha) \in \{0,1\} \label{eqn_lp1} \\
& \sum_{p\in R}  l(p,\alpha) = 1,\quad \forall\, \alpha \in \{1,...,|Sc_i|\} \label{eqn_lp2}\\
& \sum_{\alpha \in \mathcal{A}_i}  l(p,\alpha) \leq 1, \quad \forall\, p\in R  \label{eqn_lp3} \\ 
&  l(p,\alpha) = 1 \Rightarrow M_{rbt}(p) \cap M_{pos}((P_{i,\alpha},t_i)) \neq \emptyset \label{eqn_lp4} 
\end{align}
where $l(p,\alpha)$ represents the individual assignment of robot $p\in R$ to timed position $(P_{i,\alpha},t_i) \in Sc_i$, and is $1$ if the assignment is done, and $0$ otherwise. 

 \vspace{0.5em}
  Note that for employing the \textit{Distributed-Hungarian} algorithm, we view the above stated LSAP in graph theoretic terms, as the equivalent \textit{Minimum Weight Bipartite Matching Problem (P)} from \ref{LSAP} (we assume that the
  underlying time-varying communication graph, induced as the robots execute
  their paths, satisfies Assumption \ref{ass_comm}). Moreover,
  the robots solve assignments between \textit{future} consecutive time
  instants, while \textit{simultaneously} executing routes that they have
  already determined (see Figure \ref{fig_dynrob}). 

\begin{figure*}
    \begin{center}
      \includegraphics[width = 15cm, height =
      8cm]{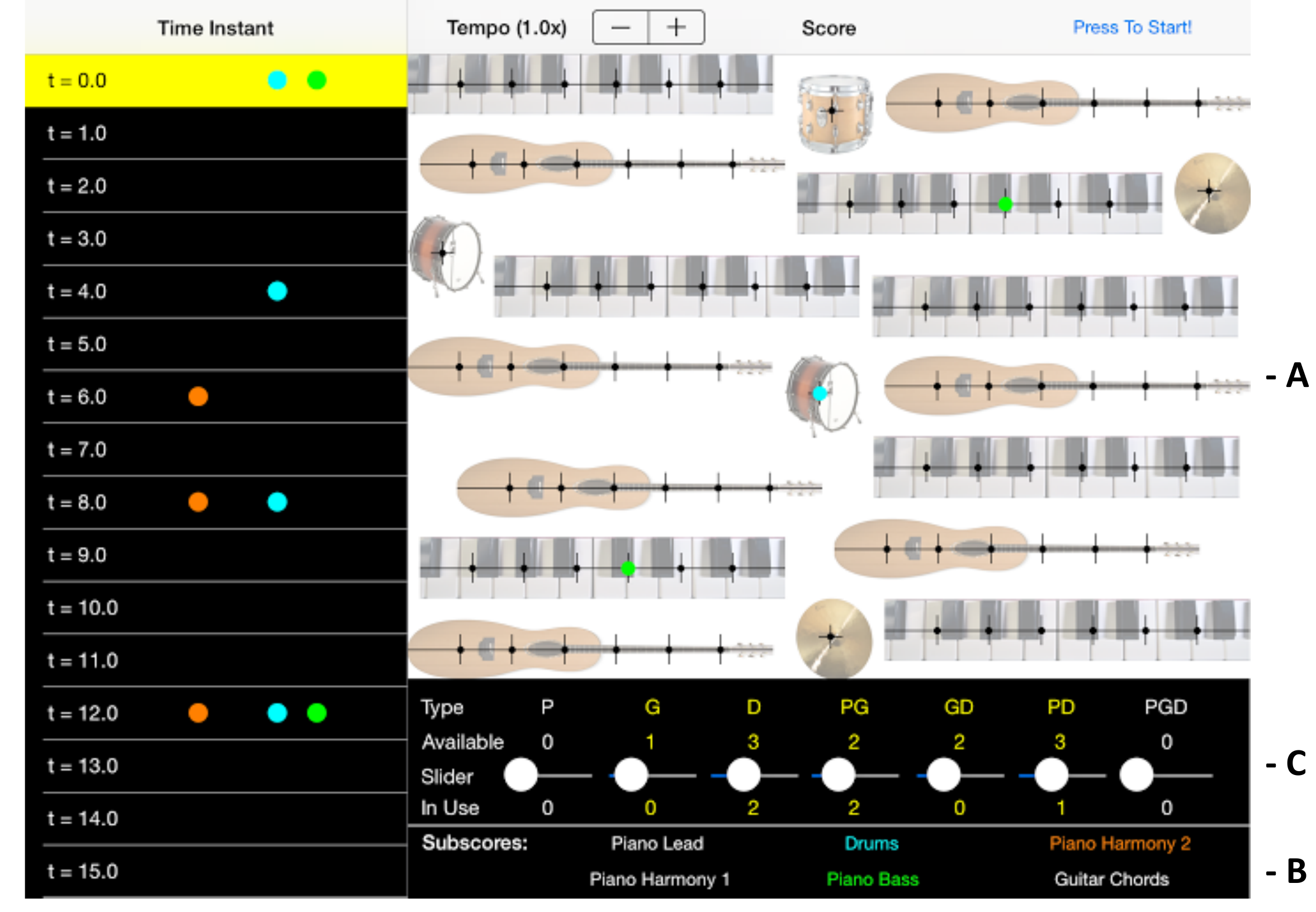} %
      \caption{The tablet interface for modifying and broadcasting changes to
        the Score. \textbf{A}: A simulated example of the \textit{Robot
          Orchestral Floor} comprising of unique positions corresponding to
        piano and guitar notes, as well as drum beats. \textbf{B}: Pre-defined
        sub-scores for an instrumental version of ``The Final
        Countdown''. \textbf{C}: A feature for either selecting, or setting the
        available number of robots $(R, M_{rbt})$.}  %
      \label{fig_iPad} %
    \end{center} %
  \end{figure*}

 \vspace{0.5em}
\item{\textit{Dynamic Aspect:}} 
A user can dynamically modify the Score as follows:
\begin{enumerate}[i]
\item{Add a timed position with a corresponding skill set (add a note of an instrument (piano or guitar), or a beat of a drum, to be played at a particular time instant).}
\item{Remove a timed position (remove a note of an instrument (piano or guitar), or a beat of a drum, from a particular time instant).}
\item{Modify the skill set of a timed position (substitute the instrument of a note (from a piano to a guitar, or vice versa), or replace one kind of drum with another, at a particular time instant).}
\end{enumerate}

Since the routes of the robots are determined through piece-wise assignments between robot positions and timed
positions at successive time instants in the Score, the instant a dynamic
modification is received, each robot chooses the time instant in the Score up
until which, its previously-determined routes need not be modified, and begins
recalculating its route \textit{from such a time instant onwards} (while
executing its trajectory on the previously determined route). In Figures
\ref{fig_dyn1} - \ref{fig_dyn3}, we provide examples of three
different cases that can occur, when a particular dynamic modification is
received. 

 \vspace{0.5em}
As mentioned previously, a user issues dynamic modifications through a user-interface. We assume that such an interface has knowledge of the Score and the robots, i.e. ($Sc_i, R, M_{pos},M_{rbt}$), and is able to broadcast the modifications to the robots in real time. 
Since we assume that the timed positions in the Score are sufficiently apart in time, the interface does not allow a user to modify the Score unless the modification occurs after a pre-specified (conservative) time duration (depicted by the red regions in Figures \ref{fig_dyn1} - \ref{fig_dyn3}). Moreover, the interface does not allow modifications that violate feasibility (in that, the available number of robots do not fall short), as per our results in \cite{chopra:pacc2014a}\footnote{Velocity constraints are not considered here, since
    the user has no knowledge of the positions of the robots, and hence, cannot
    ascertain feasibility in that respect.}.
\end{enumerate}

\subsection{The Multi-Robot Orchestra}
\label{sec_impl}
\begin{figure*}
\begin{center}
\subfloat[A user interacting with simulated robots through the orchestral floor tablet interface.]{
\includegraphics[scale = 0.33]{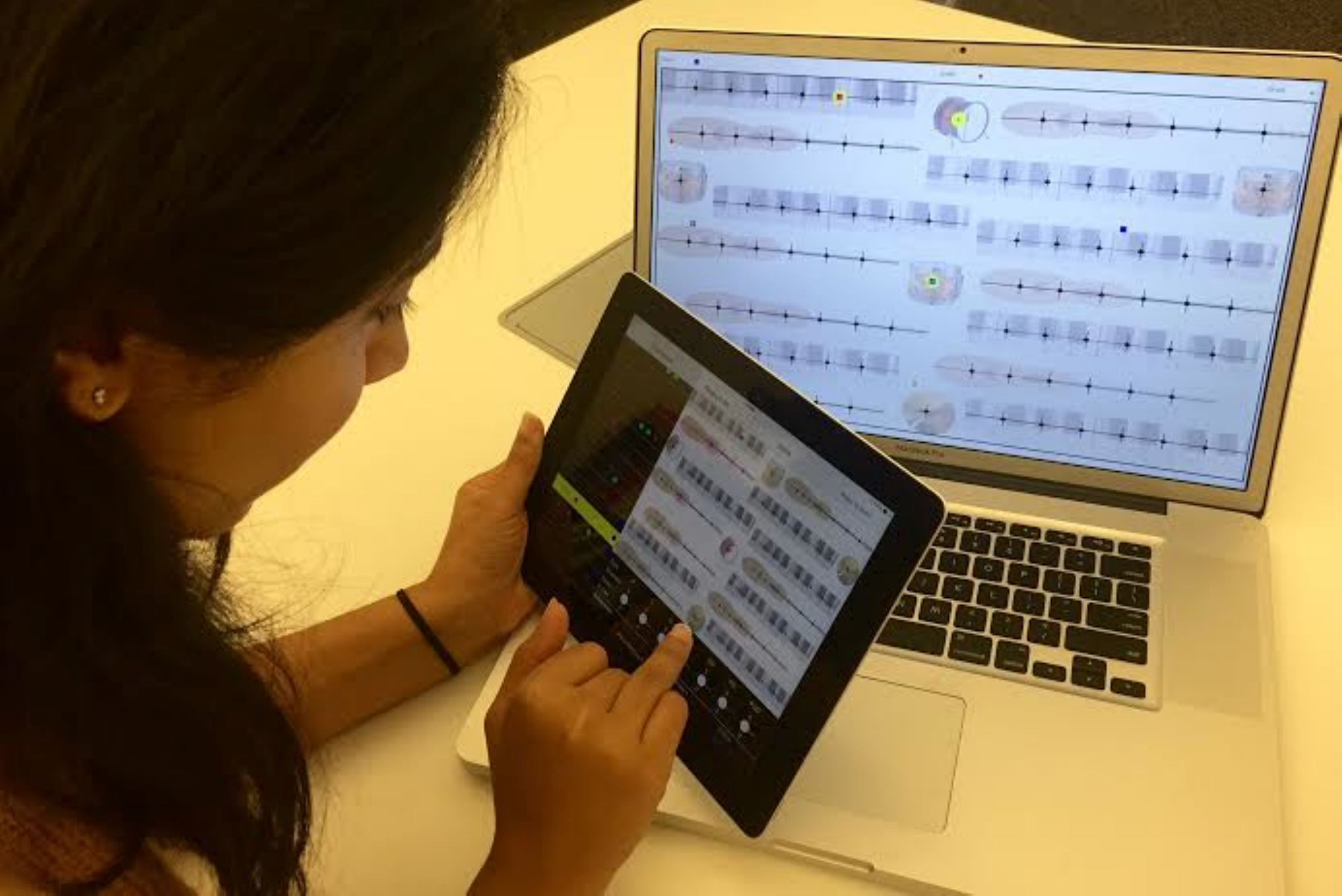}
}
\hfill 
\subfloat[An instance of simulated robots (geometric shapes - diamonds, circles and squares), performing the Score by executing dynamic spatio-temporal routing (the black lines denote the underlying dynamic communication network, required for a distributed implementation).]{
\includegraphics[scale = 0.39]{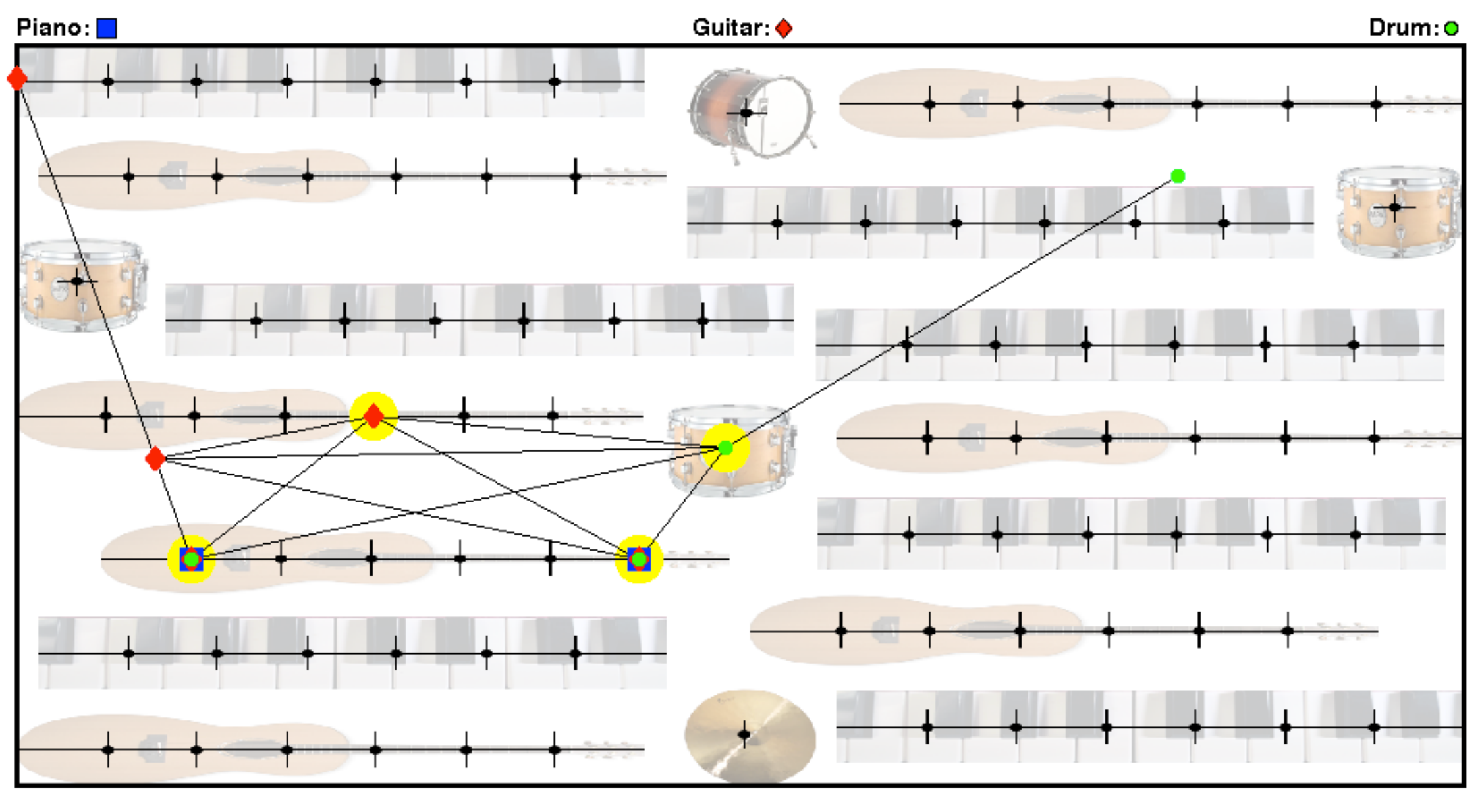}
}   
\hfill
\subfloat[A user interacting with actual robots through the orchestral floor tablet interface.]{
\includegraphics[scale=0.35]{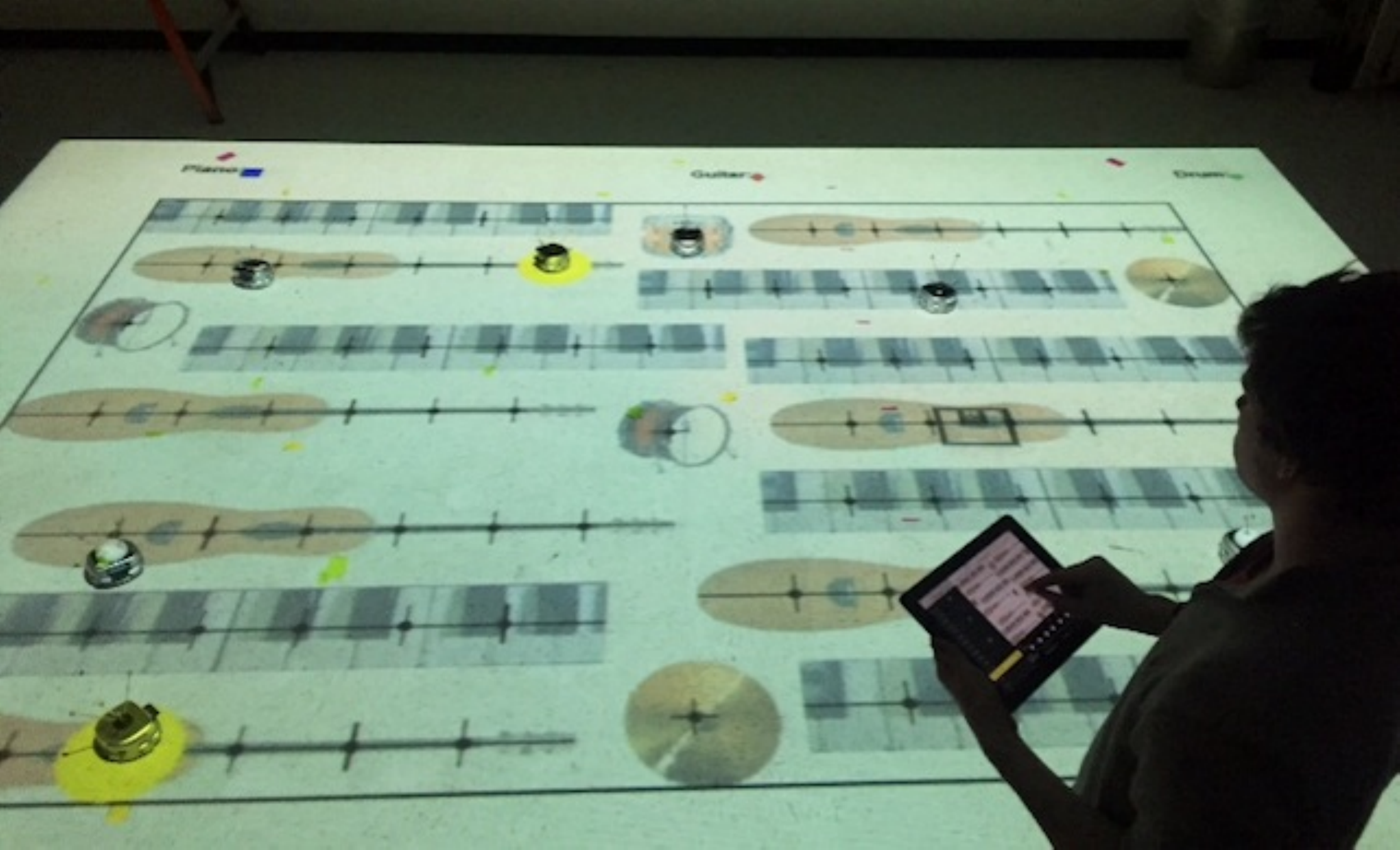}
\label{subfig_sim1}
}
\hfill
\subfloat[An instance of actual robots, performing the Score by executing dynamic spatio-temporal routing.]{
\includegraphics[scale=0.35]{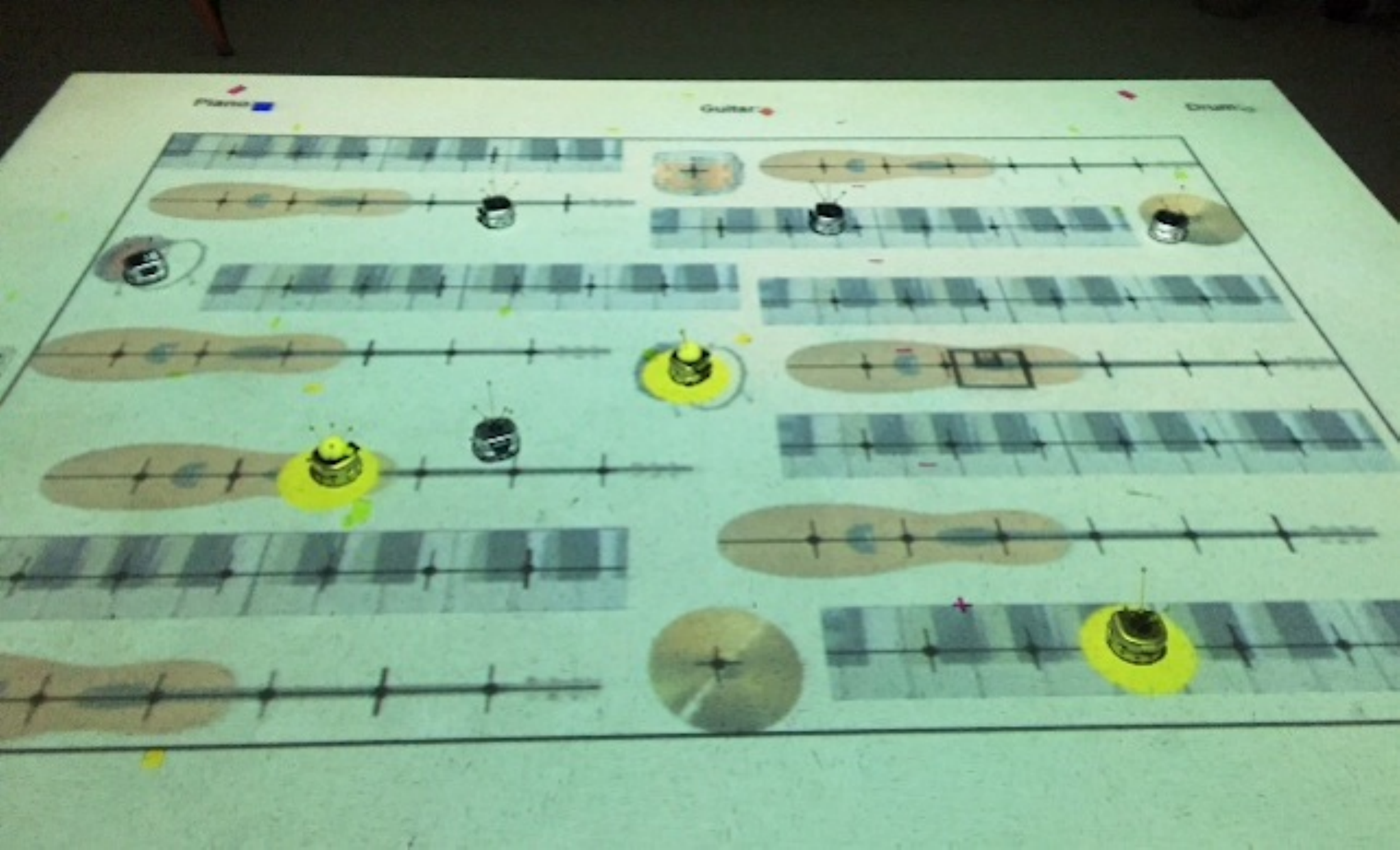}
\label{subfig_hw1}
}
\caption[]{Simulation (\url{https://www.youtube.com/watch?v=L-PSPy9O_BE}), and hardware (\url{https://www.youtube.com/watch?v=z7SiivWvZLc}) implementation of the multi-robot orchestra, performing  ``The Final Countdown'' on the \textit{Robot Orchestral Floor}, with a user ``conducting'' (modifying) the Score through the tablet interface. In both cases, when a particular robot reaches a timed position on the orchestral floor, its is highlighted by a light (yellow) circle, and the corresponding sound of the note/beat is generated.}  %
\label{fig_iPad_sim}                                %
\end{center}                                 %
\end{figure*}

In this section, we apply the theory developed so far, towards enabling multiple robots to execute different musical pieces (presented to them as Scores). To this end, we simulated a version of the \textit{Robot Orchestral Floor} in MATLAB, instrumented to include piano, guitar and drum sounds (see A in Figure \ref{fig_iPad}). In addition to the simulated floor, we developed a graphical user-interface (GUI) (Figure \ref{fig_iPad}) that allows a user to create, and administer changes to a Score on the simulated floor. The user-interface is developed on a tablet, that broadcasts the changes issued by a user, to the robots executing the Score.

For convenience, we created beforehand, a heterogeneous Score comprising of piano and guitar notes, and drum beats associated with the popular song ``The Final Countdown'' by the Swedish band ``Europe''. We divided the Score into multiple single-instrument sub-scores. For instance, we separated the piano notes into individual sub-scores corresponding to the piano lead, piano bass, and second and third harmonies (see B in Figure \ref{fig_iPad}). The motivation behind the creation of such sub-scores was to enable a user to ``add, delete or modify'' the Score through these structures, in an intuitive and immediately recognizable manner. In addition to the sub-scores, we included the option of adding and removing individual timed positions, and switching instruments (pianos to guitars and vice versa, drums from one type to another). 

To execute an example of dynamic spatio-temporal routing, either the user selects, or is given the number of robots available for use (see C in Figure \ref{fig_iPad}). Moreover, the user creates an initial Score using the iPad interface. We assume that all robots are initially positioned along a vertical edge of the floor. Once the user hits the start button, the iPad broadcasts this Score to the team of robots. From this point onwards, the routes of the robots are determined and executed in real time, while the iPad broadcasts changes to the Score, as and when a user decides to modify it. 

We implemented the multi-robot orchestra in both simulation and hardware environments (Figures \ref{fig_iPad_sim}). The hardware implementation was conducted in the Georgia Robotics and Intelligent Systems Laboratory, where the indoor facility is equipped with a motion capture system which yields real time accurate data for all tracked objects, and an overhead projector is used for embedding algorithm/environment-specific information (see \cite{omidshafiei:2015a} for a similar hardware setup). We used Khepera III miniature robots by K Team as our hardware ground robots. In both cases (simulation and hardware), the instant a robot reached a timed position (a note of an instrument, or a beat of a drum, specified at a particular time instant) on the orchestral floor, it was encircled by a light (yellow) circle and the corresponding sound of the note (or beat in the case of drums) was generated. In this manner, we enabled multiple robots to perform a real-time rendition of ``The Final Countdown''.

\section{Conclusion and Future Directions}
This paper provides a distributed version of the \textit{Hungarian Method} for solving the well known Linear Sum Assignment Problem (LSAP). The proposed algorithm allows a team of robots to cooperatively compute the optimal solution to the LSAP, without any coordinator or shared memory. We prove that under a synchronous implementation, all robots converge to a \textit{common} optimal assignment within $\mathcal{O}(r^3)$ iterations. By running simulation experiments over multiple instances of the LSAP with varying problem sizes, we show that the average number of iterations for convergence is much smaller than the theoretic worst-case bound of $\mathcal{O}(r^3)$. Moreover, we show that the computational load per robot is minor in comparison to the centralized $\mathrm{Hungarian\_Method}$, since the robots perform only sub-steps of the centralized algorithm, at each iteration of the proposed algorithm. 

To demonstrate the theory developed in this paper, we extend our proposed algorithm to solving a class of ``spatio-temporal'' multi-robot routing problems, considered under a distributed and dynamic setting. In essence, the robots find online, sub-optimal routes by solving a sequence of assignment problems iteratively, using the proposed distributed algorithm for each instance. As a motivating application and concrete experimental test-bed, we develop the ``multi-robot orchestral'' framework, where spatio-temporal routing is musically interpreted as ``playing a series of notes at particular time instants'' on a so-called orchestral floor (a music surface where planar positions correspond to distinct notes of different instruments). Moreover, we allow a user to act akin to a ``conductor'', modifying the music that the robots are playing in real time through a tablet interface. Under such a framework, we perform simulations and hardware experiments that showcase our algorithm in a practical setting.  

An interesting future direction that we are currently exploring, is the interpretation of the Hungarian Method as a primal algorithm (as opposed to its native, dual form), and its subsequent redesign in a distributed setting. Essentially, the primal algorithm would provide a constantly improving \textit{feasible} assignment at every iteration. We hypothesize that such an algorithm, when redesigned in a distributed setting, could be more robust and simple than the current approach described in this paper, though slower to converge. Distributed implementations of both primal and dual versions of the Hungarian Method would yield interesting technical comparisons in the assignment literature.   

\small

 \normalsize
\end{document}